\documentclass[12pt]{article}
\usepackage{epstopdf,graphicx,epsfig,rotating}
\usepackage{rotating}

\usepackage{times}
\usepackage{epstopdf, graphicx}
\usepackage{amssymb,amsmath}
\usepackage{scicite}
\usepackage{ulem}

\def\be{\begin{equation}}
\def\ee{\end{equation}}
\def\bea{\begin{eqnarray}}
\def\eea{\end{eqnarray}}
\def\tol#1#2#3{\hbox{\rule{0pt}{15pt}${#1}^{+{#2}}_{-{#3}}$}}
\def \phn {\hphantom{0}}
\newenvironment{sciabstract}{%
\begin{quote} \bf}
{\end{quote}}

\renewcommand\refname{References and Notes}
\def \qso {SDSSJ0841$+$3921\,}
\def \fgqso {SDSSJ084158.47+392121.0}
\def \bgqso {SDSSJ084159.26+392140.0}
\def \rqso {FIRSTJ084158.6+392114.7}
\def \rphys {$R_\perp$}
\def \mrphys {R_\perp}
\def \mkms      {\rm\ km\ s^{-1}}
\def\mlnull{L_{\nu_{\rm LL}}}
\def \kpc       {{\rm\ kpc}}
\def \cgssb {{\rm\,erg\,s^{-1}\,cm^{-2}\,arcsec^{-2}}}
\def\sci#1{{\; \times \; 10^{#1}}}
\def \arcsec    {^{\prime\prime}}
\def \sqarcsec    {\box^{\prime\prime}}
\def \msol      {{\rm\ M}_\odot}
\def \kms            {{\rm km~s}^{-1}}
\def \tkms    {$\kms$}

\newcommand\ion[2]{#1$\,${\scshape{#2}}} 
\newcommand{\mnhi}{N_{\rm HI}}
\newcommand{\nhi}{$N_{\rm HI}$}
\def\cm#1{\, {\rm cm^{#1}}}
\def \cgssb {{\rm\,erg\,s^{-1}\,cm^{-2}\,arcsec^{-2}}}
\newcommand{\mlya}{${\rm Ly\alpha}$}

\newcommand{\lya}{{\rm Ly}\alpha}

\newcommand{\oiii}{[\ion{O}{iii}]}
\newcommand{\mzfg}{z_{\rm fg}}

\newcommand{\mzagnw}{z_{\rm AGN2}}
\newcommand{\mzagnr}{z_{\rm AGN3}}

\newcommand{\zzfg}{2.0412}
\newcommand{\zzbg}{2.2138}
\newcommand{\zzagno}{2.05476}
\newcommand{\zzagnw}{2.058} 
\newcommand{\zzagnr}{2.050} 

\def\aap{A \& A}
\def\aj{AJ}
\def\apj{ApJ}
\def\apss{Ap\&SS}
\def\apjl{ApJL}
\def\apjs{ApJS}
\def\apjsupp{ApJS}
\def\araa{ARAA}
\def\mnras{MNRAS}
\def\nat{Nature}
\def\pasp{PASP}
\def\prd{PhRvD}
\def\aapr{A\&A Rev.}
\def\physrep{Physics Reports}
\def\nar{New A Rev.}
\def\aaps{A\&AS}

\newcommand{\nebasz}{$37\arcsec$}
\newcommand{\neblsz}{$310\,$kpc}
\newcommand{\neblum}{2.1\times 10^{44}\,{\rm erg\,s^{-1}}}
\newcommand{\sbulim}{5.1\times 10^{-18}} 
\newcommand{\sbannu}{1.3\times 10^{-17}} 
\newcommand{\sbsig}{1.7\times 10^{-18}} 

\newcommand{\mxaoff}{18\arcsec} 
\newcommand{\mxloff}{150} 

\newcommand{\numlae}{75} 

\newcommand{\nhval}{20.4}  
\newcommand{\nhival}{19.2}  
\newcommand{\nhsig}{20.4 \pm 0.4}  
\newcommand{\nvolmn}{2.0}  
\newcommand{\uval}{-2.7 \pm 0.3}  
\newcommand{\xfrac}{-1.2 \pm 0.3}  
\newcommand{\rnotval}{29.3}  
\newcommand{\rnoterr}{4.9}  
\newcommand{\gammaval}{1.5}  

\newcommand{\figimages}{1}
\newcommand{\figcluster}{2}
\newcommand{\figslits}{3}
\newcommand{\figabs}{4}

\newcommand{\sfigchimaps}{S1}
\newcommand{\sfigagn}{S2}
\newcommand{\sfiglineratio}{S3}
\newcommand{\sfigcontour}{S4}
\newcommand{\sfigcorr}{S5}
\newcommand{\sfignearir}{S6}
\newcommand{\sfigabs}{S7}
\newcommand{\sfigqpq}{S8}
\newcommand{\sfigcldyHI}{S9}
\newcommand{\sfigcldyH}{S10}

\newcommand{\tabagn}{S1}
\newcommand{\tabagnphot}{S2}
\newcommand{\tablines}{S3}
\newcommand{\tablae}{S4}
\newcommand{\tabhzrg}{S5}
\newcommand{\tababs}{S6}

\newcounter{lastnote}

\title{\bf Quasar Quartet Embedded in Giant Nebula Reveals
  Rare Massive Structure in Distant Universe}

\author{Joseph F. Hennawi$^{1,\ast}$, J. Xavier Prochaska$^{2}$,Sebastiano Cantalupo$^{2,3}$,\\
  Fabrizio Arrigoni-Battaia$^{1}$\\
  \normalsize{$^{1}$Max-Planck-Institut f\"ur Astronomie, K\"onigstuhl, Germany} \\
  \normalsize{$^{2}$University of California Observatories-Lick Observatory, UC Santa Cruz, CA}\\
  \normalsize{$^{3}$ETH Zurich, Institute of Astronomy, Zurich, Switzerland}\\
  \\
  \normalsize{$^\ast$To whom correspondence should be addressed; E-mail: joe@mpia.de}
}
\date{}

\begin{document}

\maketitle
\begin{sciabstract}  

All galaxies once passed through a hyper-luminous quasar phase powered by 
accretion onto a supermassive black hole. But because these episodes 
are brief, quasars are rare objects typically separated by 
cosmological distances.  In a survey for Lyman-$\alpha$ emission at
redshift $z\approx 2$, we discovered a physical association of four 
quasars embedded in a giant nebula.  Located within a substantial overdensity
of galaxies, this system is likely the progenitor of a massive galaxy  
cluster.  The chance probability of finding a quadruple-quasar is
estimated to be $\sim 10^{-7}$, implying a physical connection between
Lyman-$\alpha$ nebulae and the locations of rare protoclusters. Our 
findings imply that the most massive structures in the distant 
Universe have a tremendous supply ($\simeq 10^{11}$ solar masses) of cool
dense (volume density $\simeq 1\,{\rm cm}^{-3}$) gas, in conflict with 
current cosmological simulations.
\end{sciabstract}

Cosmologists do not fully understand the origin of supermassive black
holes (SMBHs) at the centers of galaxies, and how they relate to the
evolution of the underlying dark matter, which forms the backbone for
structure in the Universe. In the current paradigm, SMBHs grew in
every massive galaxy during a luminous quasar phase, making distant
quasars the progenitors of the dormant SMBHs found at the centers of
nearby galaxies.
Tight correlations between
the masses of these local SMBHs and both their host galaxy
\cite{mtr+98} and dark matter halo masses \cite{ferrarese02} support
this picture, further suggesting that the most luminous quasars at
high-redshift should reside in the most massive galaxies.  It has 
also been proposed that quasar activity is triggered by the frequent
mergers that are a generic consequence of hierarchical
structure
formation\cite{Bahcall97,HR93}.
Indeed, an excess in the number of
small-separation binary quasars \cite{Djor91,BINARY}, as
well as the mere existence of a handful of quasar triples
\cite{Djor07,farina13}, support this hypothesis.  If quasars are
triggered by mergers, then they should preferentially occur in rare
peaks of the density field, where massive galaxies are abundant and the frequency of
mergers is highest \cite{LaceyCole93}.

Following these arguments, one might expect that at the peak of their
activity $z \sim 2-3$, quasars should act as signposts for
protoclusters, the progenitors of local galaxy clusters and the most
massive structures at that epoch.  However, quasar clustering
measurements \cite{trainor12,White12} indicate that quasar
environments at $z \sim 2-3$
are not extreme: these quasars are hosted by dark matter halos with
masses $M_{\rm halo}\sim 10^{12.5}\msol$ (where $\msol$ is the mass of
the Sun), too small to be the progenitors of local clusters
\cite{Fanidakis13}. But the relationship between quasar activity and
protoclusters remains unclear, owing to the extreme challenge of
identifying the latter at high-redshift.  Indeed, the total comoving
volume of even the largest surveys for distant galaxies at $z\sim 2-3$
is only $\sim 10^{7}\,{\rm Mpc}^3$, which would barely contain a
single rich cluster locally.

%
Protoclusters have been discovered around a
rare-population of active galactic nuclei (AGN) powering large-scale
radio jets, known as high-redshift radio galaxies (HzRGs)
\cite{Venemans07}.  The HzRGs routinely exhibit giant $\sim 100\,$ kpc
nebulae of luminous Ly$\alpha$ emission $L_{\rm Ly\alpha} \sim
10^{44}\,{\rm erg\,s^{-1}}$.
Nebulae of comparable
size and luminosity have also been observed in a distinct
population of objects known as `Ly$\alpha$ blobs' (LABs)
\cite{francis96,Steidel00}.
The LABs are also frequently
associated with AGN activity
\cite{dey05,overzier13,prescott15}, although lacking
powerful radio jets, and
appear to reside in
overdense, protocluster environments\cite{Steidel00,prescott08,Yang09}.
Thus
among the handful of protoclusters \cite{Chiang13} known, most appear to share
two common characteristics: the presence of an active SMBH and a giant
Ly$\alpha$ nebula.


We have recently completed a spectroscopic search for extended
$\lya$\ emission around a sample of 29 luminous quasars at $z \sim 2$,
each the foreground (f/g) member of a projected quasar pair
\cite{QPQ4}.  Analysis of spectra from the background (b/g) members in
such pairs reveals that quasars exhibit frequent absorption from a
cool, metal-enriched, and often optically thick medium on scales of
tens to hundreds kpc \cite{QPQ1,QPQ2,QPQ3,QPQ4,QPQ5,QPQ6,QPQ7}.  The
UV radiation emitted by the luminous f/g quasar can, like a
flashlight, illuminate this nearby neutral hydrogen, and power a
large-scale Ly$\alpha$-emission nebula, motivating our search. 
By construction, our survey selects for exactly the two criteria which
seem to strongly correlate with protoclusters: an active SMBH 
and the presence of a large-scale
Ly$\alpha$ emission nebula.

Of the 29 quasars surveyed, only \qso exhibited
extended large-scale ($\gtrsim 50$\,kpc) $\lya$\ emission above our
characteristic sensitivity limit of $6\sci{-18}\cgssb$ ($2\sigma$).  
We designed a custom narrow-band filter tuned to the wavelength of
\mlya\ at the f/g quasar redshift $z=2.0412$ ($\lambda_{\rm center} =
3700$\AA, ${\rm FWHM}_\lambda = 33$\AA), and imaged the field with the
Keck/LRIS imaging spectrometer for 3hrs on UT 12 November 2012.  The
combined and processed images reveal \mlya\ emission from a giant
filamentary structure centered on the f/g quasar and extending
continuously towards the b/g quasar (see Fig.~\figimages).  This
nebulosity has an end-to-end size of \nebasz\ corresponding to
\neblsz\, and a total line luminosity $L_{\rm Ly\alpha}= \neblum$,
making it one of the largest and brightest nebula ever discovered.

The giant nebula is only one of the exceptional properties of
SDSSJ~0841$+$3921. Our images reveal three relatively compact
candidate Ly$\alpha$ emitting sources with faint continuum magnitudes
$V\simeq 23-24$, embedded in the Ly$\alpha$ filament and roughly
aligned with its major axis. Follow-up spectroscopy reveals that the
sources labeled AGN1, AGN2, and AGN3 are three AGN at the same
redshift as the f/g quasar (see right panel of Fig.~\figimages and
\cite{supp2}), making this system the only quadruple AGN known.  Adopting
recent measurements of small-scale quasar clustering \cite{Kayo12}, we
estimate that the probability of finding three AGN around a quasar
with such small separations is $\sim 10^{-7}$ \cite{supp3}.  Why then
did we discover this rare coincidence of AGN in a survey of just 29
quasars?
Did the giant nebula mark the location of a protocluster with dramatically
enhanced AGN activity?

To test this hypothesis, we constructed a catalog of
Ly$\alpha$-emitting galaxies (LAEs), and computed the cumulative
overdensity profile of LAEs around SDSSJ0841$+$3921, relative to the
background number expected based on the LAE luminosity
function\cite{ciardullo12} (see Fig.~\figcluster). To perform a
quantitative comparison to other giant Ly$\alpha$ nebulae, many of
which are known to coincide with protoclusters, we measured the giant
nebulae-LAE cross-correlation function for a sample of eight systems
-- six HzRGs and two LABs -- for which published data was available in
the literature \cite{supp4}.  In Fig.~\figcluster, we compare the
overdensity profile around SDSSJ0841$+$3921 to this giant nebulae-LAE
correlation function. On average, the environment of HzRGs and LABs
hosting giant Ly$\alpha$ nebulae (red line) is much richer than that of
radio-quiet quasars \cite{trainor12} (blue line),
confirming that they indeed reside in protoclusters. Furthermore, the
clustering of LAEs around \qso has a steeper overdensity profile, and
exceeds the average protocluster by a factor of $\gtrsim 20$ for
$R < 200\,{\rm kpc}$ and by $\sim 3$ on scales of $R\simeq 1\,{\rm
  Mpc}$.  In addition to the overdensity of four AGN, the
high number of LAEs surrounding \qso make it one of the
most overdense protoclusters known at $z\sim 2-3$.

The combined presence of several bright AGN, the Ly$\alpha$ emission
nebula, and the b/g absorption spectrum, provide an unprecedented
opportunity to study the morphology and kinematics of the protocluster
via multiple tracers, and we find evidence for extreme motions
\cite{supp6}. Specifically, AGN1 is offset from the precisely
determined systemic redshift \cite{supp5} of the f/g quasar by $+1300
\pm 400\,\kms$. This large velocity offset cannot be explained by
Hubble expansion -- the miniscule probability of finding a quadruple
quasar in the absence of clustering $P\sim 10^{-13}$ \cite{supp3} and
the physical association between the AGN and giant nebula demand that
the four AGN reside in a real collapsed structure -- and thus provides
an unambiguous evidence for extreme gravitational motions.  In
addition, our slit spectra of the giant Ly$\alpha$ nebula reveal
extreme kinematics of diffuse gas (Fig.~\figslits), extending over a
velocity range of $-800$ to $+2500\,\kms$ from systemic. Furthermore,
there is no evidence for double-peaked velocity profiles
characteristic of resonantly-trapped \mlya, which could generate large
velocity widths in the absence of correspondingly large gas motions.
Absorption line kinematics of the metal-enriched gas, measured from
the b/g quasar spectrum at impact parameter of $R_\perp = 176\,\kpc$
(Fig.~\figabs), show strong absorption at $\approx +650\,\kms$ with a
significant tail to velocities as large as $\simeq 1000\,\kms$. It is
of course possible that the extreme gas kinematics, traced by
Ly$\alpha$ emission and metal-line absorption, are not gravitational
but rather arise from violent large-scale outflows powered by the
multiple AGN. While we cannot completely rule out this possibility,
the large velocity offset of $+1300\,\kms$ between the f/g quasar and
the emission redshift of AGN1 can only result from gravity.

One can only speculate about the origin of the dramatic enhancement
of AGN in the \,\,\,
\qso protocluster.  Perhaps the duty cycle for AGN
activity is much longer in protoclusters, because of frequent
dissipative interactions\cite{Djor91,BINARY}, or an abundant supply of
cold gas. A much larger number of massive galaxies could also be the
culprit, as AGN are known to trace massive halos at $z\sim 2$.
Although \qso is the only example of a quadruple AGN with such small
separations, previously studied protoclusters around HzRGs and LABs,
also occasionally harbor multiple AGN \cite{Venemans07,lehmer09}.
Regardless, our discovery of a quadruple AGN and protocluster from a
sample of only 29 quasars suggests a link between giant Ly$\alpha$
nebulae, AGN activity, and protoclusters -- similar to past work on
HzRGs and LABs -- with the exception that SDSSJ~0841$+$3921 was
selected from a sample of normal radio-quiet quasars. From our survey
and other work \cite{supp7}, we estimate that $\simeq 10\%$ of quasars
exhibit comparable giant Ly$\alpha$ nebulae.  Although clustering
measurements imply that the majority of $z\sim 2$ quasars reside in
moderate overdensities \cite{trainor12,White12,Fanidakis13}, we
speculate that this same 10\% trace much more massive protoclusters.
\qso clearly supports this hypothesis, as does another
quasar-protocluster association 
\cite{trainor12,Mostardi13}, around which a giant Ly$\alpha$ nebula
was recently discovered \cite{Martin14a,supp8}.


Given our current theoretical picture of galaxy formation in massive
halos, an association between giant Ly$\alpha$ nebulae and
protoclusters is completely unexpected.  The large Ly$\alpha$
luminosities of these nebulae imply a substantial mass ($\sim
10^{11}\,M_{\odot}$) of cool ($T \sim 10^{4}\,{\rm K}$) gas
\cite{Slug}, whereas cosmological hydrodynamical simulations indicate
that already by $z\sim 2-3$, baryons in the massive progenitors
($M_{\rm halo} \gtrsim 10^{13}\,M_\odot$) of present-day clusters 
are dominated by a hot
shock-heated plasma $T\sim 10^7\,$K \cite{fumagalli14,FG14}. These hot
halos are believed to evolve into the X-ray emitting intra-cluster
medium observed in clusters, for which
absorption-line studies indicate negligible $\lesssim 1\%$ cool gas
fractions \cite{Lopez08}.
Clues about the nature of this apparent discrepancy come from our
absorption line studies of the massive $\simeq 10^{12.5}\msol$ halos
hosting $z\sim 2-3$ quasars. This work reveals substantial reservoirs
of cool gas $\gtrsim 10^{10}\msol$
\cite{QPQ1,QPQ2,QPQ3,QPQ4,QPQ5,QPQ6,QPQ7}, manifest as a high covering
factor $\simeq 50\%$ of optically thick absorption, several times
larger than predicted by hydrodynamical
simulations\cite{fumagalli14,FG14}. This conflict most likely
indicates that current simulations fail to capture 
essential aspects of the hydrodynamics in massive halos at $z\sim 2$
\cite{QPQ6,fumagalli14}, perhaps failing to resolve the formation of
clumpy structure in cool gas \cite{Slug}. 


If illuminated by the quasar, these large cool gas reservoirs in the
quasar circumgalactic medium (CGM) will emit fluorescent Ly$\alpha$
photons, and we argue that this effect powers the nebula in \qso \cite{supp10}. But according to this logic, nearly every quasar in the
Universe should be surrounded by a giant Ly$\alpha$ nebulae with size
comparable to its CGM ($\sim 200\,{\rm kpc}$). Why then are these
giant nebulae not routinely observed?

This apparent contradiction can be resolved as follows.  If this cool
CGM gas is illuminated and highly ionized, it will fluoresce in the
$\lya$ line with a surface brightness scaling as ${\rm SB}_{\rm
  Ly\alpha} \propto N_{\rm H}n_{\rm H}$, where $N_{\rm H}$ is the
column density of cool gas clouds which populate the quasar halo, 
and $n_{\rm H}$ is the number density of hydrogen atoms inside these clouds.
Note the total cool gas mass of the halos scales as $M_{\rm cool} \propto R^2N_{\rm H}$,
where $R$ is the radius of the halo \cite{supp10}. 
Given our best estimate for the properties of the CGM
around typical quasars ($n_{\rm H}\sim 0.01\,{\rm cm^{-3}}$ and
$N_{\rm H}\sim 10^{20}\,{\rm cm^{-2}}$ or $M_{\rm cool}\simeq
10^{11}\,M_{\odot}$) \cite{QPQ4,QPQ5,QPQ6}, we expect these nebulae to
be extremely faint ${\rm SB}_{\rm Ly\alpha}\sim 10^{-19}\cgssb$, and
beyond the sensitivity of current instruments\cite{QPQ4}.  One comes
to a similar conclusion based on a full radiative transfer calculation
through a simulated dark-matter halo with mass $M_{\rm halo} \approx
10^{12.5}\,M_{\odot}$ \cite{Slug}.  Thus the factor of $\sim 100$
times larger surface brightness observed in the \qso and other
protocluster nebulae, arises from either a higher $n_{\rm H}$, $N_{\rm
  H}$ (and hence higher $M_{\rm cool}$), or both. The cool gas
properties required to produce the \qso nebula can be
directly compared to those deduced from an absorption line analysis of
the b/g quasar spectrum \cite{supp9}.

The b/g quasar sightline pierces through SDSSJ0841$+$3921 at an impact
parameter of $R_{\perp}=176$\, kpc, giving rise to the absorption
spectrum shown in Fig.~\figabs.  Photoionization modeling of these
data constrains the total hydrogen column density to be $\log_{10}
N_{\rm H} = \nhsig$ \cite{supp10}, implying a substantial mass of cool
gas $1.0\times 10^{11}\,M_{\odot} < M_{\rm cool} < 6.5\times
10^{11}\,M_{\odot}$ within $r=250\,{\rm kpc}$. Assuming that the
Ly$\alpha$ emitting gas has the same column density as the gas
absorbing the b/g sightline, reproducing the large fluorescent
Ly$\alpha$ surface brightness requires that this gas be distributed in
compact $r_{\rm cloud}\sim 40\,{\rm pc}$ clouds at densities
characteristic of the interstellar medium $n_{\rm H} \simeq 2\,{\rm
  cm}^{-3}$, but on $\sim 100\,{\rm kpc}$ scales in the protocluster.

Clues to the origin of these dense clumps of cool gas comes from their
high enrichment level, which we have determined from our absorption
line analysis \cite{supp9} to be greater than 1/10th of the solar
value. At first glance, this suggests that strong tidal interactions
due to merger activity or outflows due to powerful AGN feedback are
responsible for dispersing dense cool gas in the protocluster.
However, the large cool gas mass $\sim 10^{11}\msol$ and high
velocities $\sim 1000\,\kms$, imply an extremely large kinetic
luminosity $L_{\rm wind} \sim 10^{44.6}$ for an AGN powered wind,
making the feedback scenario implausible \cite{QPQ3}.  An even more
compelling argument against a merger or feedback origin comes from the
extremely small cloud sizes $r_{\rm cloud} \sim 40\,{\rm pc}$ implied
by our measurements.  Such small clouds moving supersonically $\sim
1000\,{\rm km\,s^{-1}}$ through the hot $T\sim 10^7\,{\rm K}$
shock-heated plasma predicted to permeate the protocluster, will be
disrupted by hydrodynamic instabilities in $\sim
5\times 10^6\,{\rm yr}$, and can thus only be transported $\sim
5\,{\rm kpc}$ \cite{Crighton14}.  These short disruption timescales
instead favor a scenario where cool dense clouds are formed in situ,
perhaps via cooling and fragmentation instabilities, but are
short-lived. The higher gas densities might naturally arise if hot
plasma in the incipient intra-cluster medium pressure confines the
clouds, compressing them to high densities \cite{mb04,mm96}.
Emission line nebulae from cool dense gas has also been observed at
the centers of present-day cooling flow clusters
\cite{Heckman89,Mcdonald10}, albeit on much smaller scales $\lesssim
50\,{\rm kpc}$. The giant Ly$\alpha$ nebulae in $z\sim 2-3$
protoclusters might be manifestations of the same phenomenon, but with
much larger sizes and luminosities, reflecting different physical
conditions at high-redshift.



The large reservoir of cool dense gas in the protocluster \qso, 
as well as those implied by the giant nebulae in
other protoclusters, appear to be at odds with our
current theoretical picture of how clusters form. This is likely
symptomatic of the same problem of too much cool gas in massive halos
already highlighted for the quasar CGM\cite{QPQ6,fumagalli14,FG14}. Progress
will require more cosmological simulations of massive halos $M \gtrsim
10^{13}\msol$ at $z\sim 2$, as well as idealized higher-resolution studies. 
In parallel, a survey for extended $\lya$ emission around
$\sim 100$ quasars would uncover a sample of $\sim 10$ giant
Ly$\alpha$ nebulae likely coincident with protoclusters, possibly
also hosting multiple AGN, and 
enabling 
continued exploration of the relationship between AGN, cool gas, and
cluster progenitors.

\clearpage
\begin{figure}[!ht]
\begin{center}
\includegraphics[width=0.9\textwidth]{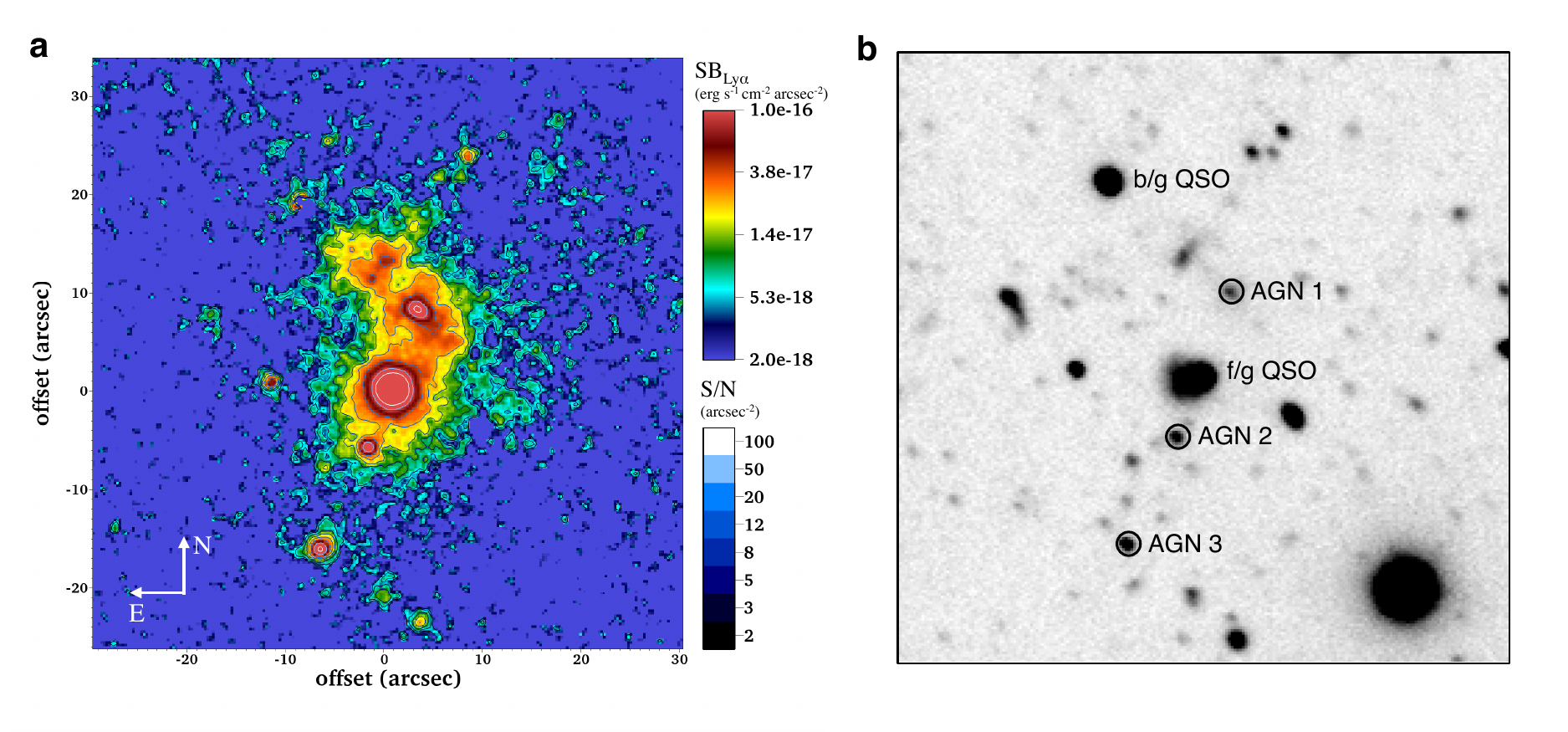}
\end{center}
\small {\bf Fig.~\figimages: Narrow and broad band images of the
  field surrounding \qso} {\small {\bf Left}: Continuum-subtracted,
narrow-band image of the field around f/g quasar. The color map and the
contours indicate the \mlya\ surface brightness (upper color bar) and
the signal-to-noise ratio per arcsec$^{2}$ aperture (lower color bar),
respectively. This image reveals a giant \mlya\ nebula on the northern
side of the f/g quasar
and several compact, bright \mlya\ emitters in addition to the f/g
quasar. Three of these have been spectroscopically confirmed as active
galactic nuclei (AGN) at the same redshift. {\bf Right:}
Corresponding $V$-band continuum image of the field presented at left
with the locations of the four AGN marked. The AGN are roughly
oriented along a line coincident with the projected orientation of the
\mlya\ nebula. We also mark the position of the b/g quasar, 
which is not physically associated with the quadruple AGN system, but
whose absorption spectrum probes the gaseous environment of the f/g
quadruple AGN and protocluster (see Fig.~\figabs).  }
\end{figure}

\clearpage
\begin{figure}[!ht]
\begin{center}
\includegraphics[width=0.8\textwidth]{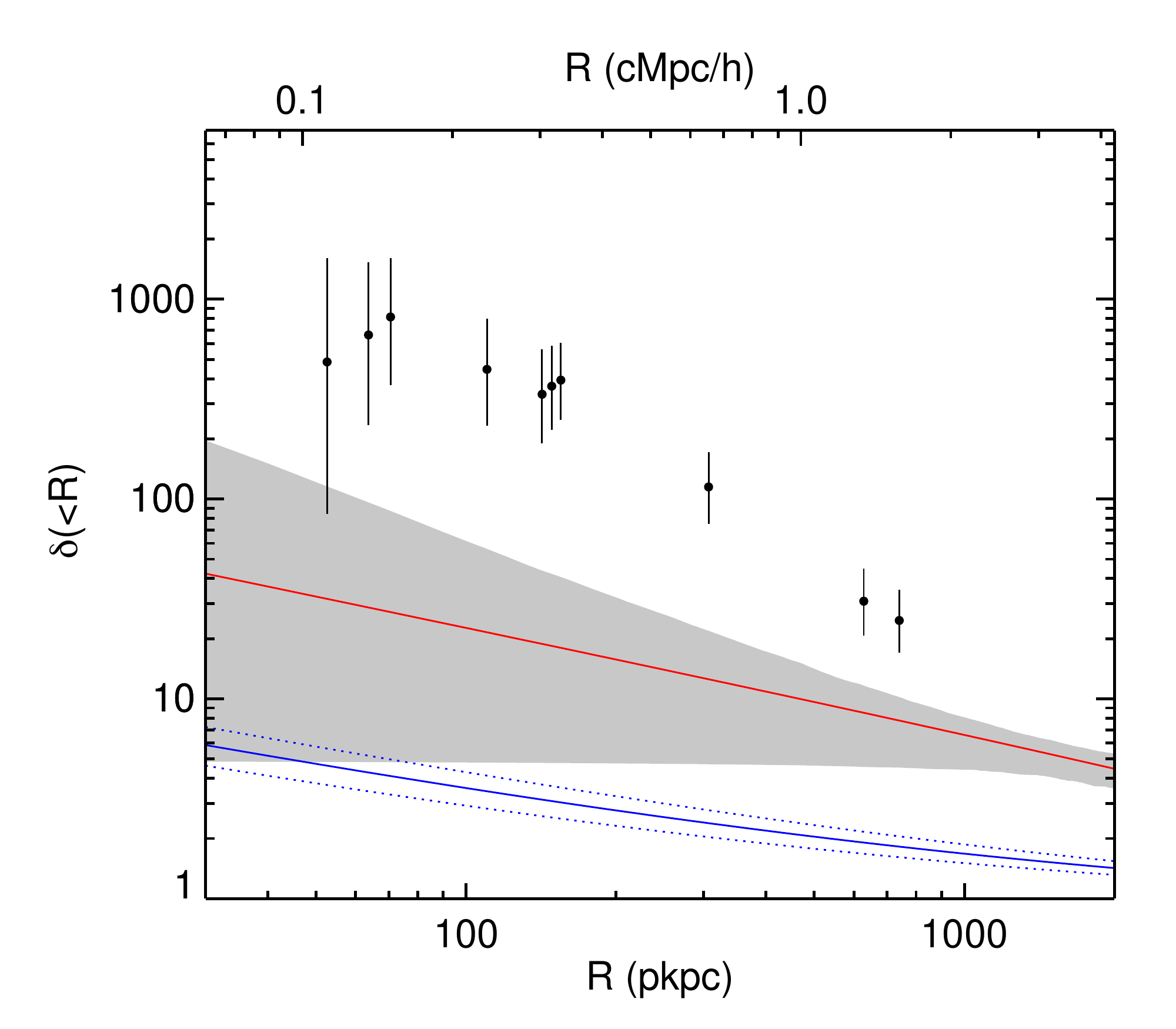}
\end{center}
\noindent
{\bf Fig.~\figcluster: Characterization of the protocluster environment
  around \qso.} {\small The data points indicate the cumulative
  overdensity profile of LAEs $\delta(<R)$ as a function of impact
  parameter $R$ from the f/g quasar in \qso, with Poisson error
  bars. The red curve shows the predicted overdensity profile, based
  on our measurement of the giant nebulae-LAE cross-correlation
  function determined from a sample of eight systems -- six HzRGs and
  two LABs -- for which published data was available in the
  literature. Assuming a power-law form for the cross-correlation
  $\xi_{\rm cross} = (r\slash r_0)^{-\gamma}$, we measure the
  correlation length $r_0 = \rnotval \pm \rnoterr\,h^{-1}{\rm \,Mpc}$,
  for a fixed value of $\gamma = \gammaval$. The gray shaded region
  indicates the $1\sigma$ error on our measurements based on a
  bootstrap analysis, where both $r_0$ and $\gamma$ are allowed to
  vary.  The solid blue line indicates the overdensity of Lyman break
  galaxies (LBGs) around radio-quiet quasars based on recent
  measurements\cite{trainor12}, with the dotted blue lines the
  $1\sigma$ error on this measurement. On average, the environment of
  HzRGs and LABs hosting giant Ly$\alpha$ nebulae is much richer than
  that of radio-quiet quasars \cite{trainor12}, confirming that they
  indeed reside in protoclusters.  \qso exhibits a dramatic excess of
  LAEs compared to the expected overdensities around radio-quiet
  quasars (blue curve).  Its overdensity even exceeds the average
  protocluster (red curve) by a factor of $\gtrsim 20$ for $R <
  200\,{\rm kpc}$ decreasing to an excess of $\sim 3$ on scales of
  $R\simeq 1\,{\rm Mpc}$, and exhibits a much steeper profile.  }
\end{figure}

\clearpage
\begin{figure}[!ht]
\begin{center}
\includegraphics[width=0.9\textwidth]{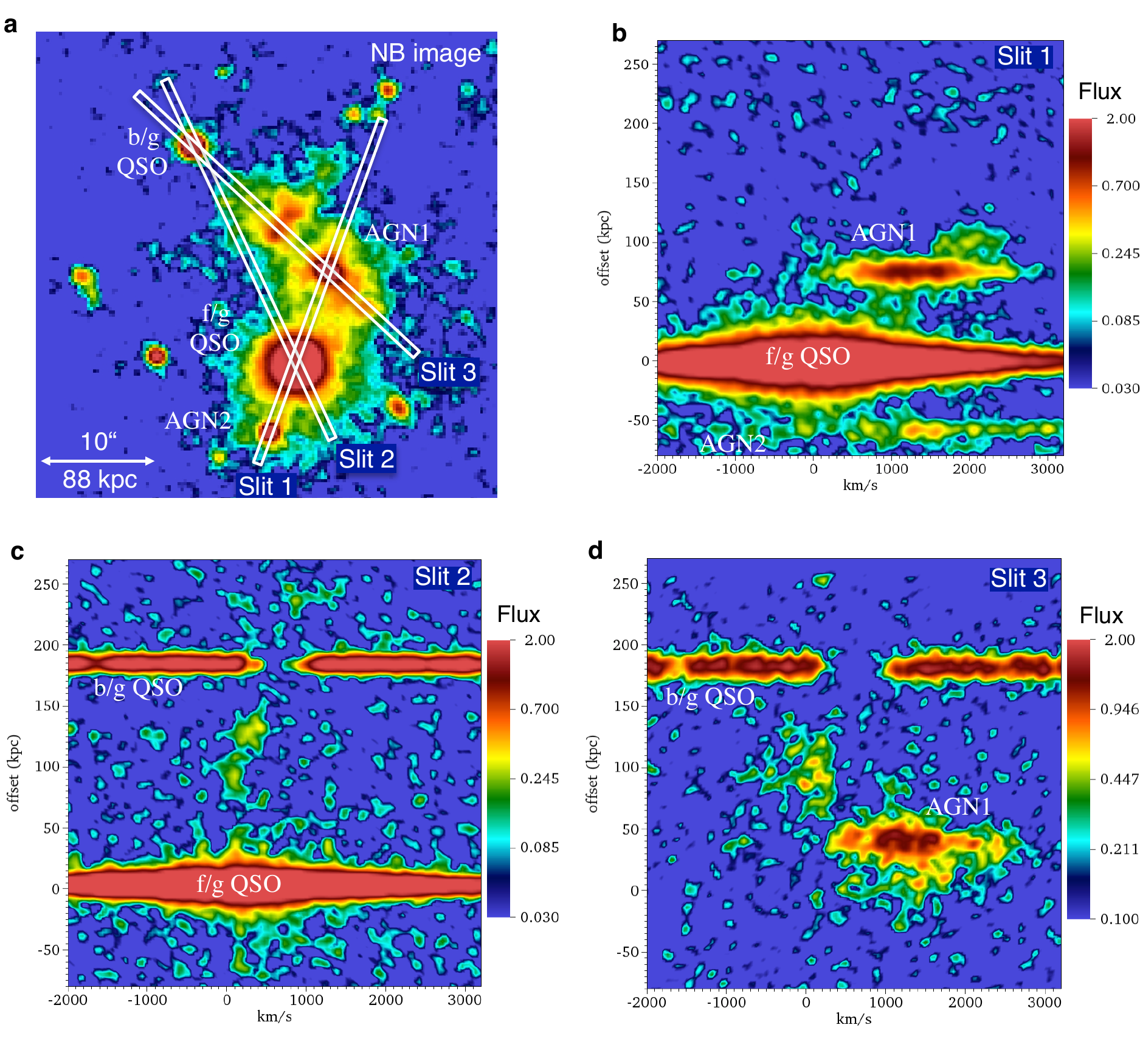}
\end{center}
\noindent
{\bf Fig.~\figslits: \mlya\ spectroscopy of the giant nebula and its
  associated AGN.} {\small {\bf Upper Left:} The spectroscopic slit
  locations (white rectangles) for three different slit orientations
  are overlayed on the narrow band image of the giant nebula. The
  locations of the f/g quasar (brightest source), b/g quasar, AGN1,
  and AGN2 are also indicated.  Two-dimensional spectra for Slit1 ({\bf Upper Right}), Slit2 ({\bf Lower Left}), and Slit3 ({\bf Lower Right}) 
  are shown in the accompanying panels. In the upper
  right and lower left panels, spatial coordinates refer to the
  relative offset along the slit with respect to the f/g
  quasar. Spectra of AGN1 are present both in Slit 1 (upper right) and
  Slit 3 (lower right) at spatial offsets 75kpc and 25kpc,
  respectively, while the \mlya\ spectrum of AGN2 is located at a
  spatial offset $-$60\,kpc in Slit 1 (upper right). The b/g quasar
  spectra are located in both Slit 2 (lower left) and Slit 3 (lower
  right) at the same spatial offset of 176\,kpc. The spectroscopic
  observations demonstrate the extreme kinematics of the system: AGN1
  has a velocity of $+1300 \pm 400\,\kms$ relative to the f/g quasar
  and the Ly$\alpha$ emission in the nebula exhibits motions ranging
  from $-800$\tkms\ (at $\approx 100$\,kpc offset in Slit~3, lower
  right) to +2500\tkms\ (at $\approx 100$\,kpc in Slit~1, upper
  right).  A $3\times 3$ pixel boxcar smoothing, which corresponds
  $120\,{\rm \kms}\times 0.8\arcsec$, has been applied to the spectra.
  In each two-dimensional spectrum, the zero velocity corresponds to
  the systemic redshift of the f/g quasar. The color bars indicate the
  flux levels in units of
  ${\rm\,erg\,s^{-1}\,cm^{-2}\,arcsec^{-2}\,\AA^{-1}}$.}
\end{figure}

\clearpage
\begin{figure}[!ht]
\begin{center}
\includegraphics[width=0.9\textwidth]{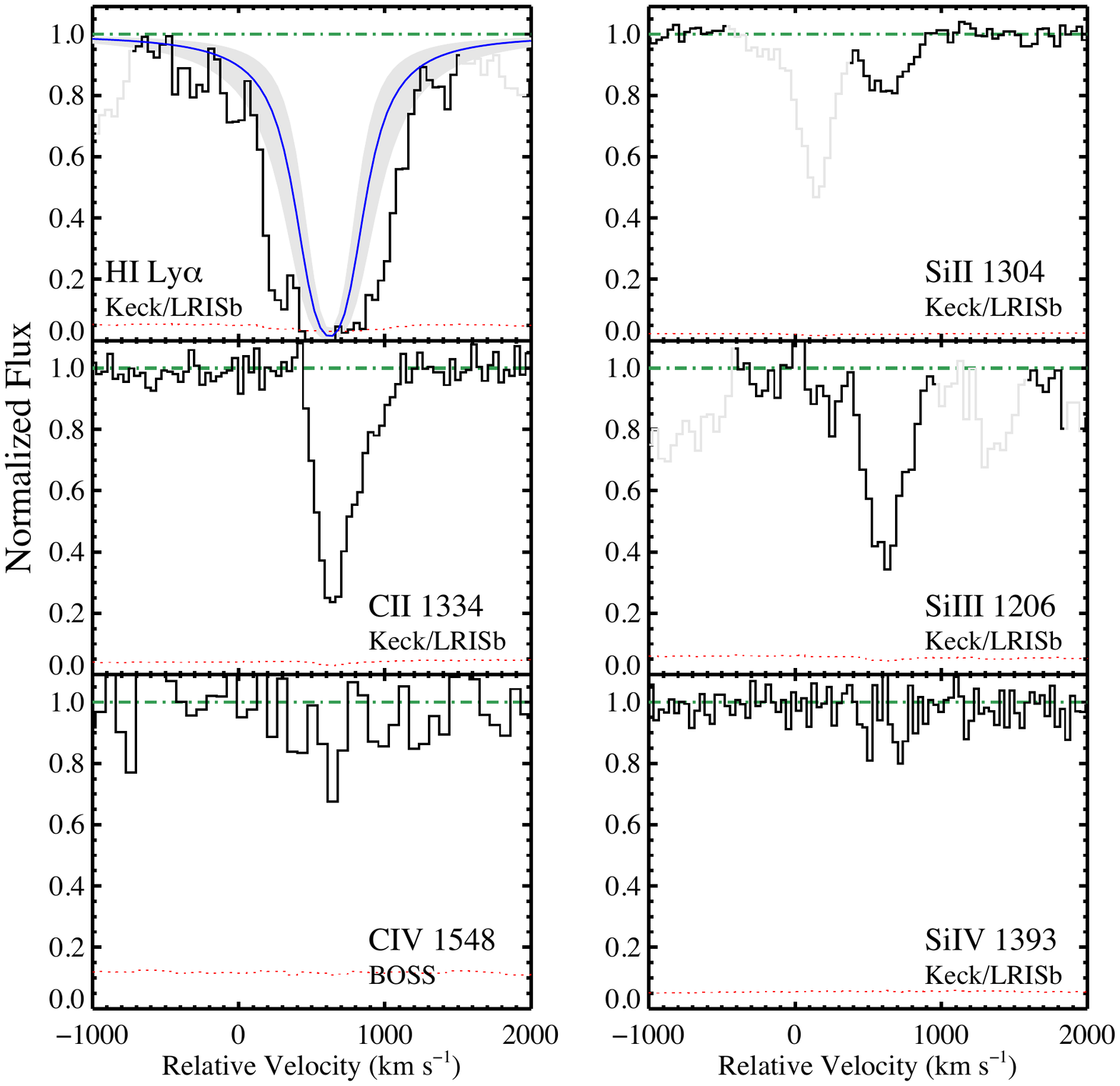}
\end{center}
\vskip -5cm
\noindent
{\bf Fig.~\figabs: Absorption line spectrum of cool gas in \qso.}
{\small Spectrum of the absorbing gas detected in the b/g quasar
  sightline at an impact parameter of 176\,kpc from the f/g
  quasar. The gas shows strong \ion{H}{i} and low-ionization state
  metal absorption, offset by $\approx 650$\tkms\ from the f/g
  quasar's systemic redshift.  The CII absorption in particular
  exhibits a significant tail to velocities as large as $\simeq
  1000\,\kms$, providing evidence for extreme gas kinematics. We
  modeled the strong HI \mlya\ absorption with a Voigt profile (blue
  curve with grey band indicating uncertainty) and estimate a column
  density $\log \mnhi = 19.2 \pm 0.3$. The strong low and intermediate
  ion absorption (SiII, CII, SiIII) and correspondingly weak high-ion
  absorption (CIV, SiIV) indicate that the gas is not highly ionized,
  and our photoionization modeling \cite{supp10} implies $\log_{10}
  x_{\rm HI} = \xfrac$ or $\log_{10} N_{\rm H} = \nhsig$.  We estimate
  a conservative lower-limit on the gas metallicity to be 1/10 of the
  solar value.}
\end{figure}


\clearpage


\vskip 1cm
\noindent
{\bf Acknowledgments}\\
  We thank the staff of the W.M. Keck Observatory for their support
  during the installation and testing of our custom-built narrow-band
  filter. We are grateful to B. Venemans and M. Prescott for providing
  us with catalogs of LAE positions around giant nebulae in electronic
  format.  We also thank the members of the ENIGMA
  group (http://www.mpia-hd.mpg.de/ENIGMA/) at the Max Planck
  Institute for Astronomy (MPIA) for helpful discussions.  JFH
  acknowledges generous support from the Alexander von Humboldt
  foundation in the context of the Sofja Kovalevskaja Award. The
  Humboldt foundation is funded by the German Federal Ministry for
  Education and Research. J.X.P. acknowledge support from the National
  Science Foundation (NSF) grant AST-1010004. The data presented here
  were obtained at the W.M. Keck Observatory, which is operated as a
  scientific partnership among the California Institute of Technology,
  the University of California and NASA. The Observatory was made
  possible by the financial support of the W.M. Keck Foundation. We
  acknowledge the cultural role that the summit of Mauna Kea has
  within the indigenous Hawaiian community.  We are most fortunate to
  have the opportunity to conduct observations from this mountain.
  The data reported in this paper are available through the Keck
  Observatory Archive (KOA).

\vskip 1cm
\noindent
{\bf Supplementary Online Material}\\
www.sciencemag.org\\
Materials and Methods\\
SOM Text\\
Figs. S1 to S10\\
Tables S1 to S6\\
References (52-133)
\clearpage

\setcounter{page}{1}

\def\be{\begin{equation}}
\def\ee{\end{equation}}
\def\bea{\begin{eqnarray}}
\def\eea{\end{eqnarray}}
\def\tol#1#2#3{\hbox{\rule{0pt}{15pt}${#1}^{+{#2}}_{-{#3}}$}}
\def \phn {\hphantom{0}}

\renewcommand\refname{References and Notes}
\def \qso {SDSSJ0841$+$3921\,}
\def \fgqso {SDSSJ084158.47+392121.0}
\def \bgqso {SDSSJ084159.26+392140.0}
\def \rqso {FIRSTJ084158.6+392114.7}
\def \rphys {$R_\perp$}
\def \mrphys {R_\perp}
\def \mkms      {\rm\ km\ s^{-1}}
\def\mlnull{L_{\nu_{\rm LL}}}
\def \kpc       {{\rm\ kpc}}
\def \cgssb {{\rm\,erg\,s^{-1}\,cm^{-2}\,arcsec^{-2}}}
\def\sci#1{{\; \times \; 10^{#1}}}
\def \arcsec    {^{\prime\prime}}
\def \sqarcsec    {\box^{\prime\prime}}
\def \msol      {{\rm\ M}_\odot}
\def \kms            {{\rm km~s}^{-1}}
\def \tkms    {$\kms$}

\def\cm#1{\, {\rm cm^{#1}}}
\def \cgssb {{\rm\,erg\,s^{-1}\,cm^{-2}\,arcsec^{-2}}}

\def\aap{A \& A}
\def\aj{AJ}
\def\apj{ApJ}
\def\apss{Ap\&SS}
\def\apjl{ApJL}
\def\apjs{ApJS}
\def\apjsupp{ApJS}
\def\araa{ARAA}
\def\mnras{MNRAS}
\def\nat{Nature}
\def\pasp{PASP}
\def\prd{PhRvD}
\def\aapr{A\&A Rev.}
\def\physrep{Physics Reports}
\def\nar{New A Rev.}
\def\aaps{A\&AS}

\noindent {\Huge \bf Supplementary Online Material}

\vskip 1cm
\noindent
{\bf This PDF file includes}\\
Materials and Methods\\
SOM Text\\
Figs. S1 to S10\\
Tables S1 to S6\\
References (52-133)

\section{Optical Observations}\label{sec:optical_obs}

\subsection{Discovery Spectra}

The source \fgqso\ (f/g quasar) was targeted as a quasar by 
the Sloan Digital Sky Survey (SDSS) through their color-selection
algorithms and observed
spectroscopically in the standard survey \cite{SDSSDR7}.  
It has a cataloged emission redshift of $z=2.046$. 
Our analysis of the SDSS imaging data revealed a second, neighboring 
source at SDSSJ084159.26+392140.0 (b/g quasar) 
with colors characteristic of a $z \sim 2$ quasar, characterizing
this system as a candidate quasar pair.
In the course of our ongoing spectroscopic campaign
to discover close quasar pairs at $z>2$
\cite{BINARY,HIZBIN},  
we confirmed this source to have $z=2.214$ implying a projected quasar
pair with a physical separation of $R_\perp = 176$\,kpc at the f/g
quasar redshift.
\bgqso\ was also
observed by the SDSS-III survey in the Baryonic Oscillating
Spectroscopic Survey campaign at a spectral resolution of $R \approx
2000$ and with wavelength coverage $\lambda \approx 3600-10,000$\AA\
\cite{BOSS}.

On UT 2007 Jan 18, we observed the quasar pair with the
Low Resolution Imaging Spectrograph (LRIS) \cite{LRIS}. 
These data were taken to study the intergalactic medium probed by
the quasar pair, to examine the HI gas associated with
the circumgalactic medium of the f/g quasar, and 
to search for fluorescent emission associated with the f/g quasar.
The latter analyses of these data have been
presented in previous works \cite{QPQ4,QPQ5,QPQ6,QPQ7}.
Summarizing the observations, we used LRIS
in multi-slit mode with a
custom designed slitmask which allowed placement of one slit on 
the known quasars and other slits on additional quasar candidates in the field.
Specifically, a slit was placed at the position angle PA=$25.8^\circ$ between
the f/g and b/g quasars, allowing them to be observed simultaneously.
LRIS is a double spectrograph with
two arms giving simultaneous coverage of the near-UV and optical.  
We used the D460 dichroic with the $1200$ lines mm$^{-1}$ grism blazed at
$3400$~\AA\ on the blue side, resulting in wavelength coverage of
$\approx 3250-4300$~\AA. The dispersion of this grism is
$0.50$~\AA\ per pixel and our $1''$ slits give a resolution with
full width at half maximum (FWHM) FWHM$\simeq 160\kms$.  
On the red side we used
the R1200/5000 (covering $\approx 4700-6000$\AA)
and R300/5000 (covering $\approx 4700-10,000$\AA)
gratings having a FWHM of $\approx 100\kms$ 
and $\approx 400\kms$ respectively. 

The science frames were complemented by a series of calibration
images: arc lamp, dome flat, twilight sky, and standard star 
spectra with the same instrument configurations.  All of these
exposures were reduced using the
LowRedux (http://www.ucolick.org/$\sim$xavier/LowRedux/) pipeline
which bias subtracts and flat fields the images, corrects for non-uniform
illumination, derives a wavelength solution, performs sky subtraction,
optimally extracts the sources, and fluxes the resultant spectra.  The
1D spectra are corrected for instrument flexure and
shifted to a heliocentric, vacuum-corrected system.

Using custom software, we also coadded the 2D spectral images to
search for diffuse \mlya\ emission surrounding the f/g
quasar \cite{QPQ4}.
This software enables us to model and subtract the
spectral PSF of sources in the 2D spectra.  Extended \mlya\ emission
will be manifest as residual flux in our 2D sky-and-PSF-subtracted
images which is inconsistent with being noise. To visually assess the
statistical significance of any putative emission feature, we define a
$\chi$ image $\chi_{\rm sky+PSF} = ({\rm DATA} - {\rm SKY} - {\rm
OBJECTS})\slash \sigma$. If our model is an accurate description of
the data, the distribution of pixel values in the $\chi_{\rm sky+PSF}$
should be a Gaussian with unit variance.  The middle row of the images in 
Fig.~\sfigchimaps\, shows this quantity for SDSSJ~0841$+$3921 for
each slit orientation.  The
lower row of these images shows $\chi_{\rm sky}= ({\rm DATA} - {\rm
SKY})\slash \sigma$.  The upper row shows a smoothed version of the
middle row, helpful for identifying extended emission. Specifically,
the smoothed images are given by $\chi_{\rm smth} = \frac{{\rm
CONVOL}[{\rm DATA} - {\rm SKY} - {\rm OBJECTS}]}{\sqrt{{\rm
CONVOL^2}[\sigma^2]}}$, where the ${\rm CONVOL}$ operation denotes
smoothing of the stacked images with a symmetric Gaussian kernel (same
spatial and spectral widths) with FWHM=$235\,\kms$
(dispersion $\sigma_{\rm smth} = 100\,\kms$), which is 5.7 pixels, or
1.4 times the spectral resolution element, and corresponds to ${\rm
FWHM}=1.5\arcsec$ spatially.  The operation ${\rm CONVOL^2}$
represents convolution with the square of the smoothing kernel. With
this definition of $\chi_{\rm smth}$ the distribution of the pixel
values in the smoothed image will still obey Gaussian statistics,
although they are of course correlated, and hence not independent.

Sky and object PSF-subtracted $\chi$-maps for all of the
slit-orientations that we used to characterize extended emission in
SDSSJ~0841$+$3921 are shown in Fig.~\sfigchimaps. Fig.~\figslits\ of
the Main text, shows just the smoothed sky-subtracted images with a
color-map chosen to accentuate the faint extended emission and a
slightly different smoothing, namely a $3\times 3$ pixel boxcar
smoothing, which corresponds $120\,{\rm \kms}\times 0.8\arcsec$.  The
$\chi$-maps in Fig.~\sfigchimaps\ enable the reader to objectively
assess the statistical significance of all emission features in the
unsmoothed data.  Note that in all of these maps, only the PSF model
of the f/g and b/g quasars have been subtracted, whereas neither AGN1
nor AGN2 were removed from the other slit-orientations (see Fig.~1 in
main text). Following the calibration procedure described
in \cite{QPQ4},
we deduce the following spectroscopic surface
brightness limits ($1\sigma$) for the Ly$\alpha$ line at the f/g
quasar redshift $z=2.0412$: Slit1, ${\rm SB}_{1\sigma} =
2.2\sci{-18}\cgssb$; Slit2, ${\rm SB}_{1\sigma} = 2.3\sci{-18}\cgssb$;
and Slit3, ${\rm SB}_{1\sigma} = 4.8\sci{-18}\cgssb$, where these SBs
are computed in windows of $700\,{\rm km\,s^{-1}}\!\times 1.0\arcsec$,
which corresponds to an aperture of $700\,{\rm km\,s^{-1}}\!\times
1.0\,{\rm arcsec}^2$ on the sky because we always used a $1.0\arcsec$
slit. The depth that we attain in our Ly$\alpha$ spectroscopy is comparable to that
achieved by our narrow-band imaging  ${\rm SB}_{1\sigma}= 1.7\sci{-18}\cgssb$
(see next section).

The original observation from the Quasars Probing Quasars (QPQ) \cite{QPQ4}
survey corresponds to slit-orientation Slit2 (see
Fig.~\sfigchimaps\ and Fig.~3 in main text). Our initial visual
inspection of the \mlya\ emission map revealed a bridge of \mlya\
emission along the slit connecting the quasar pair. Although the SB
varies with position, this structure has a characteristic ${\rm
SB}_{\lya} \simeq 10^{-17}\cgssb$ and is detected at $\simeq
4-5\sigma$ at location $R_{\perp}\simeq +100\,\kpc$ between the two
quasars. The emission extends along the slit from $-80\,\kpc$ below
the f/g quasar to $+150\,\kpc$ between the two quasars, and possibly
to $+250\,\kpc$. With an end-to-end size of $\sim 250\,\kpc$, this
represented one of the largest
\mlya\ emission nebula ever detected, and thus motivated 
the additional imaging and spectroscopic observations, which are the
focus of this manuscript.



\vskip 0.2in

\subsection{Narrow Band Imaging}
\label{sec:NB}
We purchased a custom-designed narrow-band (NB) filter from
Andover Corporation, sized to fit within the grism holder of the Keck/LRISb
camera.  The filter was tuned to $\lambda_{\rm center} = 3700$\AA,
and designed with a narrow band-pass ${\rm FWHM}_\lambda =
33$\AA\, to minimize sky background while maintaining
throughput.  
On UT 12 November 2012, we imaged the $\sim 5' \times 7'$
field-of-view surrounding \qso, offset to place the quasar pair on
the CCD with highest quantum efficiency.  We observed for a total of
3.2 hours in a series of dithered, 1280s exposures.  
Conditions were clear with sub-arcsecond atmospheric seeing. 
In parallel, we obtained 3hrs of broad-band $V$ images
with the LRISr camera.  The instrument was configured with 
the D460 dichroic and the detectors of the blue camera
were binned 2x2 to minimize read noise.

The images were reduced using standard routines within the IRAF
reduction software package.  This includes bias subtraction, flat
fielding and an illumination correction. A combination of twilight sky
flats and unregistered science frames were used to produce flat-field
images and illumination corrections in each band. The individual
frames were sequentially registered to the SDSS-DR7 catalog using the
SExtractor \cite{sextractor} and SCAMP \cite{scamp} packages.  The RMS
uncertainty in the astrometry of our registered images is
approximately 0.2$''$.  Finally, the corrected frames from each band
were average-combined using the SWarp package \cite{swarp}.

We calibrated the photometry of our images as follows.  At the
beginning and end of the night, we observed two spectrophotometric
stars (G191b2b and Feige34) with the NB filter under clear conditions.
Neither star has significant spectral features in the relevant
wavelength range covered by our NB filter.  For the broad-band images,
we observed the standard star field PG0231+051.  To compute the
zero-point for the narrow-band images, we compared the measured count
rates of Feige34 and G191b2b with the expected fluxes estimated by
convolving the standard star spectra (resolution of 1\AA) \cite{oke90}
with the normalized filter transmission curve. We calculate an
average, zero-point magnitude of 23.58\,mag, with a few percent
difference between the two stars. With this calibration, we deduce
that the $1\sigma$ limiting surface brightness of our combined NB
images is ${\rm SB}_{1\sigma}= 1.7\sci{-18}\cgssb$ for a $1.0\,{\rm
arcsec}^2$ aperture.

For the broad-band images, we compared the number of counts per second
of the five stars in the PG0231+051-field with their tabulated
$V$-band magnitudes \cite{lan92}.
The derived zero-point for the five
stars are consistent to within a few percent and we adopt the average
value: $V_{\rm ZP}=28.07$.  Because the Feige34 and the PG0231+051
fields were observed with an airmass (AM$\approx 1.2$) similar to our
science field, we did not correct the individual images before
combination.  We estimate that the correction would be on the order of
few percent. The $1\sigma$ limiting point source magnitude for our combined
broad band images is $V = XX$. 

To isolate the emission in the Ly$\alpha$ line we estimated and then
subtracted the continuum emission underlying the NB3700 filter. We
estimate the continuum using the $V$-band (not affected by the
Ly$\alpha$ line) and assuming a flat continuum slope (in frequency),
i.e.  $\beta_\lambda = -2$. 
This assumption is dictated by our dataset, i.e. we can rely only on a
deep $V$-band image to estimate the continuum, and it follows the
prescriptions of previous work \cite{cantalupo12}
which showed that $\beta_\lambda\sim-1.9$ for the most
luminous sources and higher values for the faintest. Thus, assuming a
flat continuum slope gives us a conservative estimate of the \mlya\
emission and its equivalent width.  
As the $V$-band image has slightly
worse seeing ($\sim1.3\arcsec$), we convolved the NB image to the same
seeing before subtracting the continuum.
The continuum
subtraction has been applied using the following formula
\begin{equation}
{\rm Ly\alpha} = {\rm NB3700}
- \frac{\rm FWHM_{NB3700}}{{\rm FWHM}_{V}}\frac{\rm Tr_{\rm NB3700}}{{\rm Tr}_{V}} \, V,
\end{equation} 
where \mlya\ is the final subtracted image, NB3700 is the narrow-band
image, $V$ is the $V$-band image, and ${\rm Tr_{NB3700}}$ and ${\rm Tr}_V$ are the
peak transmission values for the NB3700 and $V$-band filters,
respectively. The result of this procedure is a Ly$\alpha$ only narrow-band
image, which is shown in the left panel of Fig.~\figimages.


\vskip 0.2in

\subsection{Follow-up Spectroscopy}

After the discovery of an extended nebula in our NB imaging, we opted
to obtain additional spectroscopy of both the diffuse nebula and
several compact sources near the projected quasar pair during the same
observing run.  On UT 13 November 2012 in clear observing conditions
and seeing varying from ${\rm FWHM}= 0.6-1.0\arcsec$, we used the LRIS
spectrometer with a 1.0$\arcsec$ longslit, configured with the D460
dichroic, B1200/3400 grism, and R600/7500 grating.  We obtained two
additional orientations of the longslit as illustrated in
Fig.~\figslits: (Slit~1) along the line connecting source f/g quasar
and AGN2 with a PA=$227^\circ$; (Slit~3) along the line connecting the
b/g quasar and AGN1 with a PA=$343^\circ$. We exposed for a total of
4800s and 2400s for Slit~1 and Slit~3, respectively. The PA of
Slit~1 was chosen to be aligned with the f/g quasar and a nearby radio
source detected in the FIRST survey (see Supp.~2.2), which we
learned about just prior to the November 2012 Keck observations. This 
turned out to be an AGN at the same redshift (AGN2).  The slit
orientation also conveniently allowed us to simultaneously observe a
bright spot in the Ly$\alpha$ image, which is also the companion
AGN1. The orientation of Slit~3 was chosen to be aligned with the b/g
quasar as a position reference and two other bright spots in the
nebula, one of which was also covered by Slit~1,
i.e. AGN1. Coordinates, photometry, and other information about these
targets are listed in Table~\tabagn. Corresponding calibration frames
were obtained and these data were reduced with the same procedures
described above.  The slit orientation of our original January 2007
spectroscopy is also indicated in Fig.~\figslits\ as Slit~2, which
simultaneously observed the f/g and b/g quasars at PA=$25.8^\circ$ for
an exposure time of 1800s (but in better conditions).

Our Ly$\alpha$ image (left panel of Fig.~\figimages) reveals several
compact Ly$\alpha$-emitting sources (LAEs) with rest-frame $W_{\rm
  Ly\alpha} > 20$\AA\ in the vicinity of the f/g quasar, which are
hence very likely to be at the same redshift.  On UT 2012 Dec 14, we
targeted two of these LAEs with the DEIMOS spectrometer \cite{DEIMOS}
on the Keck~II telescope in partly cloudy conditions. Specifically, we
employed the $1.0\arcsec$ long-slit mask oriented to cover the object
labeled AGN3 in Fig.~\figimages\ (see also Table~\tabagn), and another
LAE which we designate as Target1 at RA$=$08:41:58.8,
DEC$=+$39:21:57.4, which lies off of the image in Fig.~\figslits.  The
instrument was configured with the 600ZD grating tilted to a central
wavelength of 7200\AA\ which provides coverage from $\approx
4600-9800$\AA\ with a spectral resolution FWHM$\approx 235\kms$.  We
took two exposures totaling 4500s.  These data were reduced and
extracted with the SPEC2D pipeline \cite{newman13,cooper12} and fluxed
using a spectrum of the spectrophotometric standard star Feige~34
taken that (non-photometric) night with the same instrument
configuration.  Note that given the limited blue sensitivity of
DEIMOS, these spectra did not cover the Ly$\alpha$ emission line at
$\simeq 3700$\AA. Owing to its faint continuum magnitude
($V=25.2$\,mag) the spectrum of Target1 was inconclusive, although its
large Ly$\alpha$ equivalent width $W_{\rm Ly\alpha}=97$\AA\ suggests
it is a real LAE. A setup covering Ly$\alpha$ is likely necessary to
spectroscopically confirm this source. The object AGN3 is an AGN at
the same redshift as the f/g quasar, as described in the next section.

\section{Discovery of Four AGN}
\label{sec:four_AGN}



Our follow-up spectroscopy reveals three additional AGN within $\theta
< 18\arcsec$ or $R_{\perp} < 150\,{\rm kpc}$ of the f/g quasar, with
very similar redshifts. These objects are labeled as AGN1-3 in
Fig.~\figslits, and their optical spectra are shown in
Fig.~\sfigagn. Relevant information  for all five of the AGN
associated with the nebula, namely the f/g quasar, the three AGN
discovered at the same redshift, and the b/g quasar 
are provided in Table~\tabagn. 


Motivated by the discovery of these AGN, we force-photometered the
SDSS and WISE survey images at the coordinates of each AGN, measured
from our deep Keck images. This was performed using a custom
algorithm \cite{tractor},
which simultaneously models the input
sources at their respective locations, as well as all other nearby
sources detected in the SDSS survey imaging.  This modeling fully
accounts for the potentially overlapping PSFs of the sources (a
significant issue for WISE given its FWHM$\approx 6\arcsec$ PSF),
and thus effectively de-blends the photometry. It also utilizes data
from multiple visits, thus effectively co-adding all epochs and
improving the effective depth over that in the published catalogs.
Table~\tabagnphot\ shows photometry for the five AGN.  There we list
the SDSS $ugriz$ and WISE $W1,W2,W3,W4$ measurements determined from
our forced photometry procedure, measurements or limits on the Peak
20cm radio flux $F_{\rm 20cm}$ from the FIRST survey, the LRIS
$V$-band photometry, the Ly$\alpha$ line-flux, and the rest-frame
Ly$\alpha$ equivalent width.

\subsection{AGN1: An Obscured Type-2 Quasar} 

The object AGN1 is embedded in a bright ridge of the nebular
Ly$\alpha$ emission, as shown in our Ly$\alpha$ image and the 2D
spectrum (Fig.~\figslits). Our LRIS spectra have relatively low
continuum signal-to-noise ${\rm S\slash N}\sim 1-2$ ratio, consistent
with the faint continuum magnitude of AGN1: $V=23.95\pm
0.07$. Nevertheless, we detect strong emission in several lines:
\ion{He}{ii}I$\lambda 1640$, \ion{C}{iii}]$\lambda 1908$, \ion{C}{ii}]$\lambda2328$, and a
tentative detection of \ion{C}{iv}$\lambda 1549$, along with bright
Ly$\alpha$, which our narrow-band imaging indicated should be
strong. The FWHM, line flux, and rest-frame equivalent widths (EWs) of
these lines are listed in Table~\tablines.  The presence of emission
in several high-ionization UV lines (i.e. \ion{He}{ii}  and \ion{C}{iv}) characteristic
of AGN, clearly indicates that this source is powered by a
hard-ionizing spectrum, rather than star-formation. Although
high-ionization lines like \ion{C}{iv}, \ion{He}{ii}, and \ion{C}{iii}] can be formed in
starbursts, stellar winds, the photospheres of massive stars, and the
interstellar medium, the spectra of star-forming galaxies typically
exhibit much smaller rest-frame EWs ($\lesssim 2$\AA) \cite{shapley03}
in these lines than observed in our spectra.  Several independent
lines of argument suggest that this object is actually a luminous but
obscured quasar, also referred to as a Type-2 quasar, as we elaborate on
below. From the weak \ion{He}{ii}$\lambda 1640$ line, we measure $z
= \zzagno$ which is $\approx 300 \kms$ lower than the estimates
from \mlya, and \ion{C}{iii}]$\lambda 1908$.  We estimate an uncertainty in
this redshift of $400 \kms$, which arises both because of
line-centroiding error, and possible systematic uncertainty about the
degree to which a high-ionization line like \ion{He}{ii}$\lambda 1640$ traces
the systemic frame for a Type-2 AGN.


According to unified models of AGN  \cite{Antonucci93,UrryPadovani95},
orientation is the primary determining factor governing the
ultraviolet/optical appearance of an AGN. In this context, vantage
points which are not extincted by the obscuring torus observe a
spatially unresolved power-law ultraviolet and optical continuum
believed to emerge from the SMBH accretion disk, and broad
high-ionization emission lines (FWHM$\simeq 5000-20,000\kms$) from
photoionized gas in the so called broad line region at distances $\sim
1\,{\rm pc}$ from the SMBH. However, all vantage points, observe narrow emission
lines from the spatially extended ($\sim 100{\rm pc}-10\kpc$) narrow
(FWHM$\simeq 500-1500\kms$) line region. When the
line-of-sight to the central engine is blocked by a dusty obscuring torus,
only the narrow line region is seen. AGN lacking broad emission lines,
but which nevertheless exhibit narrow high-ionization lines are
classified as Type-2 systems, while those with broad-line emission are
classified as Type-1. Alternatives to the AGN unification viewing angle picture
argue instead  that Type-1s and Type-2s  represent
different phases of quasar evolution \cite{Sanders88},
with all quasars passing through
an obscured phase before outflows expel the obscuring
material.

Characteristics of Type-2 quasars at other wavelengths are a hardened
X-ray spectrum, resulting from photoelectric absorption of soft X-rays
presumably from the same obscuring torus extincting the broad-line
region, and strong mid-IR ($\lambda \sim 10\,\mu {\rm m}$) emission,
which represents ultraviolet/optical energy emitted from the disk, but
absorbed, reprocessed, and re-emitted isotropically by the torus.  In
a subset of radio-loud systems, synchrotron radiation, believed to be
emitted isotropically from scales much larger than the torus, is also
observed. 


Luminous high-redshift radio-loud Type-2 quasars have been studied for
decades (see the review by \cite{mccarthy93})
but their radio-quiet counterparts have been much harder to identify.
Over the past decade, significant progress has been made in
identifying sizable samples of low-redshift ($z\lesssim 1$)
radio-quiet Type-2s from mid-IR \cite{Lacy04,Stern05}, 
X-ray \cite{BrandtHasinger05},
and optical narrow-line emission \cite{Zakamska03}.
However, to date the vast majority of these
Type-2s are at $z < 1$, and there are still only handfuls of
bonafied Type-2 quasars known at $z\sim 2$ \cite{Brusa10}
with bolometric
luminosities comparable to the typical SDSS/BOSS Type-1 quasar, $L_{\rm
bol}\gtrsim 10^{46}\,{\rm erg\,s^{-1}}$


The three lines of argument that we use to establish that AGN1 is a
Type-2 quasar are 1) the narrow velocity width of its emission lines 2)
the line ratios in its spectrum, and 3) its extremely red optical-to-mid-IR colors. 

Unlike the three other AGN physically associated with the nebula (f/g
quasar, AGN2, and AGN3), which exhibit broad emission lines (FWHM$\simeq
5000-10,000\kms$), Gaussian fits to the lines in AGN1 reveal
relatively narrow lines $\lesssim 1500\kms$, which are listed in
Table~\tablines.  At low-redshifts ($z < 0.3$), AGN exhibit a bimodal
distribution of H$\alpha$ emission line widths, which corresponds to
the Type-1 and Type2 populations. Based on this distribution, the
canonical dividing line between Type-1/Type-2 quasar classification is
${\rm FWHM_{\rm H\alpha} < 1200\,\kms}$ \cite{Hao05}.
However, it is
unclear how to extrapolate this classification to the rest-frame UV
lines observed at $z\sim 2$.  First, higher redshift sources are
intrinsically much more luminous, have more massive SMBHs, and hence
SMBH-galaxy scaling relations imply that they should be hosted by much
more massive galaxies. Indeed, for the handful of Type-1 quasars at
$z\sim 2$ for which the narrow-line region [\ion{O}{iii}] emission line has
been observed in the near-IR, larger line-widths FWHM$=1300-2100\kms$
are indeed observed \cite{Kim13},
contrasting with narrower line
widths observed at lower redshift \cite{Hao05}.
Furthermore, even in
the narrow-line region, one generally expects larger widths for
Ly$\alpha$ and higher ionization potential lines, because
photoionization modeling generally predicts that emission in these
lines is produced at smaller distances from the
nucleus, where velocities are likely to be larger. 
Indeed, exactly such trends were observed in the spectrum of a Type-2
quasar at $z=3.288$ discovered by \cite{Stern02},
who measured ${\rm
FWHM}_{\rm Ly\alpha}=1520\pm30\kms$, ${\rm FWHM}_{\rm
HeII}=940\pm140\kms$, and ${\rm FWHM}_{\rm CIII]}=1090\pm 140\kms$,
which are comparable to our measured values for AGN1. A near-IR
spectrum of the Stern et al. source reveals narrower widths for the
rest-frame optical lines that are typically used to classify
low-redshift Type-2s: ${\rm FWHM}_{\rm H\beta}=170\pm130\kms$ and
${\rm FWHM}_{\rm [OIII]}=430\pm 30\kms$. Note that the unusually hard
X-ray spectrum of the Stern et al. source \cite{Stern02}
and the
implied high photoelectric absorbing column $N_{\rm H} =(4.8\pm
2.1)\times 10^{23}\,{\rm cm}^2$ leave little doubt that it is a
bonafied Type-2 quasar, notwithstanding the fact that Ly$\alpha$ and
other high-ionization UV lines have line widths in excess of ${\rm
FWHM}=1200\,\kms$.  We suspect that the canonical rest-frame optical
line-width classification based on $z < 0.3$ Type-2
quasars \cite{Hao05}
needs to be revised for rest-frame UV lines in
more luminous Type-2s at $z\sim 2$.
Based on the narrow lines of AGN1 and their similarity to the Stern et
al. source \cite{Stern02},
we conclude that it is very likely to be a
Type-2 quasar.



Table~\tablines\ presents measurements of the line fluxes and
measurements or limits on rest-frame equivalent width for the
characteristic AGN lines covered by our spectrum. Based on these line
flux measurements, Fig.~\sfiglineratio\ shows the line-ratios of
AGN1 in the \ion{C}{iv}$\slash$\ion{He}{ii} vs \ion{C}{iv}$\slash$\ion{C}{iii}]
plane (left) as well as the \ion{C}{iii}]$\slash$\ion{He}{ii}
vs \ion{C}{iii}]$\slash$\ion{C}{ii}] plane (right). These are the standard
line ratios conventionally discussed in studies that use
photoionization modeling to diagnose the physical conditions in the
narrow-line
region \cite{Groves04}, 
although we note that the ratios \ion{C}{iii}]$\slash$\ion{He}{ii}
vs  \ion{C}{iii}]$\slash$\ion{C}{ii}] are not typically plotted against each
other.  In Fig.~\sfiglineratio, we compare the line ratios of AGN1 to other
Type-2 AGN compiled from the literature. Specifically, the circles are
individual measurements of HzRGs (black) from the compilation
of \cite{debreuck00},
and narrow-line X-ray sources (cyan)
and Seyfert-2s (blue) from the compilation
of \cite{NagaoNLR06}. 
In addition, triangles indicate
measurements from the composite spectra of HzRGs (orange)
from \cite{mccarthy93},
the composite Type-2 AGN spectra (purple) 
from \cite{Hainline11},
who split their population into two
samples above and below Ly$\alpha$ EW of 63\AA, and a composite
spectrum of mid-IR selected Type-2 AGN (green) from \cite{Lacy13}. 
Finally, the stars represent measurements of these line ratios from composite
spectra of Type-1 quasars, based on the analysis
of (magenta) \cite{vanden01},
(red-magenta) \cite{NagaoBLR06},
and (blue-magenta) \cite{Zheng97}. 

We caution that determination of robust emission line fluxes for
Type-1 quasars is extremely challenging. Generally the broad emission
lines, non-trivial emission line shapes, and the fact that many of the
lines of interest are severely blended, makes the separation of the
spectrum into line and continuum rather ambiguous.  To minimize the
impact of noise and variations among quasars, one typically analyzes
extremely high signal-to-noise ratio composite spectra, which also
average down quasar-to-quasar
variation \cite{Zheng97,vanden01,NagaoBLR06}.
But nevertheless, the
resulting line fluxes and hence line flux ratios are dependent on the
method adopted (see \cite{NagaoBLR06})
section 4.1 for a detailed
discussion).  These issues are particularly acute for the lines that
we consider: \ion{He}{ii}$\lambda 1640$ is heavily blended with the red wing of
\ion{C}{iv} $\lambda 1549$ and \ion{O}{iii}]$\lambda 1663$, \ion{Al}{ii}$\lambda 1670$, and a
broad \ion{Fe}{ii} complex, \ion{C}{iii}]$\lambda 1909$ is severely blended with
\ion{Al}{ii}$\lambda$ 1857 and \ion{Si}{iii}$\lambda$1892, and \ion{C}{ii}]$\lambda$2326 is
blended with a broad \ion{Fe}{ii} complex. These ambiguities are reflected in
the large variation in the line ratios determined from Type-1 quasar
composites in Fig.~\sfiglineratio\ (stars) by different
studies. Despite these challenges, Fig.~\sfiglineratio\
clearly illustrates that the line-ratios for AGN1 are consistent with
the Type-2/Seyfert-2 locus, and differ rather significantly from the
Type-1 line ratios, notwithstanding the somewhat divergent
measurements for the latter.  We thus conclude based on the emission
line ratios of AGN1 that it is very likely to be a Type-2 AGN.
 
Many studies utilizing WISE data have shown that at bright
magnitudes $W2 \simeq 15$\,mag, the WISE $W1-W2$ color efficiently selects
AGN at $z\sim 1-3$, including obscured
objects \cite{Stern12,Assef13,Yan13}.
This results from the
rising roughly power law shape of AGN SEDs in the
mid-IR \cite{Polletta07,Assef10},
and the fact that stellar
contamination is low, because the WISE $W1$ and $W2$ bands cover the
Rayleigh-Jeans tail of emission from Galactic stars, resulting in much
bluer colors which cleanly separate from redder AGN. At fainter
magnitudes $W2\simeq 17$, contamination increases because of optically
faint, high-redshift elliptical and Sbc galaxies, which tend to also
have very red $W1-W2$ colors.  In addition, obscured AGN are also
expected to have very red optical to mid-IR colors, if the mid-IR
traces reprocessed dust emission from a heavily extincted accretion
disk emitting in the optical/UV. Indeed, it has been shown that the
color-cut $r-W2> 6$ effectively separates obscured AGN from the their
much bluer unobscured
counterparts \cite{Hickox07,Hickox11,Yan13}.
The WISE
photometry in Table~\tabagnphot\ indicate that AGN1 is detected in
both $W1$ and $W2$ with $W1=16.89\pm0.08$, and $W2=15.86\pm 0.11$,
implying a $W1-W2=1.03\pm 0.14$, thus consistent with the expected
red color $W1-W2 > 0.8$ of AGN in the WISE bands. If we adopt our LRIS
V-band magnitude as a proxy for $r$ (color corrections are minimal),
then the faint $V=23.95$ continuum of AGN1 implies $r-W2=8.09\pm
0.13$, which is also consistent with the $r-W2> 6$ criterion for
obscured AGN. We thus conclude based on mid-IR and mid-IR to optical
colors, that AGN1 is consistent with being an obscured AGN.


%
%
%
%
%
%
%
%
%
%
%
%

\subsection{AGN2: A Radio Loud Broad Line AGN}

The second quasar (AGN2) is rather faint $V=23.12$
(Table~\tabagn), but its LRIS spectrum nevertheless reveals
broad \mlya, \ion{C}{iv}$\lambda 1549$, and \ion{C}{iii}]$\lambda 1909$ emission lines,
unequivocally indicating that it is a broad-line Type-1 AGN (see
Fig.~\sfigagn).  Centroiding the broad and relatively low S/N
\ion{C}{iii}]$\lambda 1908$ line, we estimate $\mzagnw = \zzagnw$ with a
$700 \kms$ uncertainty. Redshift errors for rest-frame ultraviolet
emission lines are dominated by intrinsic shifts between the emission
line and the true systemic frame, and we adopt the procedure described
in \cite{QPQ1}
to centroid the broad lines, and use the procedure
of \cite{Shen07}
to determine the redshift errors.

Motivated, in part, by the previous association of bright Ly$\alpha$
nebulae to HzRGs and/or bright radio sources, we searched the
literature and public catalogs for sources in the field
surrounding \qso.  The Faint Images of the Radio Sky at Twenty cm
survey (FIRST) \cite{FIRST}
used the Very Large Array (VLA) to produce a
map of the 20 cm (1.4 GHz) sky with a beam size of $5.\arcsec4$ and an
RMS sensitivity of about 0.15 mJy beam$^{-1}$.  The survey covers the
same 10,000 deg$^2$ sky region covered by the SDSS imaging, 
has a typical detection threshold of 1 mJy, and an astrometric accuracy of
$0.05\arcsec$. 
The FIRST catalog reports only a single source within 60$''$ of
\qso at RA=08:41:58.6 DEC=+39:21:14.7, approximately $6''$ South of
the f/g quasar and coincident with AGN2.
The FIRST peak and integrated fluxes at 20\,cm are 
$F_{\rm peak}=12.87$~mJy/beam and $F_{\rm int}=$14.81~mJy 
respectively, giving a morphological parameter $\Theta =
\log_{10}(F_{\rm int}/F_{\rm peak}) = 0.06$.  This indicates a
compact, unresolved source at 1.4\,GHz \cite{Montenegro08}.
Therefore, there is no evidence for a bright, extended radio source (e.g.\
a radio jet) associated with the protocluster system and giant \mlya\ nebula.

Follow-up VLA observations of this source were performed by \cite{Dipompeo11},
having erroneously associated it with the f/g quasar.  The spectral slope
measured from 8.4\,GHz and 4.9\,GHz observations is relatively steep,
$\alpha_{4.9}^{8.4} = -1.21$, which is generally interpreted as
evidence that the source has a large viewing angle (i.e. edge-on) \cite{Fine11}.
This may indicate that \rqso\ (AGN2) is oriented away
from us, although we re-emphasize that it exhibits a Type-1 spectrum.
These much deeper radio images of the source show no evidence for
extended structure or jets, providing further evidence that it is not
a HzRG \cite{Dipompeo11}.

\subsection{AGN3: A Faint Broad Line AGN}

Similar to AGN2, the source AGN3 is also faint $V=23.09$ but exhibits
broad emission-lines of \ion{C}{iv}$\lambda\,1549$, \ion{C}{iii}]$\lambda\,1908$, and
\ion{Mg}{ii}$\lambda\,2798$ (see Fig.~\sfigagn).  Unfortunately \mlya\ was
not covered by our DEIMOS spectrum, but the source has a large
rest-frame equivalent width ($W_{\rm Ly\alpha}=35$\AA)
indicative of strong emission.  Based on the broad lines, we also
classify this source as a broad-line (Type-1) AGN.  We determine a
redshift of $\mzagnr = \zzagnr$ from the \ion{C}{iii}]$\lambda 1908$
line with a $700 \kms$ uncertainty.


\section{Probability of Finding a Quadruple Quasar}\label{sec:clust}

The probability $dP$ of detecting three quasars around a single known 
quasar (here the f/g quasar) can be written as
\be
\begin{array}{rlr}
dP  & = n_{\rm QSO}^3 dV_2 dV_3 dV_4\left[1 \right. &  \label{eqn:cluster}\\
          &  + \ \xi(r_{12}) + \cdots  & (6 \ {\rm permutations})\\          
          & +  \ \zeta(r_{12},r_{23},r_{31}) + \cdots  & (4 \ {\rm permutations})\\ 
          & +  \ \xi(r_{12})\xi(r_{34}) + . . .  & (3 \ {\rm permutations}) \\               & + \  \eta(r_{1},r_{2},r_{3},r_{4})\left.\! \right] & 
\end{array}
\ee
where $n_{\rm QSO}$ is the number density of quasars, $d V_i$
is the infinitesimal volume element centered on the location $\vec{r}_i$
of the $i$th quasar, $r_{ij}$ is the distance between the
$i$th and $j$th quasar, and $\xi$, $\zeta$, and $\eta$ are the
two-point, three-point, and four-point correlation functions,
respectively \cite{Peebles80}.  

The total probability $P$ of finding a quadruple quasar system within a maximum
radius $r_{\rm max}$ is the integral $\int d P$ over the three
volume elements $dV_i$, for all possible configurations of the
$\vec{r}_i$. Our goal in what follows is to obtain an order of magnitude
estimate for this total probability. 

First note that on the small scales of interest to us (here $r
\lesssim 200\,{\rm kpc}$), many of the terms in
eqn.~(\ref{eqn:cluster}) can be neglected. For the higher order
correlation functions $\zeta$ and $\eta$, a scaling of the form $\zeta
\propto \xi^2$ and $\eta \propto \xi^3$ is often assumed. This arises
from the fact that for Gaussian initial conditions and a scale-free
initial power spectrum, this scaling is obeyed to second order in
Eulerian perturbation theory \cite{Fry84}.  However, this scaling is
not valid for the highly non-linear small-scales of interest to us
here. On these small scales, we follow the halo model approach
\cite{CooraySheth02}, and write the higher order functions as a sum of
terms representing contributions from the multiple possible halos. For
example the four-point function can be written as the sum \be \eta =
\eta^{\rm 1h} + \eta^{\rm 2h} + \eta^{\rm 3h} + \eta^{\rm 4h}, \ee
resulting from contributions from one to four halos. As we are
concerned with scales comparable to the virial radius of the dark
matter halo hosting, i.e. the f/g quasar $r \lesssim 200\,{\rm kpc}$,
the one halo term, which quantifies the four-point correlations when
all four points lie in the same dark matter halo, will dominate
\cite{CooraySheth02}.  It is easier to calculate this one-halo term of
the four-point function $\eta^{\rm 1h}$ in Fourier space, where one
instead works with the one halo term of the trispectrum $T^{\rm 1h}$,
and it can be shown that $T^{\rm 1h}$ scales as the Fourier transform
of the dark matter halo density profile to the fourth power $T^{\rm
  1h}\propto \rho_{\rm DM}^4$. This scaling holds likewise for $\eta$.
Thus for quasar 4-point correlations on small-scales, the halo model
indicates that we expect $\eta \propto \rho_{\rm QSO}^4$, where
$\rho_{\rm QSO}$ is the density profile of quasars in the dark matter
halos which host them.

By a similar line of argument, the small-scale two- and
three-point functions scale as $\xi \propto \rho_{\rm QSO}^2$ and
$\zeta \propto \rho_{\rm QSO}^3$, respectively. Or alternatively,
$\zeta \propto \xi^{3 \slash 2}$ and $\eta \propto \xi^2$.  On the
proper scales of interest to us here $r \lesssim 200\,{\rm kpc}$
($r_{\rm com}=600{\rm kpc}$, comoving), $\xi \gg 1$ and thus
the terms dominating eqn.~(\ref{eqn:cluster}) are the three
permutations of terms like $\xi(r_{12})\xi(r_{34})$, and $\eta$, which
are all of comparable order $\xi^2$.

Thus in order to compute the total probability of finding three
quasars around a known quasar, we must integrate the four dominant
terms in eqn.~(\ref{eqn:cluster}), over all possible configurations of
the $\vec{r}_i$, which can be realized within a spherical volume, whose
radius is set by the largest separation in the quadruple quasar
$r_{\rm max}$. These integrals all have a common form and order-of-magnitude. For
example the first three terms involving pairs of correlation functions can be
written
\be
n_{\rm QSO}^3\left(\int_{0}^{r_{\rm max}}\xi(r) 4\pi r^2 dr\right)\left(\int \xi(r_{34}) dV_3dV_4\right) \sim 
n_{\rm QSO}^3 \left(\int_0^{r_{\rm max}} \xi(r) 4\pi r^2 dr\right)^2 V, 
\ee 
where $V = 4\pi\slash 3 r_{\rm max}^3$.  Thus the final expression for
an order-of-magnitude estimate of the probability is
\be
P\sim 4 n_{\rm QSO}^3 \left(\int_0^{r_{\rm max}} \xi(r)
4\pi r^2 dr\right)^2 V \;\;, \label{eqn:clust_approx}
\ee
where the factor of four accounts for the four similar 
dominant terms of order $\sim \xi^2$ in eqn.~(\ref{eqn:cluster}). 

To calculate $n_{\rm QSO}$, we use the quasar luminosity function
of \cite{HRH07}
assuming an apparent magnitude limit $V<24$. This
limit is highly conservative, as even AGN2 and AGN3 have $V=23.1$\,mag.
Although the Type-2 AGN1 has $V=24$, its nucleus is obscured from our
perspective, and this magnitude represents host-galaxy emission or a
small-fraction $\sim 5\%$ of scattered nuclear emission. The
corresponding nuclear $V$-band emission of this AGN is likely to be
much brighter than $V=24$ given that its Ly$\alpha$ flux, coming from
the narrow-line region, is so strong, i.e. larger than that of
AGN2. To account for the fact that our luminosity function only
includes unobscured AGN, we 
divide it by a factor $(1-f_{\rm
obs}$), where $f_{\rm obs}$ is the obscured fraction of AGN. We adopt
an obscured fraction of $f_{\rm obs} = 0.5$ consistent with recent
determinations \cite{reyes08,Lusso13}.
Altogether, we estimate 
that the comoving number density of faint AGN is $n_{\rm QSO}=3.8\times
10^{-5}\,{\rm cMpc^{-3}}$. 

We assume that all four AGN reside in a physical structure with
physical size $r_{\rm max}$. This is a good assumption, given that:
the projected separations of the AGN are small, they reside in a
larger-scale overdensity of Ly$\alpha$ emitters, the f/g quasar, AGN1,
and AGN2 are all embedded in the giant Ly$\alpha$ nebula, and the morphology
of the nebular emission around the AGN suggests a physical
association. Moreover, although we argue here that the probability of
finding a quadruple AGN is very small, it would be much smaller if the
AGN had larger line-of-sight separations, in which case one cannot
invoke as large of an enhancement due to clustering. Taking the f/g quasar
location as the origin, the largest angular separation measured in the
quadruple quasar is that of AGN3 with $\theta=\mxaoff$,
corresponding to to a transverse separation $r_{\perp}=\mxloff \,{\rm
kpc}$.  We thus choose $r_{\rm max} = 250\,{\rm kpc}$, such that the
the spherical comoving volume we
consider corresponds to $V=1.9\,{\rm cMpc^{3}}$. 
The probability of finding the three AGN around the f/g quasar at random, i.e. in the
absence of clustering, is extremely small 
$P_{\rm ran}=(n_{\rm QSO}V)^3 = 3.4\times 10^{-13}$.

The small-scale two-point correlation function of quasars was first
computed by (6), and later extended to even smaller scales $\sim
5\,{\rm kpc}$ by the gravitational lensing search of \cite{Kayo12},
who found that a power law form $\xi = (r\slash r_0)^{-\gamma}$ with
$\gamma = -2$ and $r_0=5.4\pm 0.3\,h^{-1}\,{\rm cMpc}$ 
provides a good fit to the data over a large range of
scales. Plugging these numbers into eqn.~(\ref{eqn:clust_approx}), 
we finally arrive at a probability of $P = 6.9\times 10^{-8}$,
justifying our order of magnitude estimate of $P \sim 1\times 10^{-7}$.

\section{Giant Nebulae-LAE Clustering Analysis}
\label{sec:LAE}

\subsection{Constructing the LAE Catalog}

We created an LAE catalog from our images of SDSSJ0841$+$3921
using {\it SExtractor} \cite{bertin96}
in dual-mode, using the
NB image as the reference image.  In order to minimize spurious
detection we varied the parameters ${\bf \rm DETECT\_MINAREA}$
(minimum detection area) and\\
${\bf \rm DETECT\_THRESH}$ (relative
detection threshold) creating a large number of LAE candidate
catalogs.  We verify the number of spurious detections for each
catalog running ${\bf \rm SExtractor}$ on the ``negative'' NB image
obtained by multiplying the NB image by $-1$. The ratio between the
number of sources detected in the NB image and the sum of sources
detected in both the NB and ``negative'' image define a ``reliability''
criterion. We selected ${\bf \rm DETECT\_MINAREA}$ equal to 8 pixels
and ${\bf \rm DETECT\_THRESH}$ equal to 1.8$\sigma$ corresponding to
a "reliability" of $94\%$.  We used the same parameters to obtain a
catalog of sources for the $V$-band image.  In order to select LAE
candidates, we estimated the ${\rm NB}-V$ color from {\it SExtractor}
isophotal magnitudes and applied a color-cut corresponding to a
Ly$\alpha$ rest-frame equivalent width of $W_{\rm Ly\alpha}>20$\AA\,
which is the standard threshold used in the literature. In particular,
we assumed a flat continuum slope in frequency to convert the V-band
magnitude to a continuum flux at the Ly$\alpha$ wavelength. For the
objects in the catalog without broad-band detection (i.e., with
S/N$<3$), we determine a lower limit on the EW
following \cite{cantalupo12}.
Finally, we inspected each object by
eye and removed four spurious sources that were lying in proximity to a
bright star.  

The final clean catalog consisted of 61 LAE candidates above a
Ly$\alpha$ flux of limit of 5.4$\times10^{-18}$ erg s$^{-1}$
cm$^{-2}$, over our $6^{\prime} \times 7.8^{\prime}$ imaging field-of-view, and
the redshift range $z=2.030-2.057$ spanned by the narrow-band
filter. We estimated that our catalog is 50\% (90\%) complete above a
flux limit of $F_{50}=6.7\times10^{-18}$ erg s$^{-1}$ cm$^{-2}$
($F_{90}=7.4\times10^{-18}$ erg s$^{-1}$ cm$^{-2}$).
This catalog of LAEs is presented in Table~\tablae.  In the following
clustering analysis, we consider only those sources with rest-frame
$W_{\rm Ly\alpha}>20$\AA\, and a luminosity $\log_{10} L_{\rm
Ly\alpha} > 42.1$ which at $z=2.04$ corresponds to a flux limit of
$4.0\times10^{-17}$\,erg s$^{-1}$ cm$^{-2}$. Given that this flux
level is a factor of five higher than the $90\%$ completeness flux
limit of our catalog, we are certain that we are complete to such
sources. The f/g quasar lies at a distance of $2.08^{\prime}$ from the
edge of our imaging field, and for simplicity, our clustering analysis
considers only the 10 sources within this radius which corresponds to
a comoving impact parameter of $2.2\,h^{-1}\,{\rm cMpc}$. Of these,
3/10 are the spectroscopically confirmed AGN1-3. The sources included
in our clustering analysis are denoted by an asterisk in
Table~\tablae.

\subsection{Clustering Analysis of HzRGs and LABs}
We now describe how we used a compilation of LAEs around HzRGs and
LABs to estimate the giant nebulae-LAE cross correlation function.
The vast majority of $z \sim 2-3$ protoclusters in the literature have
been identified and studied via overdensities of LAEs over comparable
fields of view $6^{\prime} \times 7.8^{\prime}$ as our LRIS observations. For the HzRGs,
we used a catalog of LAE positions around HzRGs from the survey
of \cite{Venemans07}.
Specifically, we focus on six HzRGs from the
Venemans et al. study in the redshift range $z=2.06-3.13$, which are
approximately co-eval with \qso at $z=2.0412$.  In addition to the six
HzRGs from Venemans et al. \cite{Venemans07},
we also include two LABs
at $z\sim 2-3$ from the literature whose environments have been
surveyed for LAEs. Nestor et al. \cite{nestor11}
conducted a narrow
band imaging survey of the well known SSA22 protocluster field at $z =
3.10$. Their Keck/LRIS observations were centered on LAB1, whereas
LAB2, the other known LAB in this field, resides at the edge of their
imaging field. For this reason we only consider the overdensity around
LAB1.  Prescott et al. \cite{prescott08}
conducted an
intermediate-band imaging survey for LAEs around an LAB at $z=2.66$
(LABd05) discovered by Dey et al. \cite{dey05},
finding a significant
overdensity, and argued that the LAB resides in a protocluster.
The eight objects used in our
clustering analysis are listed in Table~\tabhzrg.

Given that the Venemans et al. sample \cite{Venemans07}
is the largest
compilation of LAEs around HzRGs, and that, to our knowlege, the two
LABs we consider are the only ones whose environments have been
characterized using LAEs to a depth comparable to our observations,
this sample of eight HzRGs/LABs is the only dataset currently
avaailable for conducting our clustering analysis and we believe that
they comprise a fairly representative sample. Venemans et al. applied
standard HzRG selection criteria to define their sample, and they
published results for all HzRGs, not just those that hosted dramatic
overdensities. Although LAB1 and LABd05 are larger and brighter than
more typical LABs, they are comparable in size and luminosity to the
HzRGs and to SDSSJ0841+3921, making for a fair comparison. One cause
for concern is that the distribution of LAEs around the LABs were
published specifically because these two systems resided in
overdensities, and thus there may be a publication bias against LABs
residing in lower density environments. Given that LABS only comprise
a fourth of our clustering sample, we do not believe that this
significantly biases our results, but repeating our clustering
analysis for a larger and more uniformly selected sample of LABs would
clearly be desirable.

Note that the HzRG MRC~1138$-$2262 included in our analysis is the
famous Spider Web Galaxy protocluster, which has been the subject of
extensive multi-wavelength follow-up, and is one of the most dramatic
protocluster systems known.
Venemans et al. \cite{Venemans07}
argued that all of the HzRGs in
Table~\tabhzrg\ reside in overdensities of LAEs, with the exception of
MRC~2048$-$272 and TN~J2009$−$3040, for which the abundance of LAEs
was found to be consistent with the field value, albeit with large
error bars. We nevertheless include these two sources in our
clustering analysis for several reasons.  Our objective for the
clustering analysis is to shed light on the connection between active
SMBHs, large-scale Ly$\alpha$ nebulae, and protocluster
environments. Given that it has been argued that the HzRGs as a
population trace extremely overdense protocluster environments, and
given that MRC~2048$−$272 and TN~J2009$−$3040 are both powerful HzRGs
with large Ly$\alpha$ nebulae, excluding them from the analysis would
be arbitrary, and would bias our correlation function high.
Furthermore, the background number density of LAEs used by Venemans et
al. \cite{Venemans07}
is rather outdated, the overdensity estimates
have large error bars for individual systems, and the luminosity
limits that they adopted for the LAEs in each field considered were
heterogeneous. These factors complicate the interpretation,
particularly if the clustering of LAEs is luminosity dependent.  Hence
our goal is to conduct a homogeneous analysis of the clustering of
LAEs around the HzRGs/LABs which are above a uniform Ly$\alpha$
luminosity $\log_{10}L_{\rm Ly\alpha} > 42.1$, without cherry picking
specific objects.  We adopt this luminosity limit for two reasons.
First, the NB imaging observations for all eight objects listed in
Table~\tabhzrg\ are complete for identifying LAEs above this
luminosity, mitigating uncertainties due to incompleteness. Second,
this limit corresponds to the faintest luminosity for which the the
background number density of LAEs is well
characterized \cite{ciardullo12};
working fainter would thus require a
significant extrapolation of the luminosity function, which would add
systematic uncertainties to our results. For all of the objects
analyzed (see Table~\tabhzrg), the LAE catalogs available are complete
above the Ly$\alpha$ equivalent width limit of $W_{\rm
Ly\alpha} > 20$\AA\ (the same value used to define our LAE catalog
around \qso), with the exception of the Prescott et al. observations
of the LAB LABd05. These observations used an intermediate band
filter, and thus adopt a rest-frame $W_{\rm Ly\alpha} >
40$\AA. We consistently account for this difference below when we
compute the background number density of LAEs.

Given the small size of this HzRG/LAB dataset, we use an unbinned
maximum likelihood estimator to determine the parameters $r_0$ and
$\gamma$, following the procedure outlined in \cite{Croft97,Shen10}.
Specifically, we assume that the cross-correlation function between
the giant nebulae (HzRGs/LABs) and LAEs obeys a power law form
$\xi(r) \equiv (r\slash r_0)^{-\gamma}$.  We use this correlation
function to calculate the expected number of LAEs within a comoving
cylindrical volume with transverse separation from $R$ to $R + dR$,
and half-height $\Delta Z$, which is set by the width of the given
narrow band filter $\Delta Z \equiv {\rm FWHM} \slash (2 a H(z))$,
where ${\rm FWHM} = c\Delta\lambda\slash \lambda$ is the full width
half maximum of the filter in velocity units, $a=1\slash (1 + z)$ is
the scale factor, $H(z)$ is the Hubble expansion rate, and division by
$aH(z)$ converts velocities to comoving units. We imagine dividing the
transverse separation $R$ into a set of infinitesimal bins of width
$dR$, such that each bin can contain only one or zero LAEs. Under the
assumption that the clustering of LAEs around the giant nebulae
is a
Poisson process, we can write the likelihood function as
\be
\mathcal{L} = \left[\prod_i^{N_{\rm LAE}} e^{-\mu_i}\mu_i\right]\left[\prod_{j\ne i}e^{-\mu_j}\right], 
\ee
where $\mu = 4\pi R \Delta Z n_{\rm LAE}[1 + h(R)]dR$ is the
probability of finding a pair in the interval $dR$, the index $i$ runs
over the $dR$ bins containing the $N_{\rm LAE}$ HzRG/LAB-LAE pairs in
the sample, and the index $j$ runs over all the bins for which there
are no pairs. Here $n_{\rm LAE}(>\!L_{\rm Ly\alpha})$ is the number
density of LAEs brighter than the survey limit $L_{\rm Ly\alpha}$, and
$h(R)$ is the cross-correlation function averaged over the filter
width
\be
h(R) = \frac{1}{2\Delta Z} \int_{-\Delta Z}^{\Delta Z}\xi(\sqrt{R^2 + Z^2})dZ
\ee

Taking the natural logarithm of the likelihood, we have 
\be
\ln \mathcal{L} = \sum_{i}^{N_{\rm LAE}} \ln\left[1 + h(R_i)\right] - 2n_{\rm LAE}\Delta Z \int_{R_{\rm min}}^{R_{\rm max}} 2\pi R h(R)dR \label{eqn:lnL},  
\ee
where we have dropped all additive terms which are independent of our
model parameters ($r_0$,$\gamma$), and [$R_{\rm min}$,$R_{\rm max}$]
is the range of comoving scales over which we search for HzRG/LAB-LAE
pairs. In the following clustering analysis, we adopt $R_{\rm min} =
50\,h^{-1}\,{\rm ckpc}$ to minimize confusion with LAEs embedded in
the respective nebulae, and $R_{\rm max} = 2.5\,h^{-1}\,{\rm cMpc}$,
which is well matched to the maximum impact parameter probed $R =
2.2\,h^{-1}\,{\rm cMpc}$ around the f/g quasar.  Given that the HzRGs
and LABs in Table~\tabhzrg\ all have different filter widths, as well
as different $W_{\rm Ly\alpha}$ limits and redshifts (which
changes the background density $n_{\rm LAE}$), the likelihood in
eqn.~(\ref{eqn:lnL}) is specific to a single source. However, the full
likelihood of the entire dataset given the model parameters, can be
determined by simply summing log-likelihoods (multiplying the
likelihoods) over all of the HzRGs/LABs, each of which has the form of
eqn.~(\ref{eqn:lnL}),

Our clustering analysis relies on a determination of the field LAE
luminosity function $n_{\rm LAE}$.  We use the Ciardullo et
al. \cite{ciardullo12} Schechter function fits to the luminosity
function from their own LAE survey at $z = 3.1$, and the Ciardullo et
al. fits to the LAE survey data of \cite{guaita11}
at $z=2.1$. Both of these surveys are highly
complete down to our limiting luminosity of $\log_{10} L_{\rm
Ly\alpha} > 42.1$ for LAEs with equivalent width 
$W_{\rm Ly\alpha} > 20$\AA.
Ciardullo et al. also estimated the rest-frame equivalent width
distribution, and found that LAEs follow an exponential distribution
with rest-frame scale lengths of $W_0=50$\AA\ 
and $W_0=64$\AA, at $z =
2.1$ and $z = 3.1$ respectively.  Hence, for LABd05 which employed an
$W_{\rm limit} > 40$\AA, we take the luminosity function of LAEs to
be $n_{\rm LAE}(> W_{\rm limit}, > L_{\rm Ly\alpha}) = \exp{[-(W_{\rm limit} - 20{\rm \AA})\slash W_0]} n_{\rm LAE}(> 20\AA, > L_{\rm
Ly\alpha})$, where $n_{\rm LAE}(> 20\AA, > L_{\rm Ly\alpha})$ are the
Ciardullo et al. Schechter-function fits. To determine the luminosity
function at intermediate redshifts between $z = 2.1$ and $z = 3.1$, we
linearly interpolate.

In Fig.~\sfigcontour\ we show confidence regions in the
$r_0-\gamma$ plane for the likelihood in eqn.~(\ref{eqn:lnL}). The
maximum likelihood value is $r_0= \rnotval\,h^{-1}\,{\rm cMpc}$ and $\gamma
= \gammaval$, however a strong degeneracy exists between $r_0$ and $\gamma$,
such that neither is individually well-determined. The low
signal-to-noise ratio of the clustering measurement is partly to blame
for this degeneracy, however it is also a consequence of the
functional form adopted for the correlation function $\xi(r) \equiv
(r\slash r_0)^{-\gamma}$, for which the amplitude and slope of the
correlation function are not independent parameters.

Note that errors on the parameters $r_0$ and $\gamma$ given by our
maximum likelihood estimator in eqn.~(\ref{eqn:lnL}) and illustrated
in Fig.~\sfigcontour\ underestimate the true errors.  This is because
our likelihood explicitly assumes that the positions of the LAEs are
completely independent draws from a Poisson process.  This approach
ignores correlations between the LAE positions (which are generically
present for hierarchical clustering) and fluctuations due to cosmic
variance. We opted for this unbinned estimator, because the
traditional method of computing the correlation function in impact
parameter bins, and fitting a power-law will, for such small samples,
yield results which depend sensitively on the choice of binning,
unless the covariance of the bins is included in the fit. The correct
approach would be to fit the binned correlation function using the
correct covariance matrix, whose off-diagonal elements would reflect
the correlations between LAEs, and whose amplitude would include
fluctuations due to sample variance. Given our limited sample size of
eight HzRGs/LABs, the total number of LAEs in our clustering analysis
is only \numlae, and there is no hope of computing the covariance from
the data itself. In principle one could adopt a model for the
covariance based on $N$-body simulations, but this would require a
full model for how LAEs populate dark matter halos, and would be
sensitive to the assumed masses of the protocluster halos. Such a
detailed analysis is beyond the scope of the present work. As a
compromise, we use the unbinned maximum likelihood estimator in
eqn.~(\ref{eqn:lnL}) to determine our best fit model parameters, and
we obtain estimates of the errors on parameters via a bootstrap
technique.  Specifically, we generate 1000 realizations of mock
datasets, where a random sampling of the eight HzRGs/LABs in
Table~\tabhzrg\ are selected with replacement. Given that we are
measuring a cross-correlation of LAEs around eight distinct objects,
only the LAE positions arising from the same source are correlated,
whereas the others are completely independent because of the large
respective distances between the HzRGs/LABs. As such, for each of
these 1000 realizations, we also bootstrap resample the positions of
the LAEs around each object (with replacement), which encapsulates
fluctuations due to correlations in the LAE positions and Poisson
counting fluctuations. For each bootstrap realizationn of our dataset, we compute the parameters $r_0$ and $\gamma$ via our maximum likelihood procedure,
and determine errors on these parameters from the resulting
distributions.

Based on this procedure, we define the $1\sigma$ confidence region
from the 16th and 84th percentiles of the bootstrapped parameter
distributions. We find a range $0.36 < \gamma < 1.98$ for $\gamma$,
and $15.8\,h^{-1}\,{\rm cMpc} < r_0 < 9404\,h^{-1}\,{\rm cMpc}$. The
correlation length $r_0$ is poorly constrained, because for flat
correlation function slopes $\gamma < 1$, the clustering projected
along the line-of-sight is very insensitive to $r_0$, which again
simply reflects the fact that the amplitude and slope of the
correlation function are not independent parameters. This is the same
behavior seen in the contours in Fig.~\sfigcontour.  Fixing the
slope $\gamma$ to its maximum likelihood value of $\gamma=1.5$, we
determine a $1\sigma$ confidence region for the cross-correlation length of
$r_0 = \rnotval\pm
\rnoterr\,h^{-1}\,{\rm cMpc}$.


Although we do not fit the binned correlation function to determine
model parameters, it is instructive to compute a binned correlation
function to visualize the data, our model fits, and the errors on
these fits. As such, we define a dimensionless correlation function as
\be
\chi(R_{i},R_{i+1}) \equiv \frac{\sum_j^{N_{\rm proto}} \langle PG\rangle_{ij}}{\sum_{ij}^{N_{\rm proto}} \langle PR\rangle_{ij}} - 1 \label{eqn:chi}, 
\ee
where $\langle PG \rangle_{ij}$ is the number of real HzRG/LAB-LAE
pairs in the impact parameter bin $[R_i,R_{i+1}]$ around the $j$th
object.  The quantity $\langle PR\rangle_{ij} = n_{\rm LAE}(z_j , >
W_{{\rm limit},j}) V_{ij}$ is the expected random number of
pairs, which is simply the product of the background number density of
LAEs, $n_{\rm LAE}$, and the cylindrical volume, $V_{ij}$, defined by the bin
$[R_i,R_{i+1}]$ and the half-height $\Delta Z_{j}$ of the NB filter
used to survey the $j$th object.
We compute errors on the binned correlation function via an analogous
bootstrap method, whereby $\chi(R_{i},R_{i+1})$ is recomputed from
1000 bootstrap resamplings of the sample of eight HzRGs/LABs while
simultaneously bootstrap resampling the LAE separations from each
object.

In Fig.~\sfigcorr, we show the binned correlation function
calculated via this procedure, with $1\sigma$ error bars, determined
by taking the 16th and 84th percentiles of the distribution of
$\chi(R_{i},R_{i+1})$ resulting from the bootstrap samples. For any
value of the model parameters $r_0$ and $\gamma$, we can compute the
predicted value of $\chi(R_{i},R_{i+1})$ by evaluating the
integrals
\be
\langle PG\rangle_{ij} = n_{\rm LAE}(z_j,> W_{{\rm limit},j})V_{ij} 
\int_{-\Delta Z_j}^{\Delta Z_j} \int_{R_i}^{R_{i+1}} [1 + \xi(\sqrt{R^2 + Z^2})] 2\pi R dR dZ \label{eqn:PG}, 
\ee
and then evaluating the sum in eqn.~(\ref{eqn:chi}). The uncertainty on
the predicted clustering level, arising from the uncertainty on our
model parameters, can thus be determined by evaluating the 16th to
84th percentile confidence region of the predicted
$\chi(R_{i},R_{i+1})$, using the distribution of $r_0$ and $\gamma$
from the bootstrap resampling of our model fits. The resulting
$1\sigma$ uncertainty on the predicted correlation function is shown
as the gray shaded region in Fig.~\sfigcorr, with the maximum 
likelihood value $r_0= \rnotval\,{\rm h^{-1}\,cMpc}$ and $\gamma= \gammaval$ shown by the red
curve. We see that our maximum likelihood model fit and bootstrap errors are 
consistent with the binned correlation function estimates and their respective errors. 

Given the results of the clustering analysis described in this
sub-section, we generate Fig.~\figcluster\ of the main text as
follows. Using our catalog of LAEs around the f/g quasar with
$\log_{10} L_{\rm Ly\alpha} > 42.1$ (Table~\tablae), and the LAE
luminosity function prediction for the background number density
$n_{\rm LAE}$, we can compute the cumulative overdensity profile
$\delta(< R) \equiv 1 + \xi(R_{\rm min},R) = \langle
PG\rangle \slash \langle PR\rangle$. The overdensity profile predicted
by our maximum likelihood determination of $r_0$ and $\gamma$ for the
giant nebulae-LAE correlation function are determined by evaluating
eqn.~(\ref{eqn:PG}). This is shown as the red curve in
Fig.~\figcluster. Similarly we can determine the $1\sigma$ error on
the predicted $\delta(<R)$ by evaluating it for all of our bootstrap
samples of $r_0$ and $\gamma$ and taking the 16th-84th percentile
region of the resulting $\delta(<R)$ distribution. This region is
shown as the shaded grey area in Fig.~\figcluster. For comparison we
show the expected overdensity (projected over our NB filter) of Lyman
break galaxies (LBGs) around radio-quiet quasars, based on the recent
measurements of quasar-LBG clustering at $z \simeq 2.7$ by Trainor et
al. \cite{trainor12},
who found a cross-correlation length of $r_0 =
7.3\pm1.3\,h^{-1}\,{\rm cMpc}$, assuming a fixed slope of
$\gamma=1.5$, similar to the best-fit slope deduced for our giant
nebulae-LAE correlation. The blue dotted lines indicate the $1\sigma$
error quoted by Trainor et al. \cite{trainor12}
on the cross-correlation
length with the slope held fixed. Comparing to the Trainor et al. \cite{trainor12}
measurements may seem questionable, given that: a) this is the
cross-correlation with LBGs, whereas the clustering analysis in
Fig.~\figcluster\ is concerned with LAEs; b) the quasars studied by
Trainor et al. were hyper-luminous. However, both of these differences
are likely to produce small effects on the cross-correlation strength
because a) the clustering of LAEs and LBGs are comparable at $z \sim
2.7$, and, assuming hierarchical clustering, the cross-correlation
strength is only sensitive to the square-root of any differences in
clustering strength; and b) the Trainor et al. \cite{trainor12}
results are consistent, within the errors, with the quasar-LBG
cross-correlation results of \cite{AS05},
who studied much fainter
quasars.  Indeed, both \cite{AS05}
and Trainor et al. \cite{trainor12} argue that the
clustering of LBGs around quasars is independent of luminosity,
assuaging concerns about comparing to hyper-luminous quasar
clustering.

\section{Near-IR Observations: The Systemic Redshift of the f/g Quasar}\label{sec:nearIR}

Quasar redshifts determined from the rest-frame ultraviolet emission
lines (redshifted into the optical at ${\rm z}\sim 2$) can differ by
up to one thousand kilometers per second from the systemic frame,
because of the complex motions of material in the broad line regions
of quasars \cite{Richards02,Shen07}.  To achieve higher precision, one
may analyze spectra of the narrow \oiii\ emission lines emerging from
the narrow-line region, but at $z \sim 2$, this requires near-IR
spectroscopy.  On UT 2007 April 24, we obtained a near-IR spectrum of
the f/g quasar (and also the b/g quasar) with the Near InfraRed Imager
and Spectrometer (NIRI) on the Gemini North telescope.  These data
were reduced and calibrated with the LowRedux package, which performs
wavelength calibration using the atmospheric sky lines recorded in
each spectrum.  With the chosen configuration, the spectra span
$1.40-1.94 \mu$m in the H-band.


Fig.~\sfignearir\ presents the NIRI spectra, fluxed with telluric
standard stars taken that night.  Each quasar shows the detection of
strong \oiii\ emission and modest H$\beta$ emission.  The
\oiii$\lambda$5007 lines were centroided using a flux-weighted
line-centering algorithm with pixel values weighted by a Gaussian
kernel, and this algorithm was iterated until the line center
converged to within a specified tolerance (see \cite{QPQ3}
for more
details).  A small offset of 27\tkms\, was also added to the redshifts
determined in this manner to account for average blueshift of the
\oiii$\lambda$5007 from systemic (111).  We measured
wavelength centroids of $\lambda_{\rm obs} = 1.5231 \pm 0.0001 \mu\rm
m$ and $1.6095 \pm 0.00006 \mu\rm m$ for the f/g and b/g quasars, 
respectively.  This gives systemic redshifts $\mzfg = \zzfg$ for the
f/g quasar and \zzbg\ for the b/g quasar.  We estimate an uncertainty
in this redshift of 50~\tkms\ based on error in our wavelength
calibration, line centering, and the estimated uncertainty when one uses
\oiii\ to assess the systemic redshift of a quasar (111).

\section{Kinematics}
\label{sec:Kin}

In the Main text, we described the complex kinematic motions of the
quadruple AGN system and nebula as measured with slit spectroscopy
(optical and near-IR; Figs.~\figslits,\sfignearir) and absorption-line
spectroscopy of a b/g quasar (Fig.~\figabs).  The AGN exhibit
line-of-sight velocity differences of at least $\delta v = 1300\kms$,
requiring extreme gravitational motions. Note that these large
velocities cannot be explained by Hubble expansion. The miniscule
probability of finding a quadruple quasar system in the absence of
clustering $P\sim 10^{-13}$ (see Supp.~\ref{sec:clust}), and the
physical association between the AGN and giant nebula, demand that the
four AGN reside in a real collapsed structure, and thus the relative
motions of the AGN must result from gravitational motions in a
collapsed structure.

The velocities of the nebular Ly$\alpha$ emission are also extreme,
exhibiting line-of-sight velocities relative to the systemic redshift
of f/g quasar ranging from $\delta v = -800$ to $+2500 \, \kms$.
These also trace the gravitational potential of the protocluster but
may be influenced by other physical processes (e.g.\ kinematic or
radiative feedback from the AGN). Finally, the absorption line
kinematics of cool metal-enriched gas in the protocluster halo,
measured from the b/g quasar spectrum, show strong absorption at
$\approx +650\,\kms$ with a significant tail to velocities as large as
$\simeq 1000\,\kms$.

We further emphasize two aspects here.
First, although the nebular emission exhibits large, relative velocity
offsets, the internal motions are relatively quiescent and generally
unresolved at the Keck/LRIS spectral resolution (FWHM~$\approx 160
\kms$).  There is no evidence for the ``double-peaked'' emission
characteristic of resonantly trapped \mlya.  This implies that the gas
within the nebula has modest opacity to \mlya\ photons and that the
observed radiation is not predominantly scattered photons.
These data also indicate that the gas motions are not strictly random, but
are rather organized into coherent flows, 
i.e.\ the velocity offsets observed greatly exceed the measured velocity
dispersions. Similar coherent motions have been previously observed in giant
Ly$\alpha$ nebulae in HzRGs \cite{vanojik96,villar03,villar_rev07,humphrey07}
and LABs \cite{prescott15a}. 

Second, consider the gas kinematics along the single Slit~2 
(Figs.~\figslits and \sfigchimaps).  This slit covers the nebula from near 
the f/g quasar, where one observes a negligible velocity offset
($\delta v \approx 0 \kms$), to the b/g quasar at $\approx 200$\,kpc
separation. As one travels along the slit, the velocity offset
increases to $\simeq +500\kms$ as one approaches the b/g quasar, 
where the absorption-line spectrum shows a velocity
offset of $\delta v \approx +650 \kms$ (Fig.~\figabs).
Altogether, the gas
along Slit~2 exhibits a velocity gradient of at least 500\tkms\ across
200\,kpc (projected).
These data establish the presence of large-scale, high-velocity flows
of cool gas within this protocluster halo.

\section{Estimates for the Abundance of Giant Ly$\alpha$ Nebulae Around Quasars}\label{sec:abun}



The protocluster \qso\ was discovered because of its large-scale
Ly$\alpha$ emission properties, and was selected from a well-defined
survey for extended Ly$\alpha$ emission around 29 quasars \cite{QPQ4}.
We can use the statistics of this survey to estimate the prevalence of
giant Ly$\alpha$ nebulae around quasars.  Insofar as there appears to
be a physical connection between giant Ly$\alpha$ nebulae and
protoclusters, this also provides insight into the prevalence of
protoclusters around quasars. We also present an estimate for the
frequency of occurrence of giant Ly$\alpha$ nebulae from an
independent narrow-band imaging survey of quasars conducted by our
group, and also discuss the results of similar narrow-band imaging
surveys performed by others. Although much past work has been
dedicated to characterizing Ly$\alpha$ nebulae around radio-loud
quasar samples \cite{heckman91a},
it is only recently
that sensitive observations have been undertaken to characterize
the abundance of nebulae around predominantly radio-quiet quasars.

Our slit-spectroscopic observations allow us to estimate the 
effective covering factor of extended Ly$\alpha$ emission around the
typical quasars ($L_{\rm bol}\sim 10^{46}\,{\rm erg\,s^{-1}}$) we
surveyed.  Following the discussion in
\S~4.3 of (22; see Fig.~4 of that paper), large $\sim 70\,{\kpc}$
scale nebulae can be detected down to $2 \times {\rm SB}_{1\sigma}$,
  where ${\rm SB}_{1\sigma}$ are the 1$\sigma$ surface brightness
  limits quoted in Table 1 of \cite{QPQ4}.
  In that sample a nebula of comparable brightness to
  SDSSJ~0841$+$3921 ${\rm SB}_{\lya} \simeq 10^{-17}\cgssb$ could have
  been detected at $> 2\sigma$ significance in 23/29 cases (see
  Table~1 of \cite{QPQ4}).
  Based on this, we can crudely determine the
  covering factor of such filamentary emission as being the ratio of
  the area of the filament covered by our original slit observation of
  SDSSJ\,0841$+$3921 (Slit2 in Figs.~\figslits and \sfigchimaps), to
  the total area surveyed at sensitivity sufficient to detect a ${\rm
    SB}_{\lya} > 10^{-17}\cgssb$ nebula.

  We consider an annular region around a $z=2$ quasar with an inner
  radius of $R_{\rm \perp,min} = 50\,{\rm kpc}$ ($5.8\arcsec$) and
  outer radius of $R_{\rm \perp,max}=150\,{\rm \kpc}$
  ($17.3\arcsec$). The inner radius is chosen to exclude the
  small-scale fuzz at $R_{\perp} < 50\,\kpc$, which is detected in a
  large fraction of quasars \cite{QPQ4},
  from the larger scale
  filamentary emission.  The outer radius corresponds to the maximum
  extent of the (high significance) emission detected in our discovery
  spectrum of SDSSJ\,0841$+$3921 (see Slit2 in Fig.~\figslits panel
  c, and the middle column of Fig.~\sfigchimaps), and is about the
  virial radius of the dark matter halos $M_{\rm vir}=10^{12.5}\msol$
  at $z = 2$ which host quasars \cite{White12}.
  We also observe a
  drop-off in the covering factor of optically thick absorbers at
  $R_{\perp}\simeq 200\,\kpc$, comparable to this
  $R_{\rm \perp,max}$ \cite{QPQ6}.
  We assume the filamentary emission
  in the discovery spectrum extends from $+50\,\kpc$ to $+150\,\kpc$
  at positive slit position, and $-50$ to $-80\,\kpc$ at negative slit
  positions, corresponding to an area of $15$ sq. arcsec given our
  1$\arcsec$ slit ($1130\,\kpc^2$). The total area is equal to the
  $N_{\rm QSO}=23$ quasars observed to sufficient depth, times the
  $23$ sq. arcsec area ($1740\,\kpc^2$) for which our slit intersects
  the annular region, implying a total area of $530$ sq. arcsec
  ($39922\,\kpc^2$) surveyed. For reference, the total annular area
  around a {\it single} quasar is $834$ sq. arcsec ($62832\,\kpc^2$),
  so the effective area of our survey is 64\% of the area around a
  single quasar.  Hence we deduce a covering factor of filamentary
  emission of $f_{\rm C,fil} = 0.028$.

Thus taken in aggregate, we find
that only $\simeq 3\%$ of the area between $R_{\perp} = 50-150\,{\rm
  kpc}$ around quasars is covered by Ly$\alpha$ emission with ${\rm
  SB_{Ly\alpha}}> 10^{-17}\cgssb$ , which can be compared to the much
larger covering factor of $31\%$ above this level, measured for
emission over the same area from our NB image of \qso 
(see Fig.~1a). The aggregate and individual covering
factors can be commensurated
if the  majority of quasars do not exhibit large-scale Ly$\alpha$ nebulae
above ${\rm SB_{Ly\alpha}}> 10^{-17}\cgssb$, but a small fraction
$\simeq 10\%$ have bright nebulae covering a large area, as in
\qso , and the product of this $\sim 10\%$ frequency and
$\sim 31\%$ covering factor set the aggregate $\sim 3\%$
effective covering factor for the entire population. This argument is
consistent with our intuition that, given the random orientation of
the slit, we could probably only have identified the asymmetric
extended emission around \qso\ for $\sim 50\%$ of the possible
slit orientations, and given that 23 objects were surveyed to the
required depth, this again suggests that $\sim 10\%$ of
quasars exhibit giant nebulae.

An independent estimate for the frequency of giant Ly$\alpha$ nebulae
around quasars can also be determined from a narrow band imaging
survey conducted by our group.  As part of a continuing program to
search for fluorescent Ly$\alpha$ emission powered by luminous $z\sim
2$ quasars, to date we have obtained deep narrow band imaging of a
sample of 26 objects. These survey observations are analogous to the
narrow-band imaging of SDSSJ~0841$+$3921 discussed in Supp.~1.2, in that
they use custom designed narrow-band filters, however the targets
constitute an independent sample of quasars, which are not part of the
QPQ survey of \cite{QPQ4}.
These observations have been conducted on multiple
telescopes, which we briefly summarize. For the present purposes we
define a giant Ly$\alpha$ nebula to be extended emission on a scale $
> 50\,{\rm kpc}$ from the quasar with average ${\rm
SB_{Ly\alpha}}> 10^{-17}\cgssb$.

The quasar HE0109$-$3518 at $z = 2.41$ was observed with the Very
Large Telescope/ -FOcal Reducer Spectrograph (VLT-FORS) using a custom
designed narrow band filter ($\lambda_{\rm center} = 4145$\AA, ${\rm
FWHM}_\lambda = 40$\AA) for a total exposure time of about 20 hours,
achieving a sensitivity limit ($1\sigma$) of ${\rm SB_{Ly\alpha}} =
8\sci{-19}\cgssb$ \cite{cantalupo12}.
No extended Ly$\alpha$ nebula
was discovered. We have imaged a total of eight quasars with the Keck
LRIS imaging spectrometer, of which three were observed with a filter
centered on Ly$\alpha$ redshifted to $z = 2.279$ ($\lambda_{\rm
center} = 3986$\AA, ${\rm FWHM}_\lambda = 30$\AA), and five with one
centered at $z = 2.190$ ($\lambda_{\rm center} = 3878$\AA, ${\rm
FWHM}_\lambda = 30$\AA). The exposure times for the LRIS observations
vary from 2 to 10 hours, resulting in sensitivity limits ($1\sigma$)
ranging from ${\rm SB_{Ly\alpha}}=1.2\sci{-18}\cgssb$ to $5\sci{-19}$.
These VLT/FORS and Keck/LRIS observations specifically targeted
hyper-luminous quasars $V\lesssim 17$, with the goal of understanding
fluorescent Ly$\alpha$ emission from dark-galaxies \cite{cantalupo12}
and the quasar CGM \cite{Slug},
and led to the discovery of a giant
Ly$\alpha$ nebula around the quasar UM287 at $z=2.28$. Nebulae were
not detected around any of the other quasars surveyed. Finally, using
a custom filter centered on Ly$\alpha$ at $z = 2.253$ ($\lambda_{\rm
center} = 3955$\AA, ${\rm FWHM}_\lambda = 30$\AA), we have
observed a total of 17 quasars using the Gemini Multi Object
Spectrograph (GMOS) on the telescope Gemini South. For the GMOS
observations, a subset of 3 quasars were observed with longer
integrations, typically 5 hours and achieving a depth of ($1\sigma$)
${\rm SB_{Ly\alpha}} = 1.5\sci{-18}\cgssb$; whereas the other 14
quasars were observed in a fast survey mode with typical exposure
times of about 2 hours and depth ${\rm SB_{Ly\alpha}} =
3-4\sci{-18}\cgssb$. Although the instruments used and depths achieved
are heterogeneous, all of the aforementioned observations reached
sufficient depth to detect Ly$\alpha$ nebulae with ${\rm
SB_{Ly\alpha}}> 10^{-17}\cgssb$.
Taking all of these observations together, we have observed a total
of 26 quasars and detected a single giant Ly$\alpha$ nebula around
UM287 \cite{Slug},
implying that $1/26 = 0.04$ of quasars host giant
Ly$\alpha$ nebulae. Adopting the $1\sigma$ confidence interval
appropriate for the Poisson distribution in the small number regime,
we deduce that the frequency of occurrence of giant nebulae is in the
range $3 - 9\%$.

Finally, an independent narrow band imaging survey of quasars at
$z\sim 2.7$ is being carried out as part of the Keck Baryonic
Structure Survey (KBSS), and the first results on the
distribution of LAEs around quasars were recently published by Trainor
et al. \cite{trainor13}.
Based on these narrow band imaging
observations, \cite{Martin14a}
published the discovery and additional
observations of a giant Ly$\alpha$ nebula around the quasar
HS1549$+$19 with properties comparable to that around
\qso. Trainor et al. quote a typical narrow-band imaging
depth of $m_{\rm NB}(3\sigma)\sim 26.7$ for point sources, resulting
from 5-7hr integrations on Keck LRIS using filters with FWHM$\sim
80$\AA.  This corresponds to a ($1\sigma$) ${\rm SB_{Ly\alpha}} \simeq 1.5\sci{-18}\cgssb$, which
although shallower than our own Keck LRIS survey (owing to their wider
filters), it is nevertheless sufficient to detect giant nebulae with
${\rm SB_{Ly\alpha}}> 10^{-17}\cgssb$. There are 8 objects observed in
the Trainor et al. sample, but it is unknown to us whether additional
Ly$\alpha$ nebulae are present, besides that around HS1549$+$19. As such,
one can place a lower limit on the frequency of giant Ly$\alpha$ nebulae
to be $1/8 = 0.125$, or a $1\sigma$ Poisson confidence interval of $10-29\%$. 

To summarize, three independent surveys indicate that the frequency of
occurrence of giant Ly$\alpha$ nebulae around quasars is consistent with being 
$\simeq 10\%$, although there are still limitations due to relatively small
samples, heterogeneity of the datasets,  and the differences in the
search techniques (narrow-band imaging versus spectroscopy).

\section{The Environments of Other Quasars with Giant Ly$\alpha$ Nebulae and the Impact of Ly$\alpha$ Fluorescence}       

Although quasar clustering indicates that the majority of $z\sim 2$
quasars reside in moderate overdensities, we speculate that $\sim
10\%$ trace much more massive structures, and it is this subset which
also exhibits giant Ly$\alpha$ nebulae. \qso clearly
supports this interpretation, but there are indeed other examples.

Of course, we also have NB imaging data for UM287, the quasar at
$z=2.27$ with a giant Ly$\alpha$ nebula recently discovered by our
group \cite{Slug}.
However, because UM287 is a hyper-luminous quasar,
we cannot characterize its environment via the number counts of LAEs
as we have for \qso. The reason for this is that distribution of LAEs
on Mpc scales from hyper-luminous quasars is likely to be dramatically
enhanced by Ly$\alpha$ fluorescence \cite{cantalupo12},
and it is not
straightforward to disentangle the fluorescence and clustering
enhancements. For example, consider the hyper-luminous quasar
HE0109$-$3518, the proto-typical case for Ly$\alpha$ fluorescence by
dark-galaxies studied by \cite{cantalupo12}.
The number of LAEs within
$R < 4\,{h^{-1}\,{\rm cMpc}}$ of HE0109$-$3518 is $\sim 8.5$ times larger
than the cosmic background number expected from the LAE luminosity
function, whereas the quasar-LBG correlation function (a reasonable
proxy for the quasar-LAE clustering) averaged over the volume
probed by these narrow-band observations would only predict an enhancement
of $\sim 1.5$.  One comes to similar conclusions about the number of
LAEs discovered around the hyper-luminous quasars in the KBSS sample
of \cite{trainor13}.
Progress on understanding the environment of
UM287, and its relationship to the giant Ly$\alpha$ nebulae, will
require a sample of galaxies selected via broad-band photometry
(i.e. LBG selection), to avoid contamination from fluorescence.
Nevertheless, while fluorescence precludes a simple interpretation of
LAE clustering around UM287, we note that, intriguingly, follow-up
observations also revealed a faint $V\simeq 22$ companion quasar,
indicating an overdensity of AGN similar to \qso.


Note however that enhanced LAE clustering around \qso due to
Ly$\alpha$ fluorescence is expected to be negligible, because
brightest quasar in the system, namely the f/g quasar with $V=19.8$ is
an order of magnitude fainter than the hyper-luminous quasars which
exhibit this fluorescence effect. Radiative transfer simulations of
Ly$\alpha$ fluorescence around quasars \cite{cantalupo12}
supports
this conclusion. Furthermore, the four AGN in \qso clearly constitute
a dramatic overdensity of AGN, and there is no way for Ly$\alpha$
fluorescence to enhance AGN clustering. Note that the other three AGN
do not significantly enhance the ionizing luminosity and hence cannot
boost the fluorescence signal.  The fainter Type-1 AGN AGN2 and AGN3
have $V\simeq 23$ (see Table~\tabagnphot), and thus contribute
negligibly to the total ionizing photon production compared to the f/g
quasar. Although we do not detect the continuum from the accretion
disk of the Type-2 AGN AGN1, it is detected in the WISE
bands W1 and W2 (see Table~\tabagnphot). At $z\simeq 2$ these bands
probe rest-frame $\sim 1\,\mu{\rm m}$, which are likely still
dominated by emission from the reddest part of the accretion disk. The
fact that AGN1 is $ > 1$ magnitude fainter than the f/g quasar
indicates that its UV ionizing continuum (although obscured from our
perespective) is very unlikely to be brighter than that of the f/g
quasar.

The KBSS quasar HS1549$+$19 at $z=2.85$, which has been narrow band
imaged by Trainor et al. \cite{trainor13},
harbors a giant Ly$\alpha$
nebula \cite{Martin14a}
with SB and properties comparable to that
around \qso.  Extensive imaging and spectroscopy of HS1549$+$19 has
demonstrated that it is clearly a dramatic protocluster, with a
Mpc-scale overdensity of Lyman break galaxies (LBGs) a factor of $\sim
3.5$ times larger than that around typical
quasars \cite{trainor12,Mostardi13}.
Although this quasar is
hyper-luminous, Ly$\alpha$ fluorescence is not an issue because the
protocluster overdensity is based on the clustering of broad-band
selected LBGs, and not LAEs.

Finally, one may have concerns that the dramatic enhancements of LAEs
around HzRGs and LABs, quantified by our clustering analysis in
Supp.~4 and shown in Figs.~\figcluster\ and \sfigcorr, could be partly
due to Ly$\alpha$ fluorescence.  This is a valid concern, because
there is compelling evidence that the HzRGs are obscured
quasars \cite{MileyDB08},
and their ionizing photons, although
obscured from our perspective, surely shine in other directions, and
are likely to power their giant Ly$\alpha$ nebulae. Similar
considerations hold for LABs, although a direct association with
obscured AGN has been more challenging to
establish \cite{nfm+06,smith07,webb09,colbert11,prescott15}. 
However, we think that it is
unlikely that fluorescence plays a significant role in the strong
clustering of LAEs around HzRGs and LABs for two reasons. First, there
is no evidence that the unobscured counterparts of HzRGs and LABs
(i.e. the Type-1 quasar shining in other directions) are as bright as
the hyper-luminous $V\lesssim 17$ quasars, and hence sufficient to
power LAE fluorescence on Mpc scales. Furthermore, even if the HzRGs
and LABs were shining brightly in unobscured directions, the LAE
enhancement is expected to be significantly smaller than for
unobscured quasars, because given the cylindrical volume probed by NB
imaging observations, the illuminated volume will be much larger for
an unobscured quasar pointing toward the observer and hence along the
major axis of the cylindrical survey volume, than for an obscured
quasar which illuminates a volume randomly oriented relative
to our line-of-sight direction \cite{cantalupo12}.

Thus, although based on only a handful of objects, existing
observations support a picture where $\sim 10\%$ of radio-quiet
quasars are surrounded by giant Ly$\alpha$ nebulae, which acts as a
signpost for the presence of a protocluster. A similar connection
between AGN activity, giant Ly$\alpha$ nebulae, and protocluster
overdensities has been suggested by past work on HzRGs and LABs. Our
clustering analysis in Supp.~4 (see Figs.~\figcluster\
and \sfigcorr) demonstrates that these overdensities are indeed very
large and well in excess of the environment of radio-quiet
quasars.  These protocluster overdensities are real, and are not the
result of Ly$\alpha$ fluorescence.


\section{Absorption Line Analysis}\label{sec:absline}

Given the experimental design of our QPQ survey, we are able to
explore characteristics of the gas surrounding the quadruple AGN in
unprecedented detail. Fig.~\figabs\ shows strong \ion{H}{i} \mlya\
absorption coincident with the redshift of the protocluster.  We measure
a rest-frame equivalent width $W_\lambda = 3.75 \pm 0.05\,$\AA, offset
by a velocity $\delta v = c \Delta \lambda / \lambda \approx
+650\mkms$ from the f/g quasar systemic redshift. The absence of very
strong damping wings limits the \ion{H}{i} column density to be $\mnhi
< 10^{19.7} \cm{-2}$.  We also present our favored model which has
$\mnhi = 10^{19.2 \pm 0.3} \cm{-2}$, suggesting optically thick gas at
the Lyman limit.  It is possible, in the presence of significant
line-blending, that the
\nhi\ value could be much lower, but we consider this scenario very unlikely, 
since it would imply a gas metallicity exceeding the solar abundance
(see below).

Fig.~\sfigabs\ presents spectra of the b/g quasar centered at the
wavelengths of a series of strong, commonly observed UV resonance-line
transitions (including the several transitions shown in Fig.~\figabs\
of the Main text) in the rest-frame of the f/g quasar (based on its
near-IR systemic redshift). The gas
exhibits especially strong low-ion absorption
(\ion{O}{i}~1302, \ion{C}{ii}~1334) and moderately strong intermediate
ion absorption (\ion{N}{ii}~1083, \ion{Si}{iii}~1206).  The latter
absorption occurs within the \mlya\ forest and could be partially
blended with coincident IGM absorption, but we believe such
contamination is modest because the line-profiles track one another as
well as the other low-ion transitions.  Surprisingly, there is no
statistically significant absorption at the high-ion transitions of
Si$^{+3}$ or C$^{+3}$.  Even without performing a detailed
photoionization analysis, we can already conclude that the gas is not
highly ionized, which further suggests the gas is optically thick to
ionizing radiation.
One draws a similar conclusion from the relative strengths of
\ion{O}{i}~1302 to \ion{Si}{ii}~1304, where the latter is observed to
have larger equivalent width in systems that are highly ionized (e.g.,
\cite{pcw+08,prochter10}), contrary to our measurements for this
system.

For each of the absorption lines, we have estimated rest-frame
equivalent widths with simple boxcar integration.
Ionic column densities were estimated from the metal-line
transitions assuming the linear curve-of-growth approximation and
adopting lower limits to positive detections and upper limits for
non-detections (Table~\tababs).  
Because neutral oxygen O$^0$
undergoes charge-exchange reactions with atomic hydrogen, the
ratio $N_{\rm O^0}/\mnhi$ is very nearly equal to the number ratio O/H
over a wide range of physical conditions and ionizing radiation
fields \cite{QPQ3}.
Therefore, we estimate a conservative lower limit
to the oxygen metallicity
[O/H]~$> -1$.  To estimate abundances of the other ions
and to constrain physical conditions of the gas, we must evaluate
the ionization state with photoionization models (see the next section).  

%

Our previous studies have shown that
cool gas with strong metal-line absorption from lower ionization
states is common in the environment surrounding $z \sim 2$
quasars \cite{QPQ5,QPQ7}.
In Fig.~\sfigqpq\ we show the distribution of
equivalent widths for \ion{H}{i} \mlya\ and \ion{C}{ii}~1334 from the set of
QPQ pairs with separations less than 200\,kpc.  The absorption
strength for \qso\ lies toward the upper end of the distribution
but is not extreme.

In Table~\tababs\ we include estimates for the abundances of
the observed elements assuming ionization corrections from the
photoionization model discussed below and $\mnhi = 10^{19.2}
\cm{-2}$.   
The observations require $>1/10$ solar metallicity and
several measurements suggest solar (or even super-solar) enrichment.

\section{Joint Constraints on Protocluster Gas in Absorption and Emission}\label{sec:origin}


Our preferred mechanism for powering the giant Ly$\alpha$
nebula in \qso\ is fluorescent
emission from gas in the CGM of the protocluster powered by the
radiation from the f/g quasar, with the gas residing in a population
of cool dense clouds which are optically thin ($N_{\rm HI} <
10^{17.5}$) to ionizing radiation. Because our absorption line
observations of the b/g quasar provide constraints on the gas
properties in the protocluster, we adopt a novel approach
of modeling the emission and absorption properties of this gas
simultaneously, using the Cloudy photoionization
code \cite{cloudy98}.
First, we describe our model for the
distribution of the cool clouds, and introduce simple analytical
models for the fluorescent emission to build intuition about the
scaling with gas properties. Then, we consider a sequence of
photoionization models of the gas intercepted by the b/g sightline, 
which is however undetected in emission. Our absorption line measurements
and photoionization modeling imply that the total cloud column density
must be $\log_{10} N_{\rm H} = \nhsig$. Under
the reasonable assumption that the Ly$\alpha$ emitting gas in the nebula
has the same $N_{\rm H}$ as that detected in absorption, we then show
that reproducing the emission level requires very high gas volume densities
$n_{\rm H} \simeq \nvolmn\,{\rm cm^{-3}}$. 

\subsection{Cool Cloud Model and Emission Estimates}

We consider an idealized model whereby the gas in the protocluster
halo resides in a population of cool $T\sim 10^4\,{\rm K}$ spherical
gas clouds, which have a single uniform hydrogen volume density
$n_{\rm H}$ and hydrogen column density $N_{\rm H}$. The clouds are
uniformly distributed throughout a spherical halo of radius $R$, and
their aggregate covering factor, defined to be an average over the
sphere, is $f_{\rm C}$. Four parameters ($R$,$n_{\rm H}$,$N_{\rm
H}$, $f_{\rm C}$) completely describe the distribution of the
gas (see (22) for further details), for example the total cool
gas mass can be written
\be
M_{\rm c} = 3.3\sci{11}\!\left(\frac{R}{250\,{\rm kpc}}\right)^2 
\!\!\left(\frac{N_{\rm H}}{10^{20.5}\,{\rm cm}^{-2}}\right)\!\!
\left(\frac{f_{\rm C}}{0.5}\right)\,M_{\odot} \label{eqn:M_c}
\ee

We now consider fluorescent Ly$\alpha$ emission from this population
of cool gas clouds powered by a bright central quasar. There are two
regimes where simple expressions can be obtained for the resulting
emission, namely when the gas clouds are either optically thin $N_{\rm
HI} < 10^{17.5}\,{\rm cm}^{-2}$ or optically thick $N_{\rm HI} >
10^{17.5}\,{\rm cm}^{-2}$ to Lyman continuum photons.  For the
optically thin case, the gas clouds are highly ionized, and the
average Ly$\alpha$ SB from the halo will be
\be
{\rm SB}_{\rm Ly\alpha} = 1.2\times 10^{-17} \left(\frac{1+z}{3.0}\right)^{-4}
\!\!\left(\frac{f_{\rm C}}{0.5}\right) 
\!\!\left(\frac{n_{\rm H}}{1.0\,{\rm cm^{-3}}}\right)
\!\!\left(\frac{N_{\rm H}}{10^{20.5}\,{\rm cm^{-2}}}\right)\cgssb\label{eqn:SB_thin}. 
\ee
Note that because the gas is already highly-ionized, the SB is independent of the luminosity of 
the central ionizing source provided that it
is bright enough to maintain the gas as optically thin. 

If the clouds are optically thick to ionizing radiation, they will no
longer emit \mlya\ proportional to the volume that they occupy because
of self-shielding effects. Instead, a thin highly ionized skin will
develop around each cloud, and nearly all recombinations and resulting
\mlya\ photons will originate in this skin. The cloud will then behave
like a `mirror', converting a fraction $\eta_{\rm thick} = 0.66$ of
the ionizing photons it intercepts into \mlya\ photons \cite{GW96}.
In this regime, it can be shown (see \cite{QPQ4})
that the fluorescent
SB, averaged over the halos is
\bea
{\rm SB}_{\lya}&=&\frac{\eta_{\rm thick}h\nu_{\lya}}{4\pi(1+z)^4}f_{\rm C}
\Phi(R\slash \sqrt{3})\label{eqn:SB_thick}\\
&=& 8.8\times 10^{-17}\left(\frac{1+z}{3.0}\right)^{-4}\!\!
\left(\frac{f_{\rm C}}{0.5}\right)\!\!
\left(\frac{R}{100\,{\rm kpc}}\right)^{-2}\nonumber\\
& \times & 
\left(\frac{L_{\nu_{\rm LL}}}{10^{30.5}\,{\rm erg\,s^{-1}\,Hz^{-1}}}\right)\cgssb\nonumber,  
\eea
where $h\nu_{\rm Ly\alpha}$ is the energy of a Ly$\alpha$ photon and $\Phi$ $({\rm phot\,s^{-1}\,cm^{-2}})$ is the ionizing photon
number flux 
\be 
\Phi= \int_{\nu_{\rm LL}}^{\infty}
\frac{F_{\nu}}{h\nu} d\nu = \frac{1}{4\pi r^2}\int_{\nu_{\rm
    LL}}^{\infty} \frac{L_{\nu}}{h\nu} d\nu.   \label{eqn:Phi}
\ee
To obtain the second equality in eqn~(\ref{eqn:SB_thick}), we assume
that the quasar spectral energy distribution obeys the power-law form
$L_{\nu} = L_{\nu_{\rm LL}} (\nu\slash \nu_{\rm LL})^{\alpha_{\rm
UV}}$, blueward of $\nu_{\rm LL}$ and adopt a slope of $\alpha_{\rm UV}=
-1.7$ consistent with the measurements of \cite{lusso15}.
The quasar
ionizing luminosity is then parameterized by $L_{\nu_{\rm LL}}$, the
specific luminosity at the Lyman edge. As we describe in detail 
below, we estimate that f/g quasar has
$\log_{10} L_{\nu_{\rm LL}} = 30.5$, which is the value
adopted in eqn.~(\ref{eqn:SB_thick}).  In the optically thick regime,
we see that the SB now depends on the luminosity of
the quasar, as well as on the covering factor $f_{\rm C}$, but is
independent of the gas properties (i.e. $n_{\rm H} and N_{\rm H}$). 

The smooth morphology of the emission in the Ly$\alpha$ nebula implies
a covering factor of $f_{C}\gtrsim 0.5$. Even if the emitting clouds
have angular sizes much smaller than the PSF of our NB imaging, the
morphology of the nebular emission would appear much more clumpy for
lower covering factors. We have explicitly verified this by generating
simulated images of mock Ly$\alpha$ emission distributions with the required
SB and a range of covering factors $f_{\rm C}$, and then convolving them with our
NB-imaging PSF \cite{FAB14}.
In what follows we will always assume a covering
factor of $f_{\rm C} = 0.5$ for both the gas we detect in emission, as
well as that detected in absorption.  This value is well motivated: a)
for the emission it is based on the smooth morphology of the nebula;
b) for the absorption it is approximately the covering factor of
optically thick absorbers on $R \lesssim 200\,{\rm kpc}$ scales in the
quasar CGM \cite{QPQ6}.

The average SB of the giant Ly$\alpha$ nebula at $R\simeq 100\,{\rm
  kpc}$ is ${\rm SB}_{\lya} =\sbannu$\\
$\cgssb$, which we compute
via the emission from the area within our
2$\sigma$ SB contour which intersects the the annulus $R =[75,125]\,$kpc. 
This is several times lower than the 
expectation from eqn.~(\ref{eqn:SB_thick}) for the optically thick
regime. Thus our simple analytical scaling relations already appear to
favor a situation where the clouds are optically thin. We return to
this comparison below using photoionization models.

\subsection{Details of Cloudy Modeling}

The expressions in eqns.~(\ref{eqn:SB_thin}) and (\ref{eqn:SB_thick})
provide reasonable estimates for the fluorescent SB, but they neglect
other sources of Ly$\alpha$ emission such as collisional excitation
(i.e. cooling radiation), and pumped or `scattered' Ly$\alpha$ photons
arising from the diffuse continuum produced by the gas itsel. Note
that we do not attempt to model scattering of Ly$\alpha$ photons
produced by central f/g quasar itself.  Radiative transfer simulations
of radiation from a hyper-luminous quasar through a simulated gas
distribution (41), have shown that the contribution of
scattered Ly$\alpha$ line photons from the quasar do not contribute
significantly to the Ly$\alpha$ surface brightness of the
nebula. Given that this calculation was for a hyper-luminous quasar a
factor of $\sim 10$ times brighter than the f/g quasar, scattered line
photons from the quasar are expected to contribute negligibly to the
nebular emission., and they also do not treat the temperature
dependence of the emission.  To fully account for these effects, we
construct photoionization models for a slab of gas being illuminated
by a quasar using the Cloudy software package \cite{cloudy98}.
Cloudy
predicts the emergent spectrum from the gas, allowing us to compute
its SB, which we then dilute by the covering factor $f_{\rm C}=0.5$.

We model the quasar spectral-energy distribution (SED) using a
composite quasar spectrum which has been corrected for IGM
absorption \cite{lusso15}.
This IGM corrected composite is important
because it allows us to relate the $g$-band magnitude of the f/g
quasar to the specific luminosity at the Lyman limit $L_{\nu_{\rm
LL}}$. For energies greater than one Rydberg, we assume a power law
form $L_{\nu} = L_{\nu_{\rm LL}} (\nu\slash \nu_{\rm LL})^{\alpha_{\rm
UV}}$ and adopt a slope of $\alpha_{\rm UV}= -1.7$, consistent with
the measurements of \cite{lusso15}.
We determine the normalization
$L_{\nu_{\rm LL}}$ by integrating the Lusso et al. \cite{lusso15}
composite spectrum
against the SDSS filter curve, and choosing the amplitude to give the
correct $g$-band magnitude of the f/g quasar (see Table~\tabagnphot),
which gives a value of $\log_{10} L_{\nu_{\rm LL}} = 30.5$. We extend
this UV power law to an energy of 30 Rydberg, at which point a
slightly different power law is chosen $\alpha = -1.65$, such that we
obtain the correct value for the specific luminosity at 2 keV
$L_{\nu}({\rm 2\,keV})$ implied by measurements of $\alpha_{\rm OX}$,
defined to be $L_\nu(2\,{\rm keV})\slash L_\nu(2500\,{\rm \AA}) \equiv
(\nu_{\rm 2\,keV}\slash \nu_{2500\,{\rm \AA}})^{\alpha_{\rm OX}}$. We
adopt the value $\alpha_{\rm OX} = -1.5$ measured by \cite{strateva05}
for SDSS quasars of comparable luminosities. An X-ray slope of
$\alpha_{\rm X}=-1$, which is flat in $\nu f_{\nu}$ is adopted in the
interval of 2-100\,kev, and above 100\,keV, we adopt a hard X-ray
slope of $\alpha_{\rm HX} = -2$. For the rest-frame optical to mid-IR
part of the SED, we splice together the composite spectra
of \cite{lusso15}, \cite{vanden01}, and \cite{RichardsSED06}.
These
assumptions about the SED are essentially the standard ones used in
photoionization modeling of AGN (e.g. \cite{Baskin14}).
We also include the contribution from the radiation field of the ambient
UV background radiation field using the spectrum of \cite{HM12},
but
it is completely sub-dominant relative to the f/g quasar at the
distances that we consider.

These Cloudy models require as an input the ionizing photon number
flux $\Phi(r)$ (see\\
eqn.~(\ref{eqn:Phi})) at the surface of the cloud.
Because this varies with radius as $r^{-2}$, in principle modeling the
halo emission would require integrating over a distinct
photoionization model at each radius. For simplicity, we restrict
attention to just a few radii. For our model of the optically thick
absorber, we set $r = 250\,{\rm kpc}$, which we believe is 
the likely location of the absorbing gas for several reasons. First, the nebula
itself has an end-to-end size of $\simeq 500\,{\rm kpc}$ roughly centered on
the f/g quasar, such that $r=250\,{\rm kpc}$ is a reasonable radius for
the absorber given the measured impact parameter $R_{\perp}= 176\,{\rm kpc}$
of the b/g quasar sightline. Second, given this impact parameter, the characteristic  velocity separation of the absorbing
gas from systematic $v_{\rm max} = 600\,{\rm km s^{-1}}$ implied by the metal-line
absorption (see Fig.~\sfigabs), we can compute the conditional
probability distribution $P( < r | R_\perp, v_{\rm max})$ that
the distance between the quasar and the absorber is less than
$r$, which is determined by the quasar-LLS
correlation function which we have measured \cite{QPQ2,QPQ6}.
For the problem
at hand, we find that $P( < r | R_\perp, v_{\rm max}) = 0.43$ for
$r  = 250\,{\rm kpc}$, again indicating that this is a reasonable distance (see
a similar discussion in \S~4.1 of \cite{QPQ3}).  
According to eqn.~(\ref{eqn:SB_thick}), the SB arising from
such optically thick gas scales with $\Phi$, and hence will decrease
as $r^{-2}$.   For the gas producing the Ly$\alpha$
emission, we model three radii $r = [50,100,250]\,{\rm kpc}$. 
We will argue that the emitting gas in
the nebula is optically thin, and in this regime the SB is
independent of $\Phi$ (see eqn.~(\ref{eqn:SB_thin})). Thus considering a
few radii to describe the emission is a reasonable approximation
provided the gas remains optically thin at the radii in question.

The Cloudy model of the cool cloud is completely specified by the
value of $\Phi$, the SED shape, the metallicity of the gas $Z$, the
hydrogen volume density $n_{\rm H}$, and a `stopping criterion', which
sets the total column density of the cloud.  Because we assume the f/g
quasar and UVB are the only sources of ionizing photons, $\Phi$ and
the SED shapes are fixed.  For the metallicity, we adopt a value of
$\log (Z\slash Z_\odot)=-0.5$ consistent with our lower-limit ($\log
(Z\slash Z_\odot) > -1$) on the metallicity of the optically thick
absorber detected in the b/g quasar sightline
(see Supp.~9). We have confirmed that the results
discussed below are insensitive to the assumed gas metallicity.  We
have also verified that our results are insensitive to the exact value
of the UV slope $\alpha = -1.7$, within the $1\sigma$ error measured
by Lusso et al. $\alpha = -1.7\pm 0.6$.  Our models thus constitute a
one-dimensional sequence parameterized by the volume density of the
clouds $n_{\rm H}$.  Photoionization models are known to be
self-similar in the ionization parameter $U = \Phi\slash n_{\rm H} c$,
which is the ratio of the number density of ionizing photons to
hydrogen atoms. Since $\Phi$ has been fixed, our sequence of models is
thus also a sequence in $U$, spanning the relevant range of ionization
conditions.  Our stopping criterion depends on what is known about the
gas that we are attempting to model, which we elaborate on next.

\subsection{Photoionization Model of the Absorbing Gas}
\label{sec:cldyHI}

Here we compare photoionization models to the properties of the
optically thick absorber in the spectrum of the b/g quasar, lying at
at an impact parameter of $R_{\perp}= 176\,{\rm kpc}$ from the f/g
quasar.  We follow the standard approach \cite{QPQ3}
and focus our
analysis on multiple ionization states of individual elements to avoid
uncertainties related to intrinsic abundances. Specifically, our
column density measurements in Table~\tababs\ yield the following
limits on ionic ratios:\\
C$^{+}\slash $C$^{3+} > 0.8$, Si$^{+}\slash$
Si$^{3+} > 1.3$, and Si$^{+}\slash $Si$^{2+} < 0.7$. In addition, we
measured the neutral column density to be $\mnhi = 10^{19.2}\cm{-2}$,
which we use as the stopping criterion of the Cloudy model. In other
words, Cloudy will continue to add slabs of material until this
neutral column is reached, guaranteeing that we satisfy this
constraint. Finally, we do not detect Ly$\alpha$ emission from the
vicinity of the background sightline, allowing us to set an upper
limit of ${\rm SB}_{\rm Ly\alpha} < \sbulim \cgssb$ from the cool
gas. This SB limit was computed via simulations whereby a fake
circular source with radius of $3\arcsec$ was placed at the location
of the b/g quasar. This exercise showed that only sources with
${\rm SB}_{\rm Ly\alpha} > 3\times {\rm SB}_{1\sigma}$ would be clearly
detectable, where ${\rm SB}_{1\sigma} = \sbsig \cgssb$ is our
1$\sigma$ SB limit for 1.0 sq. arcsec apertures.  Note that the
individual clouds intercepted by our sightline may have extremely
small sizes $\sim 10\,{\rm pc}$ making it impossible to angularly
resolve emission from a single cloud. However, we have argued that
these clouds have a large covering factor of $f_{\rm C}=0.5$ such that
we would be able to detect their aggregate emission near the location
of the b/g sightline (see Fig.~1 in \cite{QPQ4}
for more details).

Our sequence of photoionization models is compared to the
observational constraints in Fig.~\sfigcldyHI. The three different
ionic ratios considered paint a consistent picture for the ionization
state, and we constrain the ionization parameter to be $\log_{10} U
= \uval$, where we conservatively include an error of 0.3 dex to
account for both statistical and systematic uncertainties in the
photoionization modeling. Accordingly, the neutral fraction is
constrained to be $\log_{10} x_{\rm HI} = \xfrac$, and the total
hydrogen column $\log_{10} N_{\rm H} = \nhsig$, where the error on
$N_{\rm H}$ includes the quadrature sum of modeling uncertainty in the
neutral fraction, and the uncertainty in our measurement of $N_{\rm
HI}$ (see Supp.~9).
Plugging this determination of the total hydrogen column density into
eqn.~(\ref{eqn:M_c}) we obtain a total cool gas mass of $M_{\rm c} =
2.4\sci{11}\,M_\odot$ within a spherical halo $R = 250\,{\rm kpc}$, or
the range $1.0\sci{11}\,M_\odot < M_{\rm c} < 6.5\sci{11}\,M_\odot$ implied
by the $1\sigma$ measurement and modeling error on $N_{\rm H}$. Note that given the 
self-similarity of the photoionization models with ionization parameter
$U$, these results are essentially independent of the value of $\Phi$ 
that we have assumed in our models, and hence the distance of the gas from the quasar. 
Lastly, we have calculated the ionization corrections to the ionic column
densities measured for the absorbing gas and provide limits on the elemental
abundances implied by each metal absorption line measurement in Table~\tababs.


According to the photoionization models, if indeed the absorbing gas
lies at a distance $r=250\,{\rm kpc}$ and is illuminated by the f/g quasar, 
the Ly$\alpha$ emission would be ${\rm SB}_{\rm Ly\alpha} 
\simeq 3-4\times 10^{-18}\cgssb$, which lies just below 
our detection limit ${\rm SB}_{\rm Ly\alpha} = \sbulim \cgssb$, and
is hence consistent with our non-detection of extended emission near the
b/g quasar sightline. Indeed,
because the gas is optically thick, and essentially absorbs all
ionizing photons that impinge on it, the ${\rm SB}_{\rm Ly\alpha}$ is
nearly independent of the volume density, and as expected from
eqn.~(\ref{eqn:SB_thick}), depends primarily on $\Phi$ and our assumed
covering factor $f_{\rm C}=0.5$. The weak dependence on $n_{\rm H}$
arises because, at the lowest ionization parameters and hence highest
neutral fractions,  as much as $\sim 20\%$ of the Ly$\alpha$ is
produced by collisional excitation (green curve labeled collisions in
Fig.~\sfigcldyHI). This is because collisional excitation of neutral hydrogen
requires a large neutral fraction, and it is also exponentially sensitive to
temperature. The low $n_{\rm H}\lesssim 10^{-2}\,{\rm cm^3}$
and high ionization parameter $\log_{10} U \gtrsim -1$ models are
highly ionized $x_{\rm HI} \lesssim 10^{-3}$, and have higher
temperatures $T\gtrsim 17,000\,{\rm K}$, both of which tend to suppress
collisional excitation. As a result of the very weak
dependence of ${\rm SB}_{\rm Ly\alpha}$ on $n_{\rm H}$ in the
optically thick regime, our non-detection of emission does not allow
us to put interesting constraints on the volume density $n_{\rm H}$
for the absorbing gas.

Note that the fact that our photoionization model predicts
Ly$\alpha$ emission from the absorbing gas just below our detection limit
results from
our assumed distance of $r=250\,{\rm kpc}$, since for optically thick
gas ${\rm SB}_{\rm Ly\alpha} \propto \phi \propto r^{-2}$ (see
eqn.~\ref{eqn:SB_thick}). Had we assumed a distance equal to the
impact parameter $R_\perp=176\,{\rm kpc}$ the predicted emission would
be a factor of two higher ${\rm SB}_{\rm Ly\alpha}\sim
10^{-17}\cgssb$, and hence detectable. As such it is informative to
briefly discuss all possible scenarios that would result in a
non-detection of Ly$\alpha$ from the absorbing gas. These are: 1) the gas is
illuminated by the f/g quasar, but it lies at a distance $r \gtrsim
250\,{\rm kpc}$ such that its fluorescent Ly$\alpha$ emisssion is
below our detection threshold; 2) the gas is illuminated by the quasar
and is at a distance comparable to the impact parameter, but the
covering factor of optically thick gas is much lower than we have
assumed $f_{\rm C} = 0.5$; 3) the absorbing gas is not illuminated by
the f/g quasar but is instead shadowed, perhaps because from its
vantage point the quasar is extincted by the same obscuring medium
invoked in AGN unification models \cite{Antonucci93,UrryPadovani95}. 

Unfortuately, our current observations cannot distinguish between
these three scenarios, however we favor scenario 1), that the gas lies
at a distance $\simeq 250\,{\rm kpc}$ from the quasar, as shown in
Fig.~\sfigcldyHI.  While it is possible that the covering factor of
optically thick gas is indeed lower than the $f_{\rm C}=0.5$ that we
have assumed, we consider this unlikely for two reasons. First, this
would be in conflict with the statistical properties of optically
thick absorption for the aggregate population of
quasars \cite{QPQ6}.
Although, \qso\ is atypical given that it is a
protocluster and hosts a giant nebula, we already argued that the
smooth morphology of the giant nebula also argues for a high covering
factor $f_{\rm C}\gtrsim 0.5$ of the emitting gas, to prevent the
emission from appearing too clumpy. Under the reasonable assumption
that the absorbing and emitting gas have the same origin, one expects
their covering factors to be comparable.

At first glance, the possibility that the absorbing gas is shadowed
from the quasar radiation seems quite plausible. Indeed, we have
previously argued in the QPQ series that the anisotropic clustering of
LLSs around quasars implies that the optically thick gas in the quasar
CGM is often shadowed \cite{QPQ4,QPQ1,QPQ2}.
However, we robustly
determine the ionization parameter of the absorbing gas to be
$\log_{10} U = \uval$. If obscuration shadows the gas from the quasar,
then the radiation field would be expected to be the UV background,
which given our ionization parameter, would then imply a gas density
$n_{\rm H}\sim 10^{-3}\,{\rm cm}^{-3}$. In the next section, we will
see that the detection of bright Ly$\alpha$ emission implies that the
density of the emitting gas is three orders of magnitude higher
$n_{\rm H}\sim 1\,{\rm cm^{-3}}$. As indicated in the middle panel of
Fig.~\sfigcldyHI, if the gas is illuminated by the quasar at a
distance $r=250\,{\rm kpc}$, then gas density implied by our
ionization parameter is then $n_{\rm H}\simeq 0.7\,{\rm cm^{-3}}$,
which is then consistent with the value deduced from modeling the
emission. Thus again invoking the very reasonable assumption that the
absorbing and emitting gas have the same origin, the fact that we
infer a comparable $n_{\rm H}$ for the absorbing gas as required for
the emitting gas, seems to favor a scenario where the absorbing gas is
illuminated by the f/g quasar, but lies at a distance $r\gtrsim
250\,{\rm kpc}$, such that the fluorescent Ly$\alpha$ is below our
threshold. Future deeper Ly$\alpha$ observations could test this
hypothesis if they can reach ${\rm SB}_{\rm Ly\alpha}\simeq
10^{-18}\cgssb$ levels, such that one could then detect emission from
the absorbing gas out to a distance as far as $r\simeq 500\,{\rm
kpc}$.

\subsection{Photoionization Model of the Emitting Gas}

Here we consider photoionization models for the emitting gas under the
assumption that it has a comparable column density $N_{\rm H}$ to the
gas we detect in absorption. In modeling the emission we consider gas at
$r=100\,{\rm kpc}$ which is the characteristic scale of the emission 
nebula. We assume that the total hydrogen column density does not vary
strongly between these distances. This hypothesis is supported by a
photoionization modeling study of a large sample of quasar CGM
absorbers, which indicates that on average the total column density
$N_{\rm H}$ varies weakly (if at all) with impact parameter on such
scales. We run the same Cloudy photoionization models as described
previously, but with the modification that the stopping criterion is
now chosen to be that the total hydrogen column $N_{\rm
H}=10^{\nhval}\,{\rm cm^{-2}}$, deduced in the previous section. 


In Fig.~\sfigcldyH\ we show the results of our sequence of
photoionization models for the Ly$\alpha$ emitting gas. The black
curve illustrates the dependence of ${\rm SB}_{\rm Ly\alpha}$ on
volume density $n_{\rm H}$ for our fiducial choice of the radiation
field intensity, where the gas lies at a distance of $r=100\,{\rm
kpc}$. The upper x-axis indicates the corresponding ionization
parameter $U$ at this distance. The red and blue curves are for
stronger and weaker radiation fields, corresponding to $r=50\,{\rm
kpc}$ and $r=250\,{\rm kpc}$, respectively.  For low values of $n_{\rm
H}$ the gas remains optically thin $N_{\rm HI}\lesssim 10^{17.5}\,{\rm
cm^{-2}}$, and the ${\rm SB}_{\rm Ly\alpha}$ exhibits a nearly linear
dependence on the volume density $n_{\rm H}$ and is nearly independent
of the ionizing radiation intensity, as predicted by
eqn.~(\ref{eqn:SB_thin}). The gas becomes self-shielding and optically
thick once $N_{\rm HI}\gtrsim 10^{17.5}\,{\rm cm^{-2}}$, at which
point the ${\rm SB}_{\lya}$ plateaus at a single value depending only on $\Phi$
(since we have fixed the covering factor $f_{\rm C}=0.5$) and hence on
distance $r$, as predicted by eqn.~(\ref{eqn:SB_thick}).

These models clearly indicate that, given the column density of cool
material $N_{\rm H}=10^{\nhval}\,$\\
${\rm cm^{-2}}$, the volume density
must be $n_{\rm H} \simeq \nvolmn\,{\rm cm^{-3}}$, in order to
reproduce the observed average emission level at $100\,{\rm kpc}$ of
${\rm SB}_{\lya} =\sbannu \cgssb$ (indicated by the horizontal dashed line in
Fig.~\sfigcldyH). Provided that the emitting gas is optically thin
$N_{\rm HI}\lesssim 10^{17.5}$, this conclusion is essentially
indepdenent of the radiation field and hence the assumed distance of
the gas from the f/g quasar. In this regard, the relatively flat
radial ${\rm SB}_{\lya}$  profile observed for the giant nebula is readily
explained. It naturally arises because order of magnitude variations
in the radiation field, corresponding to gas at distances ranging from
$r = 50-200\,{\rm kpc}$, produce nearly the same level of Ly$\alpha$
emission, provided that $N_{\rm H}$ does not vary significantly across
the nebula, as we have assumed.

A corollary of our absorption-line constraint on $N_{\rm H} \simeq
10^{\nhval}\,{\rm cm^{-2}}$ and our emission constraint on $n_{\rm
H}\simeq \nvolmn\,{\rm cm^{-3}}$ is that the implied absorption path
length through the cool clouds is $l_{\rm cloud} \equiv N_{\rm
H}\slash n_{\rm H} \sim 40\,{\rm pc}$, implying that the emitting
gas is distributed in extremely compact dense clouds.

\begin{figure}[!ht]
\begin{center}
        \includegraphics[width= \textwidth]{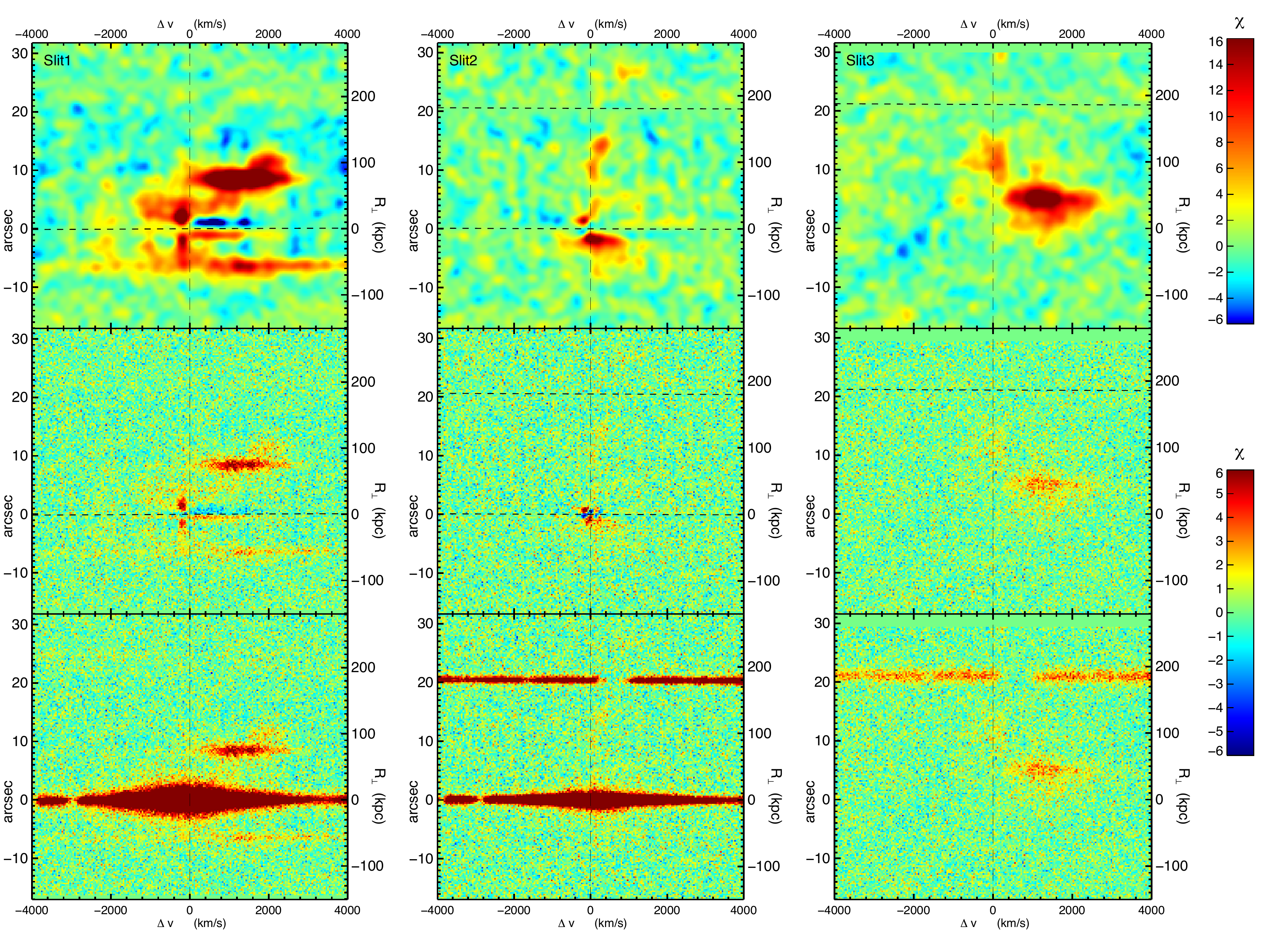}
\end{center}
{\bf Fig.~\sfigchimaps:} {\small {\bf Two dimensional spectrum
plotted as a $\chi$-map for the three slits in Fig.~\figslits.} The
lower and middle rows of each image show $\chi_{\rm sky}$ (sky-subtracted only)
and $\chi_{\rm sky+PSF}$ (sky and PSF subtracted), respectively as
defined in \S~\ref{sec:optical_obs} (see (22) for additional details).
The upper row shows the smoothed map $\chi_{\rm smth}$, (sky and PSF
subtracted, then smoothed), which is helpful for identifying extended
emission. A symmetric Gaussian kernel (same spatial and spectral
widths in pixels) was adopted, with dispersion $\sigma_{\rm smth} =
100\,\kms$ (FWHM $=235\,\kms$), and $1\sigma$ spatial smoothing of
$0.65\arcsec$ (FWHM=$1.5\arcsec$). In the absence of extended
emission, the distribution of pixel values in the the sky and PSF
subtracted images should be Gaussian with unit variance. The lower and
middle rows are displayed with a linear stretch ranging from
$-6\sigma$ to $6\sigma$ indicated by the color-bar at middle right.
The upper row has a stretch from $-6\sigma$ to $16\sigma$ is
adopted, and a $\gamma$ value chosen to decrease the range of color at
the highest significance, as indicated by the color-bar at upper
right. The horizontal dashed lines indicates the spectral traces for
the f/g and b/g quasar, and the vertical dashed line indicates $\Delta
v=0$, corresponding to the systemic redshift of the f/g quasar. Note
that the b/g quasar spectrum in Slit3 (right) has much lower ${\rm
S}\slash{\rm N}$ ratio than in Slit2 (center) because of much poorer
seeing.}
\end{figure}    
\noindent

\clearpage
\begin{figure}[!ht]
\begin{center}
\includegraphics[width=0.95\textwidth]{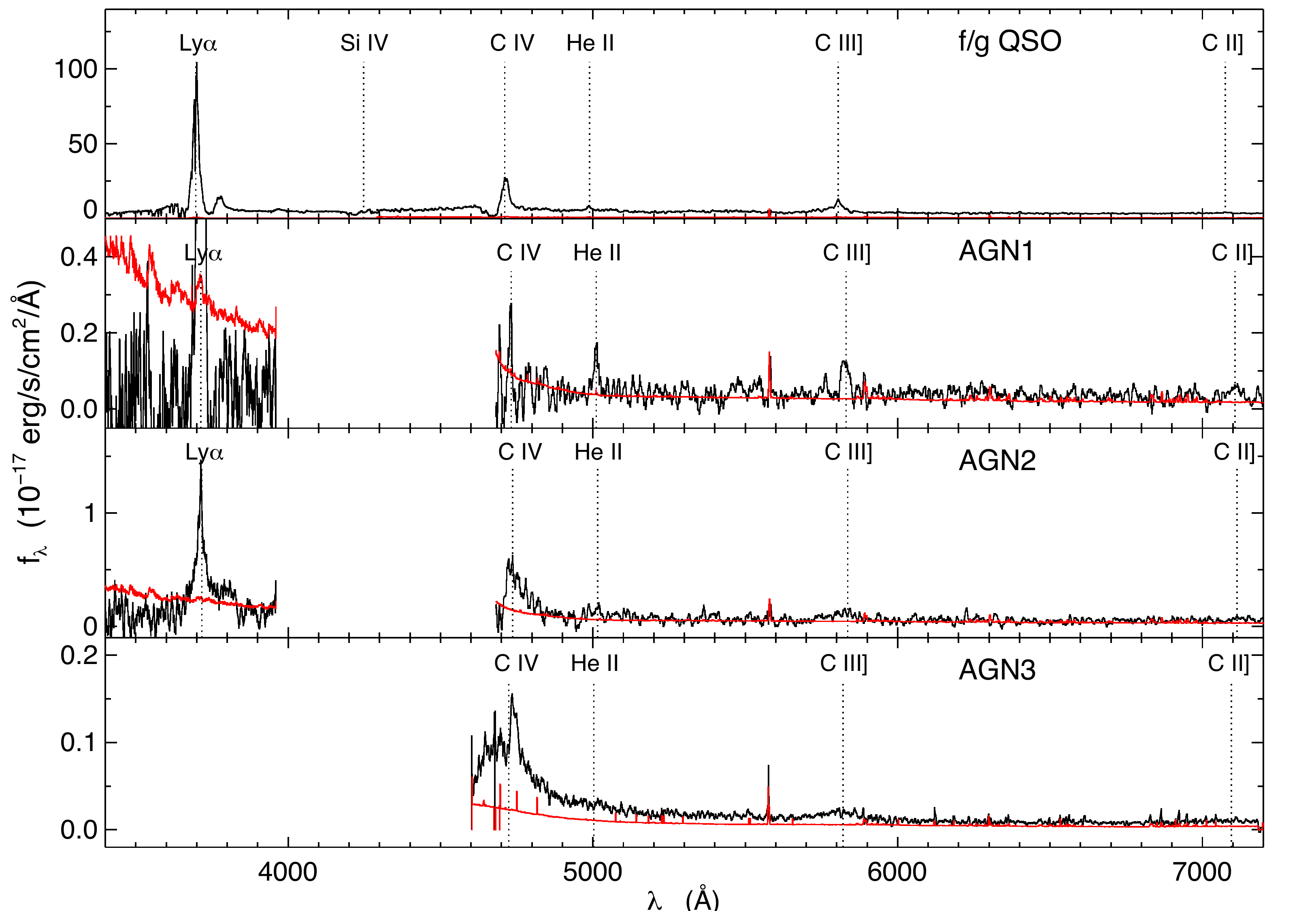}
\end{center}
\noindent
{\bf Fig.~\sfigagn:}
{\small 
{\bf Optical spectra of the four AGN at the same redshift.} From top to
bottom: f/g quasar, AGN1, AGN2, and AGN3. The gaps in the coverage of the
spectra of AGN1 and AGN2 are due to the LRIS configuration chosen to
cover both \mlya, \ion{C}{iv}, and \ion{He}{ii}. The spectrum of AGN3 does not
cover the \mlya\ line due to the limited blue sensitivity of
DEIMOS. Note the different scale on the y-axis to better show the
different emission lines, which are marked with vertical dotted lines and
labeled. 
}
\end{figure}

\clearpage
\begin{figure}[!ht]
\begin{center}
\includegraphics[width=0.45\textwidth]{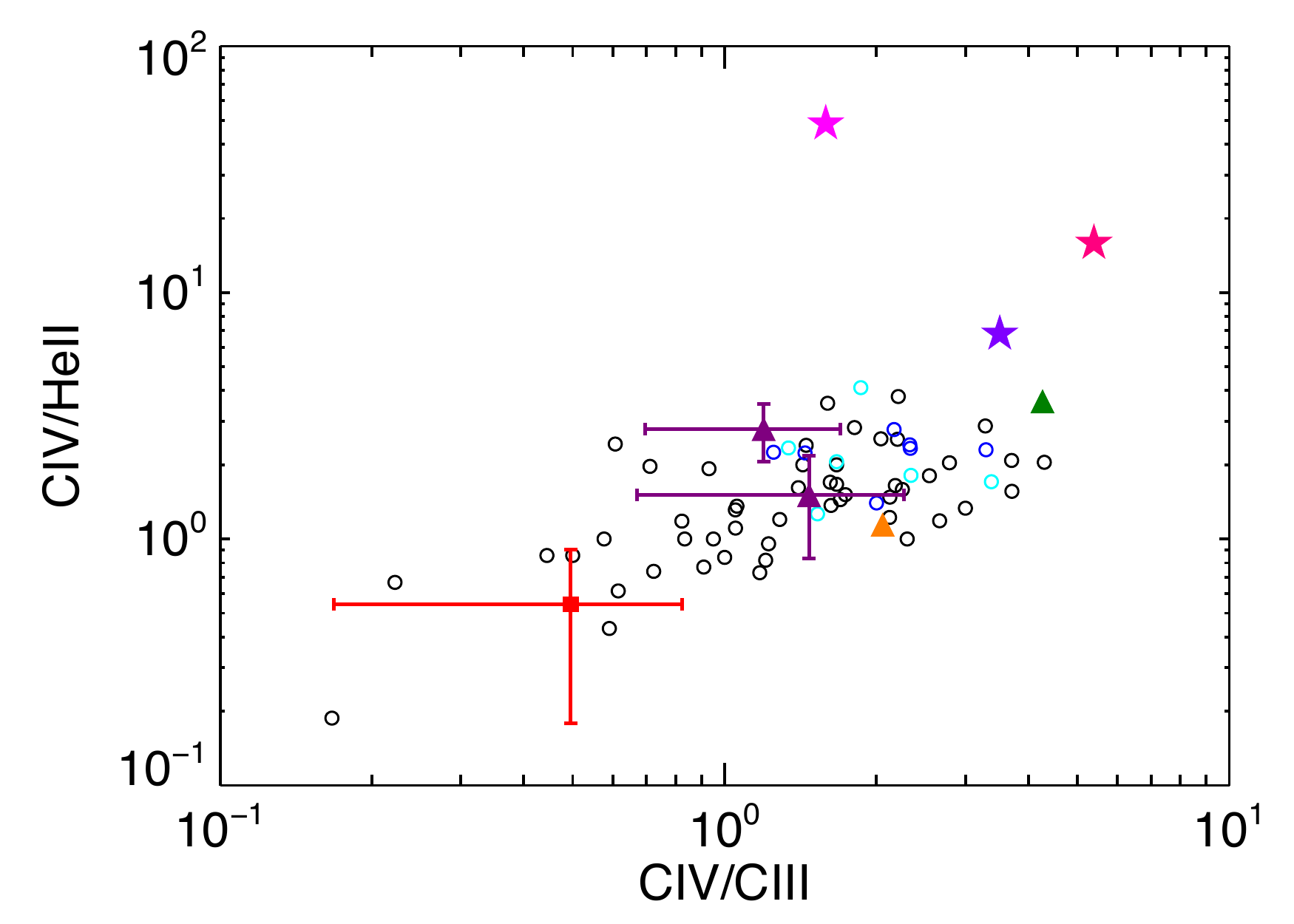}
\includegraphics[width=0.45\textwidth]{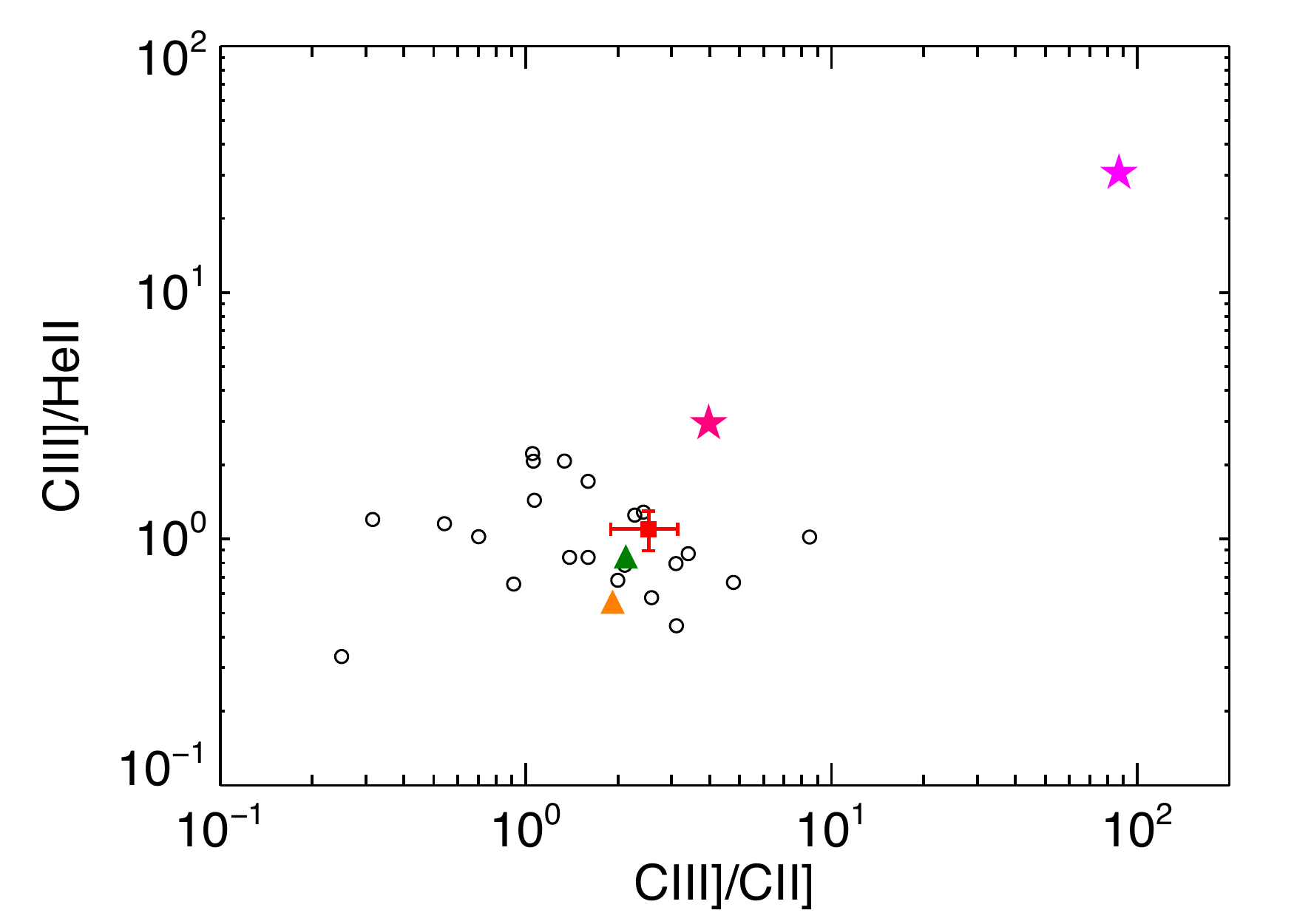}
\end{center}
\noindent
{\bf Fig.~\sfiglineratio:} {\small {\bf Comparison of the
Line-ratios of AGN1 to other Type-2 AGN.}  The line-ratios
of AGN1 (red square) are compared to other Type-2 AGN compiled
from the literature
in the \ion{C}{iv}$\slash$\ion{He}{ii}
vs \ion{C}{iv}$\slash$\ion{C}{iii}] plane (left) as well as
the \ion{C}{iii}]$\slash$\ion{He}{ii}
vs \ion{C}{iii}]$\slash$\ion{C}{ii}] plane (right).
The circles are
individual measurements of HzRGs (black) from the compilation
of \cite{debreuck00},
and narrow-line X-ray sources (cyan)
and Seyfert-2s (blue) from the compilation
of \cite{NagaoNLR06}.
Triangles indicate
measurements from the composite spectra of HzRGs (orange)
from \cite{mccarthy93},
the composite Type-2 AGN spectra (purple) 
from \cite{Hainline11},
who split their population into two
samples above and below Ly$\alpha$ EW of 63\AA, and a composite
spectrum of mid-IR selected Type-2 AGN (green) from \cite{Lacy13}. 
The stars represent measurements from composite
spectra of Type-1 quasars, based on the analysis
of (magenta) \cite{vanden01},
(red-magenta) \cite{NagaoBLR06},
and (blue-magenta) \cite{Zheng97}. 
The line-ratios for AGN1 lie in the locus of points spanned
by other Type-2 objects in the literature. 
}
\end{figure}


\clearpage
\begin{figure}[h!]
\begin{center}
\includegraphics[width=0.75\textwidth]{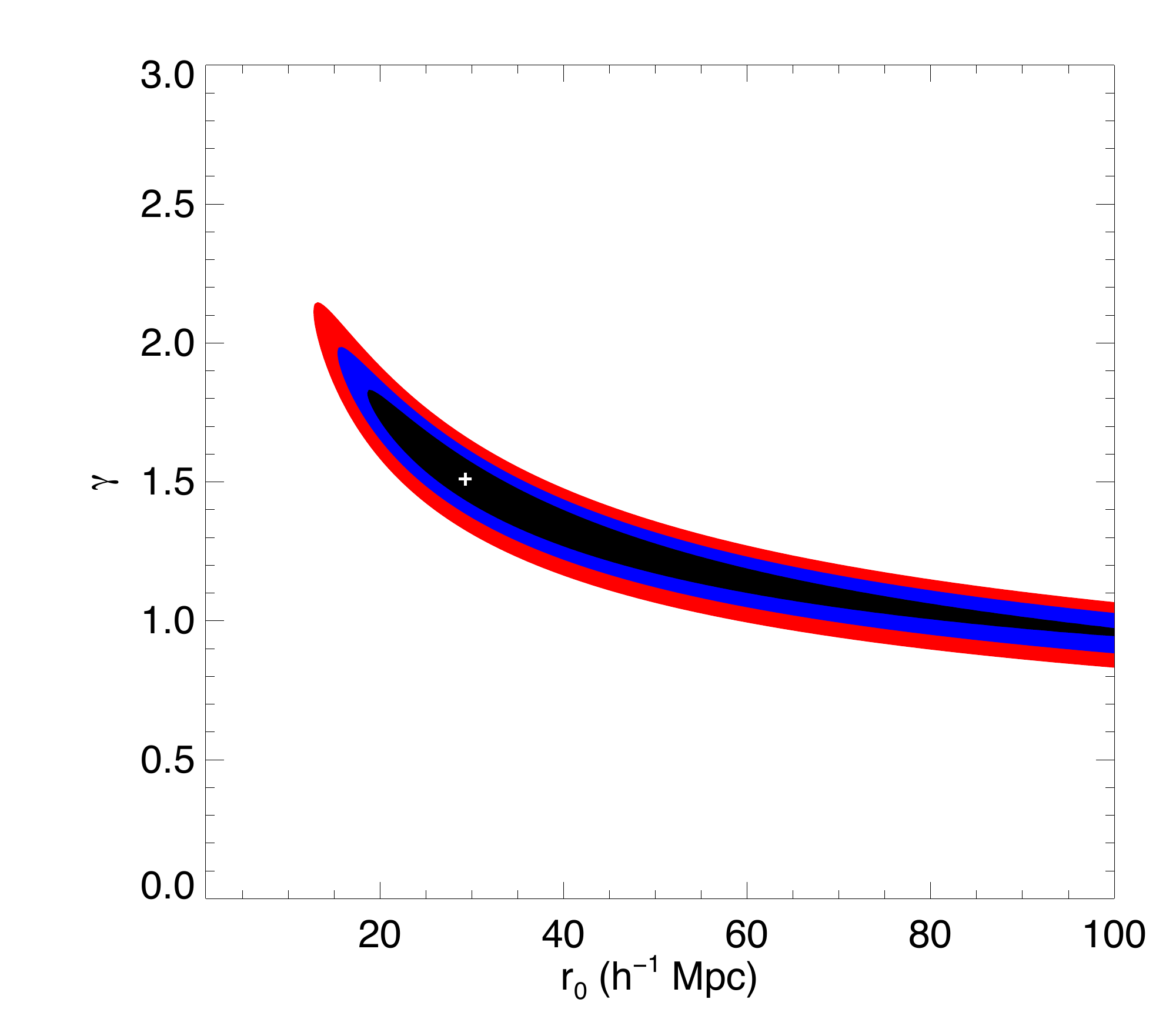}
\end{center}
\noindent
{\bf Fig.~\sfigcontour:} {\small {\bf Confidence regions on correlation
function parameters from maximum likeklihood clustering analysis.}
The 1\,,2\,, and 3$\sigma$ confidence regions are illustrated by the
black, blue, and red contours, respectively.  The maximum likelihood
value is $r_0= \rnotval\,h^{-1}\,{\rm Mpc}$ and $\gamma =\gammaval$, which is
indicated by the white cross. Due to the low signal-to-noise ratio of
the clustering measurement, and the form adopted for the correlation
function $\xi(r) \equiv (r\slash r_0)^{-\gamma}$ (such that amplitude
and slope are not independent parameters), a strong degeneracy exists
between $r_0$ and $\gamma$, such that neither is individually well
determined.}
\end{figure}

\clearpage

\begin{figure}[!ht]
\begin{center}
\includegraphics[width=0.75\textwidth]{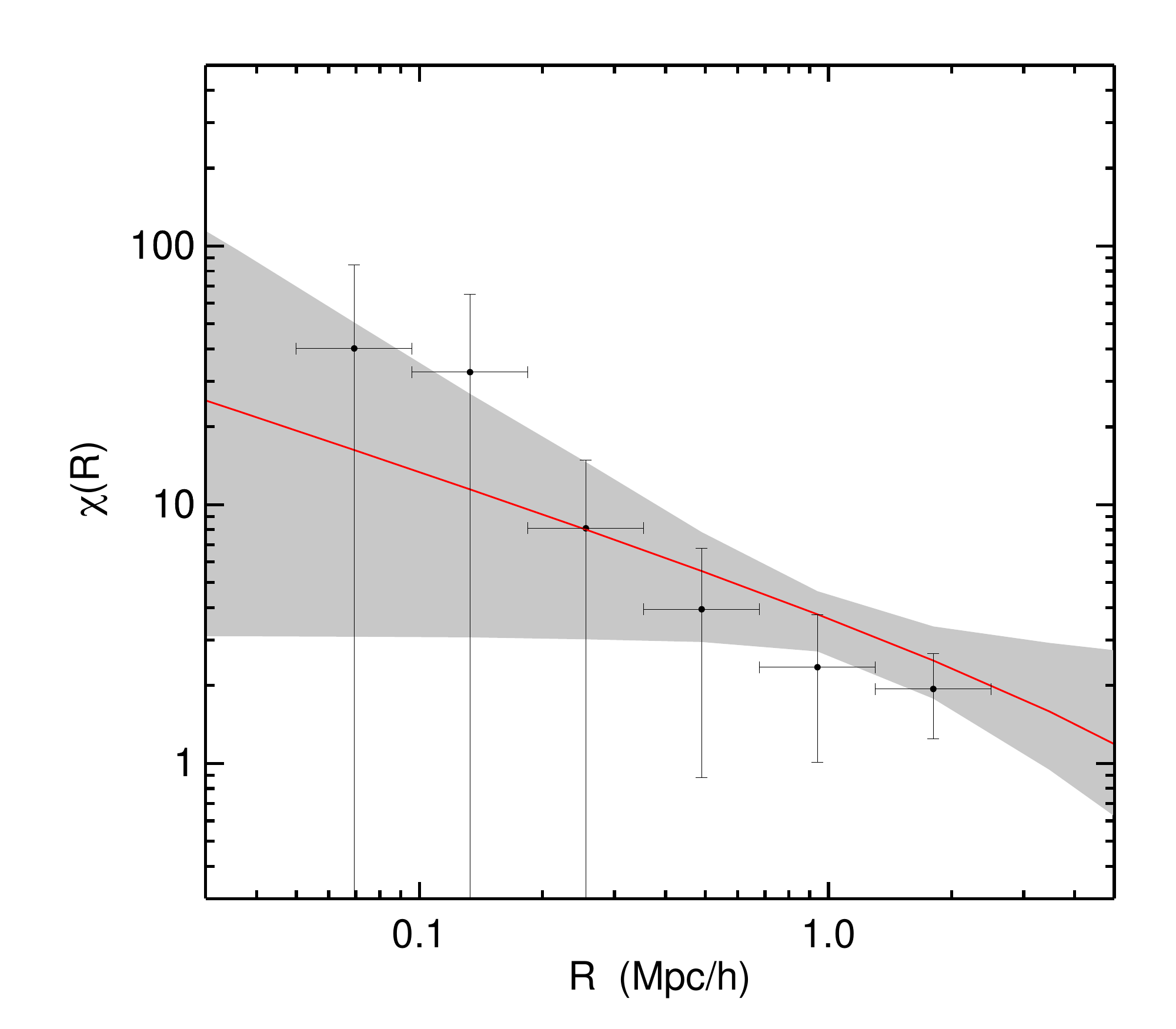}
\end{center}
\noindent
{\bf Fig.~\sfigcorr:} {\small {\bf Binned correlation function of
LAEs around HzRGs and LABs.}  The dimensionless correlation function
$\chi$ defined by eqn.~(\ref{eqn:chi}) is indicated by the data
points, with $1\sigma$ error bars, determined by taking the 16th and
84th percentiles of the distribution of $\chi$ resulting from our
bootstrap procedure. The solid red line indicates the best-fit
protocluster-LAE cross correlation function
($r_0= \rnotval\,h^{-1}\,{\rm Mpc}$ and $\gamma =\gammaval$)
determined by our unbinned maximum likelihood procedure.  The gray
shaded region indicates the $1\sigma$ error on our cross-correlation
function fit, implied by the erorrs on the parameters $r_0$ and
$\gamma$. This region is determined by evaluating $\chi$ for each
sample of the boostrap error distributions of $r_0$ and $\gamma$, and
taking the 16th to 84th percentile confidence region of $\chi$. Our best fit
correlation function (red line), and its quoted errors (gray shaded) clearly
provides an acceptable representation of the binned data points and their respective
errors.}
\end{figure}

\clearpage
\begin{figure}[!ht]
\begin{center}
\includegraphics[width=0.9\textwidth]{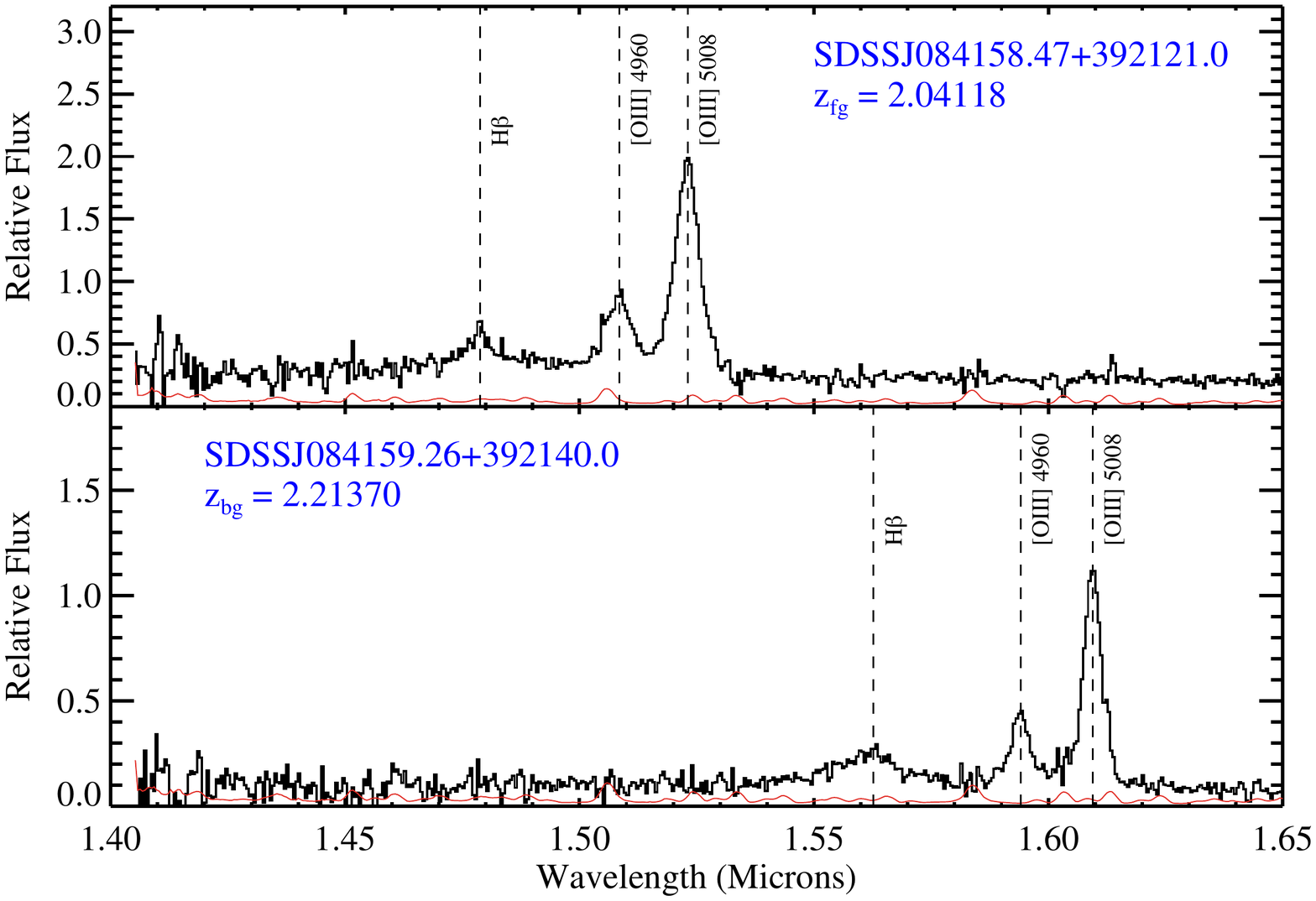}
\end{center}
\vskip -1in
\noindent
{\bf Fig.~\sfignearir:}
{\small 
{\bf Near-IR spectrum of the f/g quasar SDSSJ084158.47+392121.0 (top)
  and of the b/g quasar BOSSJ084159.26+392140.0 (bottom) obtained with
  the NIRI spectrometer on the Gemini-North telescope.} Each quasar
shows the detection of strong [\ion{O}{III}] emission and modest H$\beta$
emission. The [\ion{O}{iii}]$\lambda$5008\AA\ doublet lines were centroided,
giving observed wavelength centroids of
$\lambda_{\rm obs} = 1.5231\pm0.0001\mu$m and $1.6095\pm0.00006\mu$m
for the f/g and b/g quasars respectively. This gives systemic
redshifts $z_{\rm fg} = 2.0418$ for the f/g quasar and 2.21370 for the
b/g quasar. The estimated uncertainty on this redshift measurements is
approximately 50 km s$^{-1}$, taking into account both the error in our
wavelength calibration and the estimated uncertainty when one uses
[\ion{O}{iii}]to assess the systemic redshift of a quasar.  This
uncertainty on the measurement of the redshift is far lower than the
width of our custom narrow-band filter ($\sim 2600 \kms$).
} 
\end{figure}

\clearpage

\begin{figure}[!ht]
\begin{center}
\includegraphics[width=0.75\textwidth]{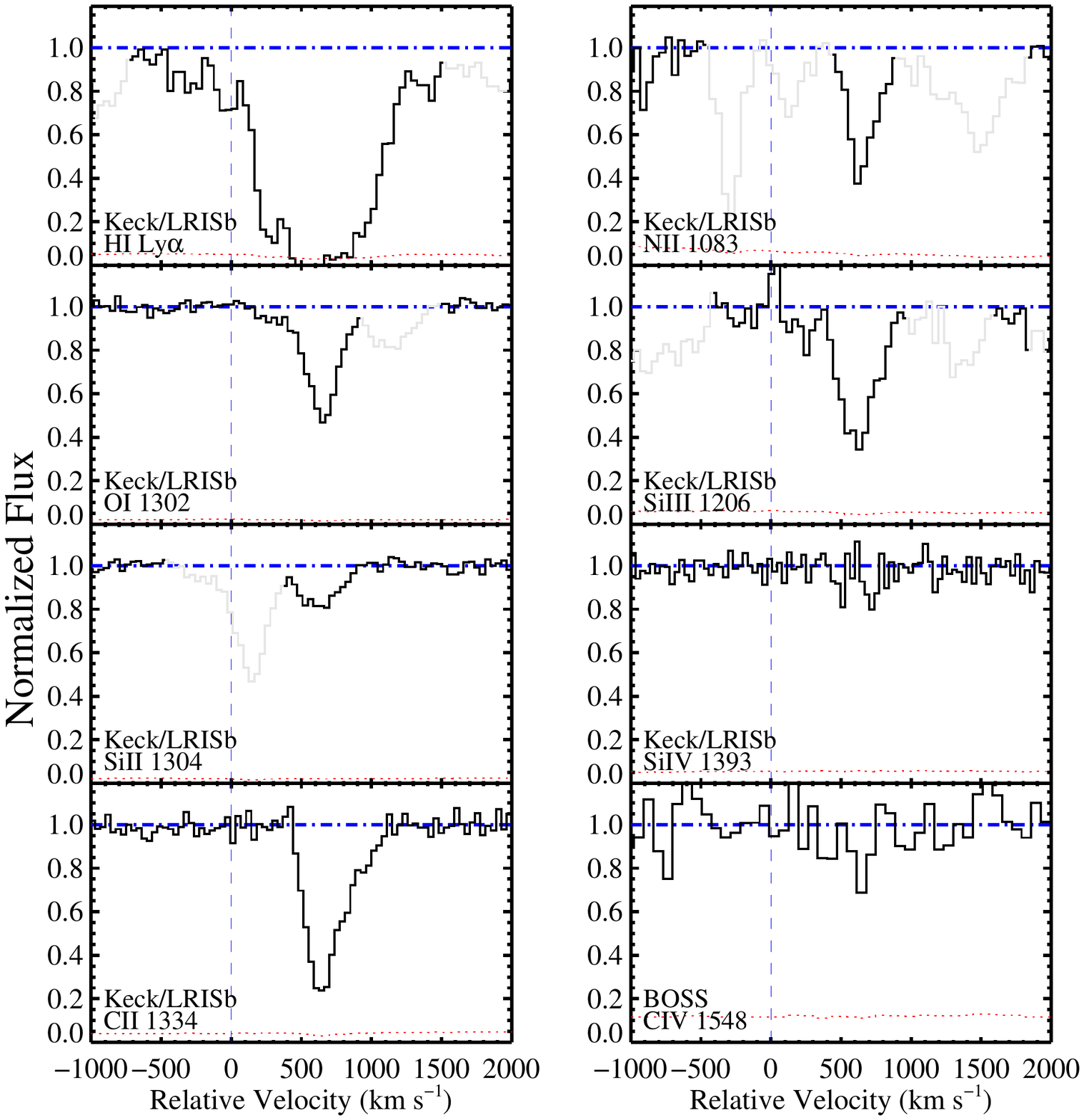}
\end{center}
\vskip -1.2in
\noindent
{\bf Fig.~\sfigabs:} {\small {\bf One-dimensional spectra of the b/g
quasar centered at the wavelengths of a series of commonly observed
UV-line transitions in the rest frame of the f/g quasar.}  The gas
exhibits especially strong low-ion absorption (\ion{O}{i}$\lambda
1302$, \ion{Si}{ii}$\lambda 1304$, \ion{C}{ii}$\lambda 1334$) and
moderately strong intermediate ion absorption (\ion{N}{ii}$\lambda 1083$, \ion{Si}{iii}$\lambda 1206$).
The latter absorption occurs within the \mlya\ forest and could be
partially blended with coincident IGM absorption, but such
contamination should be modest given that the line-profiles track one
another and also the low-ion transitions.  In contrast, there is no
statistically significant absorption corresponding to the \ion{Si}{iv}$\lambda 1393$ and
\ion{C}{iv}$\lambda 1548$ transitions indicating the gas is not highly ionized.  See
Table~\tababs\ for a summary of the properties of these metal
absorption lines.  }
\end{figure}

\clearpage

\begin{figure}[!ht]
\begin{center}
\includegraphics[width=0.75\textwidth]{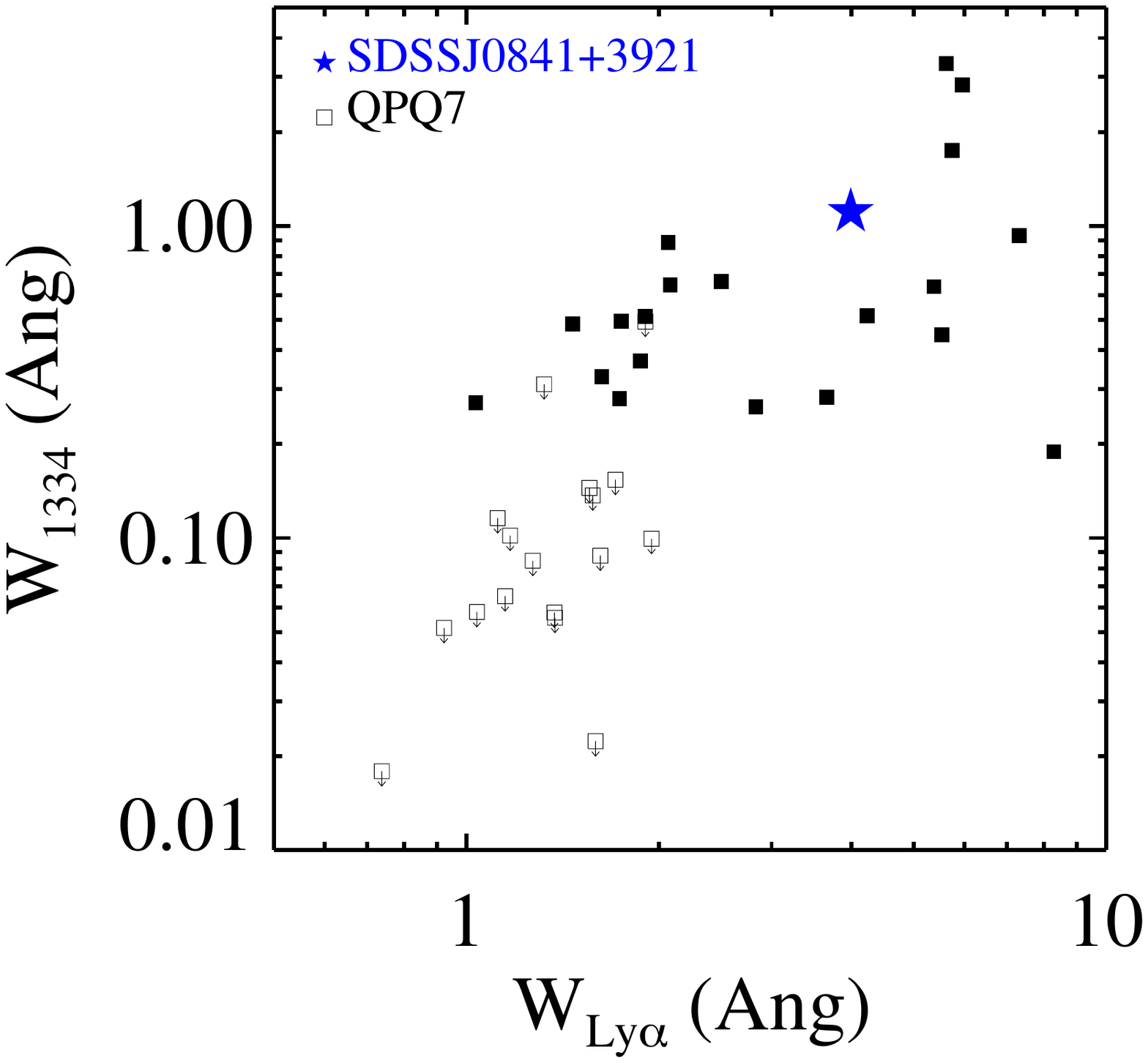}
\end{center}
\noindent
{\bf Fig.~\sfigqpq:}
{\small 
{\bf Equivalent widths of cool gas absorption in the QPQ dataset.} 
Measurements of the equivalent widths of the \ion{C}{ii}~1334
transition against the measurement for \ion{H}{i}~\mlya.  The
measurements are drawn from our QPQ survey \cite{QPQ6,QPQ7}
and the pair
that led to the serendipitous discovery of the quadruple AGN system is
marked separately.  The gas probed by the background quasar (at an
impact parameter of 176\,kpc) shows strong low-ion absorption
characterstic of the full population.  This indicates a highly
enriched and optically thick gas.
}
\end{figure}

\clearpage

\begin{figure}[!ht]
\begin{center}
\includegraphics[width=0.6\textwidth,bb=72 72 504 650]{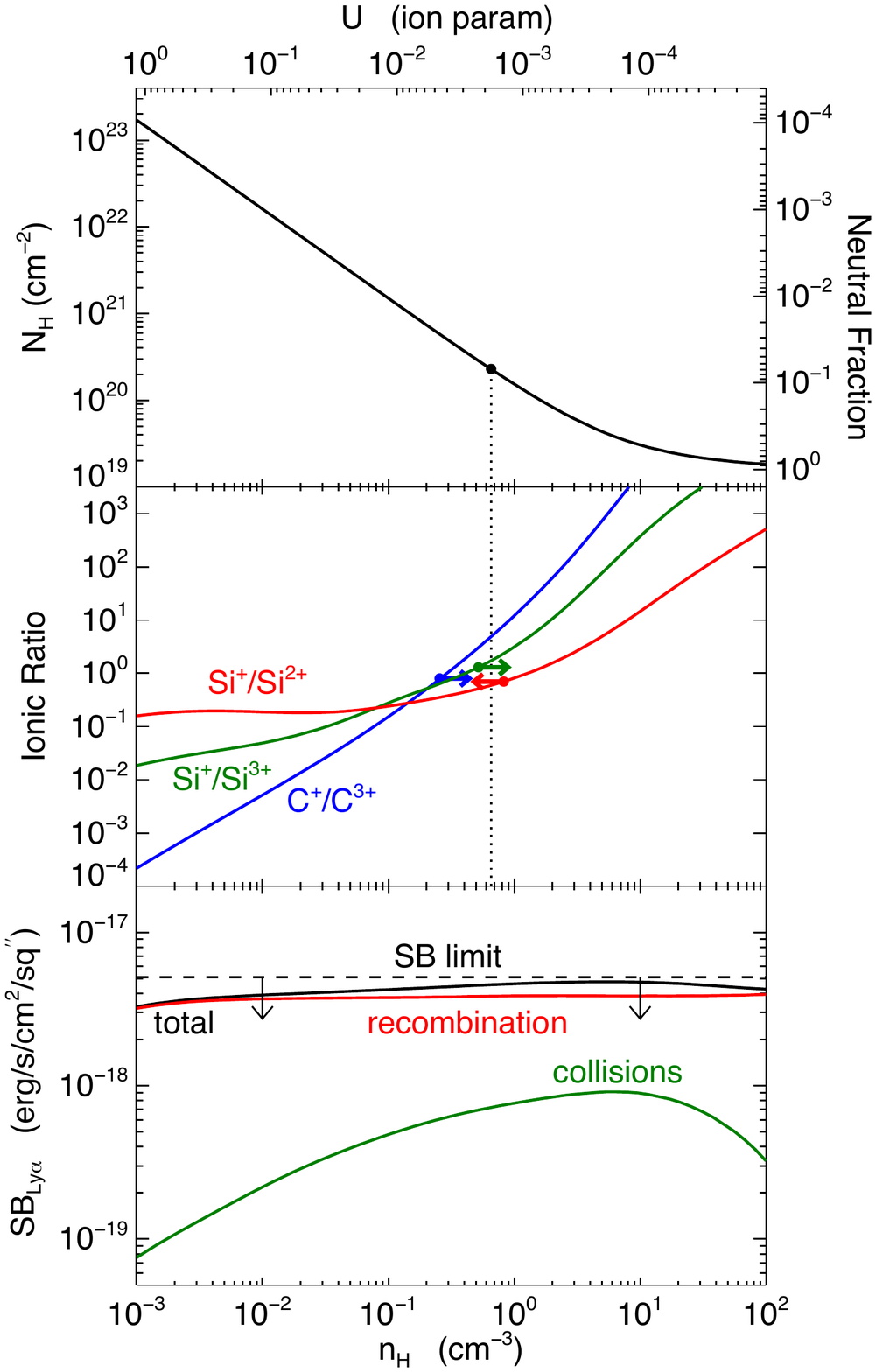}
\end{center}
\vskip -0.3in
\noindent
{\bf Fig.~\sfigcldyHI:} {\small {\bf Photoionization model of the
optically thick absorber in b/g quasar spectrum.}  The gas is assumed
to be at $r=250\,{\rm kpc}$ and exposed to the radiation field of the
f/g quasar. The models are a one-dimensional sequence parameterized by
the hydrogen volume density $n_{\rm H}$ on the x-axis, which given
that the radiation field is fixed, is also a proxy for the ionization
parameter $U$ shown on the upper x-axis.  Upper panel: Variation of
the total hydrogen column $N_{\rm H}$ with $n_{\rm H}$/$U$, given that
we require the model to reproduce our measured neutral column density
$N_{\rm HI}=10^{\nhival}$. The y-axis on the right shows the
corresponding neutral fraction. Our measured $\log_{10} U$ (middle
panel) in turn implies $\log_{10} N_{\rm H} = \nhsig$, indicated by
the vertical dotted line.  Middle Panel: Variation of three different
ionic ratios (${\rm Si^{+}\slash Si^{2+}}$ in red; ${\rm Si^{+}\slash
Si^{3+}}$ in green; ${\rm C^{+}\slash C^{3+}}$ in blue) with $n_{\rm
H}$/$U$. Our measured lower/upper limits on these ratios from
Table~\tababs\ are illustrated by the colored arrows. We obtain a
consistent photoioinization solution implying $\log_{10} U = \uval$,
indicated by the vertical dotted line.  Lower Panel: Predicted SB of
extended Ly$\alpha$ emission from the absorbing gas as a function of
$n_{\rm H}$/$U$. The red and green curves show the respective
contributions from recombinations and collisional excitation, and the
black curve shows the total Ly$\alpha$ SB. The horizontal dashed line
with arrows indicates our detection limit for an extended source near
the b/g quasar sightline ${\rm SB}_{\rm Ly\alpha} < \sbulim \cgssb$.
If the absorbing gas is at $r=250\,{\rm kpc}$ as we have assumed, our
models are marginally consistent with a non-detection of extended
Ly$\alpha$ emission at that location. If the gas lies at larger
distances $r > 250\,{\rm kpc}$, the predicted SB will decreaes as
$r^{-2}$.  }
\end{figure}

\clearpage

\begin{figure}[!ht]
\begin{center}
\includegraphics[width=0.6\textwidth,bb=72 72 504 500]{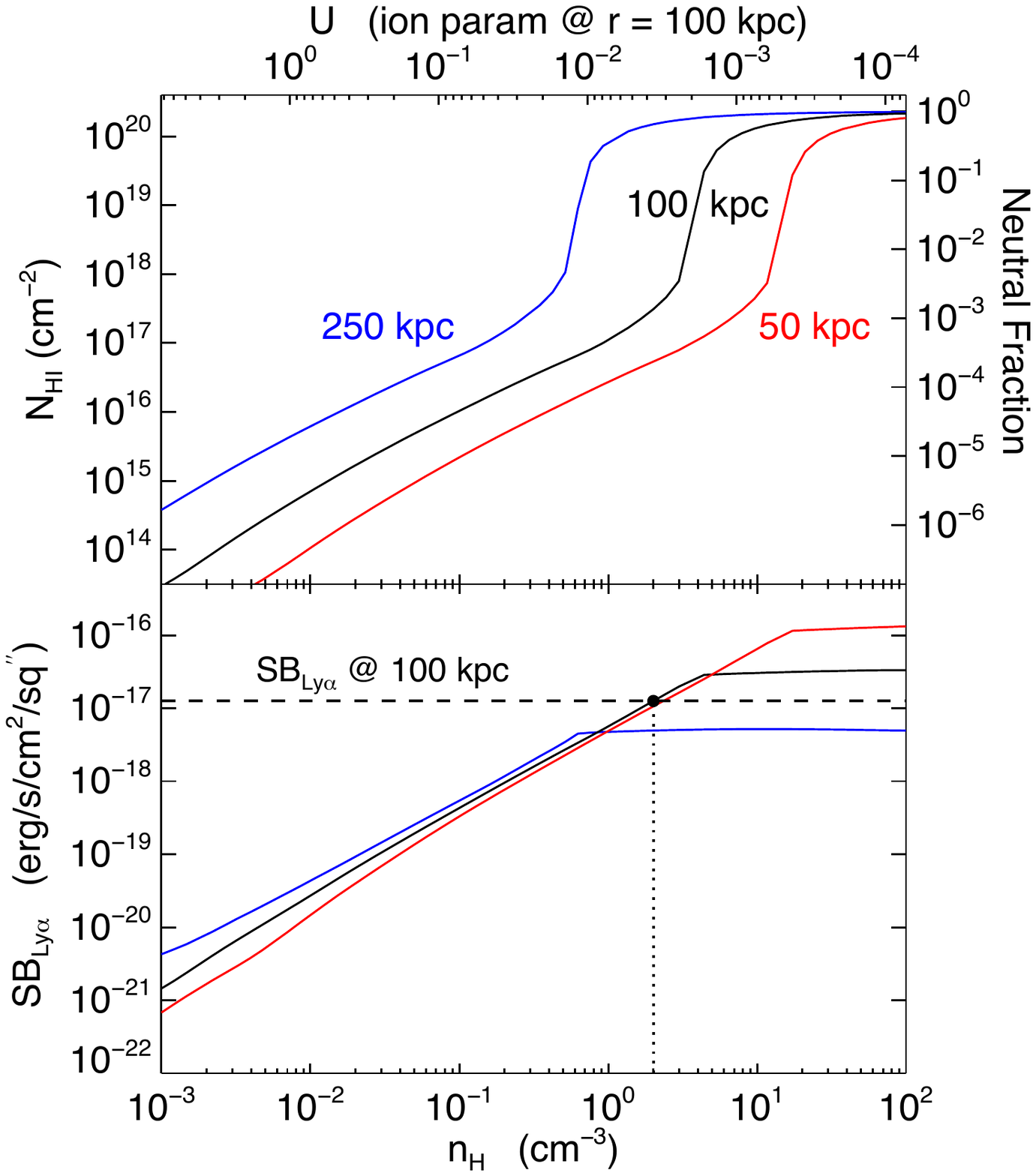}
\end{center}
\noindent
\vskip -0.2in
{\bf Fig.~\sfigcldyH:} {\small {\bf Photoionization model of
the Ly$\alpha$ emitting nebular gas.} The models assume a total
hydrogen column density $N_{\rm H}=10^{\nhval}\,{\rm cm^{-2}}$,
deduced from our photoionization modeling of the optically thick
absorber in the b/g sightline (see Fig.~\sfigcldyHI
and \S~\ref{sec:cldyHI}). The gas is assumed to be exposed to the
radiation field of the f/g quasar at distances of $r=50$, $100$, and
$250\,{\rm kpc}$, indicated by the red, black, and blue curves,
respectively.  The models are parameterized by the hydrogen volume
density $n_{\rm H}$ on the x-axis. The ionization parameter $U$
corresponding to the gas at $r=100\,{\rm kpc}$ (black curves) is
denoted on the upper x-axis. Upper panel: Variation of the neutral
hydrogen column $N_{\rm HI}$ with $n_{\rm H}$. The y-axis on the right
shows the corresponding neutral fraction. Lower panel: Variation of
the Ly$\alpha$ SB with $n_{\rm H}$. For low $n_{\rm H}$ the gas
remains optically thin $N_{\rm HI}\lesssim 10^{17.5}\,{\rm cm^{-2}}$,
and the SB ${\rm SB}_{\rm Ly\alpha}$ varies linearly with $n_{\rm H}$
and is nearly independent of the ionizing radiation intensity. At
higher $n_{\rm H}$ the gas begins self-shielding $N_{\rm HI}\gtrsim
10^{17.5}\,{\rm cm^{-2}}$, and the SB plateaus at a single value
depending only on $\Phi$ and hence on distance $r$. The models
indicate that a value of $n_{\rm H} \simeq \nvolmn\,{\rm cm^{-3}}$, 
indicated by the vertical
dotted line, is required in order to reproduce the observed average
emission level at $100\,{\rm kpc}$ of ${\rm SB}_{\lya}
=\sbannu \cgssb$, indicated by the horizontal dashed line in
Fig.~\sfigcldyH. This conclusion is essentially
independent of the radiation field and hence the assumed distance of
the gas from the f/g quasar.}
\end{figure}


\clearpage

{\footnotesize
\begin{tabular*}{\textwidth}{@{\extracolsep{\fill}}lcccccc}
\hline
\hline
Name & RA & DEC & $z$ & $\sigma_z$ & $\theta$ & $R_{\perp}$\\
     &    &     &  & (${\rm km\,s^{-1}})$ & ($\arcsec$) & (kpc)\\
\hline
f/g QSO & 08~41~58.47 & $+$39~21~21.0 & 2.0412 & \phn\phn  50 & --- & --- \\
AGN1   & 08~41~58.24 & $+$39~21~29.1 & 2.055\phn & \phn 400 & \phn 8.6 & \phn 71 \\
AGN2   & 08~41~58.66 & $+$39~21~14.7 & 2.058\phn & \phn 700 & \phn 6.6 & \phn 55 \\
AGN3   & 08~41~59.10 & $+$39~21~04.6 & 2.050\phn & \phn 700 & 18.0 & 150 \\
b/g QSO & 08~41~59.25 & $+$39~21~40.0 & 2.2138 & \phn\phn  50 & 21.1 & 176 \\
\hline
\end{tabular*}
\newline
\newline

\noindent
{\bf Table~\tabagn: AGN properties.}
{\small
The coordinates, redshift $z$, redshift error $\sigma_z$, angular $\theta$ and
transverse separation $R_{\perp}$ from the f/g quasar are listed for the four AGN in the
protocluster, as well as the b/g quasar.
}

\clearpage
\begin{sidewaystable}

{\tiny
\begin{tabular*}{\textwidth}{@{\extracolsep{\fill}}lccccccccccccc}
\hline
\hline
Name & $u\pm \sigma_u$ & $g\pm \sigma_g$ & $r\pm \sigma_r$ & $i\pm \sigma_i$ & $z\pm \sigma_z$ & 
W1$\pm \sigma_{\rm W1}$ & W2$\pm \sigma_{\rm W2}$ & W3$\pm \sigma_{\rm W3}$ & W4$\pm \sigma_{\rm W4}$ & 
$F_{20\, {\rm cm}}$ & $V\pm \sigma_V$ & $f_{\rm Ly\alpha}$ & $W_{\rm Ly\alpha}$\\
\hline
f/g QSO & 19.76$\pm$0.02 & 19.75$\pm$0.01 & 19.58$\pm$0.01 & 19.27$\pm$0.01 & 19.19$\pm$0.03 & 15.85$\pm$0.03 & 14.65$\pm$0.04 & 10.79$\pm$0.07 &  8.33$\pm$0.36 & $<$ 1.02 & 19.76$\pm$0.03 &  1709.0 & \phn 72.6 \\
AGN1   &  $>$23.11 & 24.04$\pm$0.31 &  $>$23.74 &  $>$23.24 &  $>$21.86 & 16.89$\pm$0.08 & 15.86$\pm$0.11 &  $>$12.58 &  $>$ 8.36 & $<$ 1.02 & 23.95$\pm$0.07 & \phn  210.1 & 110.8 \\
AGN2   &  $>$23.11 &  $>$24.34 &  $>$23.40 &  $>$22.93 &  $>$21.86 & 18.19$\pm$0.26 &  $>$17.10 &  $>$12.58 &  $>$ 8.35 & \phn12.87 & 23.12$\pm$0.06 & \phn\phn   41.1 & \phn 37.0 \\
AGN3   &  $>$23.11 & 22.71$\pm$0.09 & 22.73$\pm$0.16 & 22.18$\pm$0.15 &  $>$21.86 &  $>$18.52 &  $>$17.12 &  $>$12.58 &  $>$ 8.36 & $<$ 1.02 & 23.09$\pm$0.06 & \phn\phn   39.8 & \phn 34.7 \\
b/g QSO & 20.00$\pm$0.02 & 19.39$\pm$0.00 & 19.71$\pm$0.01 & 19.39$\pm$0.01 & 19.39$\pm$0.04 & 16.40$\pm$0.05 & 15.76$\pm$0.10 & 12.00$\pm$0.21 &  $>$ 8.36 & $<$ 1.02 & 19.96$\pm$0.03 & \phn\phn\phn   -7.0 & \phn\phn -6.2 \\
\hline
\end{tabular*}
\newline
\newline

}
\noindent
{\bf Table~\tabagnphot: AGN photometry.}  {\footnotesize SDSS optical
($ugriz$) and WISE mid-IR (W$1-4$) magnitudes are determined from
forced photometry of the corresponding images. SDSS and LRIS magnitudes are on
the AB scale, whereas WISE magnitudes are in the Vega system. For
measurements with ${\rm S\slash N} < 3$ we quote lower limits on the
magnitudes.  Peak 20cm radio flux $F_{\rm 20cm}$ (mJy) are listed for
objects with a match in the source catalog from the VLA/FIRST survey
(Becker et al. 1995). For objects without FIRST matches we list the
$5\sigma$ upper limit based on the sky-rms coverage maps. We also list
Keck LRIS $V$ band magnitudes, Ly$\alpha$ line fluxes $f_{\rm
Ly\alpha}$ ($10^{-17}$ erg s$^{-1}$ cm$^{-2}$), and rest-frame
Ly$\alpha$ equivalent width (\AA).}
\end{sidewaystable}

\clearpage

{\footnotesize
\begin{tabular*}{\textwidth}{@{\extracolsep{\fill}}lcccccc}
\hline
\hline
Line & FWHM &  Flux & $W_{\lambda}$\\
     &  ($\kms$) & $(10^{-17}\,{\rm erg\,s^{-1}\,cm^{-2}})$ & 
 (\AA)\\
\hline
Ly$\alpha$............ & $1575\pm 36$ & $76.5\pm 1.5$ & $>118.2$\\
NV............. & ---& $-0.4\pm 1.4$ & ---\\
CIV............ & ---& $ 1.3\pm 0.8$ & ---\\
HeII........... & $1154\pm 276$ & $ 2.4\pm 0.4$ & $> 15.8$\\
CIII].......... & $1434\pm 171$ & $ 2.6\pm 0.3$ & $ 22.6\pm 2.2$\\
CII]........... & ---& $ 1.0\pm 0.2$ & $ 11.2\pm 2.6$\\
\hline
\end{tabular*}

\noindent
{\bf Table~\tablines: Emision line measurements for the Type-2 quasar
AGN1.}  {\small Line fluxes were measured from the one-dimensional
spectrum of the AGN, and errors determined from the $1\sigma$ noise
vector. For lines detected at ${\rm S\slash N} > 3$, rest-frame
equivalent widths $W_{\lambda}$ are computed.  If the continuum lies
above the $1\sigma$ noise vector these are listed as values, otherwise
we quote lower limits. The FWHM for emission lines for which our
discovery spectra had sufficient ${\rm S\slash N}$ for a measurement
(Ly$\alpha$, \ion{He}{ii}, and \ion{C}{iii}]) are also listed.  }


{\tiny
\begin{tabular*}{\textwidth}{@{\extracolsep{\fill}}cccccccccc}
\hline
\hline
Tag			  &
RA	                  &
DEC                       &
$f_{\rm Ly\alpha}$           &
$f_{\lambda}$       &
$W_{\rm Ly\alpha}$        &
$R_{\perp}$                    \\
&
&
&
($10^{-17}\,{\rm erg s^{-1} cm^{-2}}$) &
($10^{-19}\,{\rm erg s^{-1} cm^{-2} \AA^{-1}})$ &
(\AA) &
(kpc) \\
\hline
f/g QSO      & 08:41:58.470  & +39:21:20.97	&    1708.76	   &   774.23	  &	 72.61       &  	0	   \\
AGN2*	     & 08:41:58.685  & +39:21:15.00	&      41.13	   &	36.57	  &	 36.98       &         50	   \\		   
*	     & 08:41:57.982  & +39:21:25.88	&      12.07	   &	 2.04	  &	 195.0       &         58	   \\
	     & 08:41:58.872  & +39:21:14.45	&	4.14	   &	 3.38	  &	 40.28       &         62	   \\
AGN1*	     & 08:41:58.238  & +39:21:28.87	&     210.08	   &	62.35	  &	110.80       &         65	   \\
	     & 08:41:57.535  & +39:21:17.72	&	1.33	   &	 0.41	  &	106.14       &         88	   \\
*	     & 08:41:58.543  & +39:21:34.03	&      21.45	   &	14.42	  &	 48.91       &        102	   \\
*	     & 08:41:58.262  & +39:21:37.57	&      14.07	   &	 3.97	  &	116.50       &        130	   \\
AGN3*	     & 08:41:59.107  & +39:21:04.67	&      39.81	   &	37.76	  &	 34.67       &        139	   \\
*	     & 08:41:58.286  & +39:21:39.20	&	4.33	   &	 0.95	  &	148.98       &        143	   \\
	     & 08:41:58.238  & +39:20:57.05	&	3.42	   &	 5.10	  &	 22.06       &        187	   \\
	     & 08:41:59.035  & +39:21:46.27	&	1.72	   &	 1.09	  &	 52.02       &        203	   \\
	     & 08:41:57.019  & +39:21:48.45	&	1.13	   &	 0.46	  &	 80.45       &        251	   \\
Target1*	     & 08:41:58.824  & +39:21:57.42	&      17.22	   &	 5.86	  &	 96.66       &        285	   \\
	     & 08:41:59.762  & +39:21:55.52	&	0.69	   &	 0.14	  &	157.38       &        293	   \\
	     & 08:41:55.754  & +39:21:46.27	&	0.86	   &	 0.99	  &	 28.70       &        314	   \\
	     & 08:41:54.958  & +39:21:14.99	&	1.30	   &	 0.49	  &	 87.57       &        321	   \\
	     & 08:41:56.198  & +39:21:55.51	&	0.78	   &	 0.82	  &	 31.32       &        338	   \\
	     & 08:41:53.926  & +39:21:33.75	&	0.72	   &	 0.52	  &	 45.52       &        422	   \\
	     & 08:41:54.982  & +39:20:41.54	&	0.85	   &	 0.65	  &	 43.33       &        440	   \\
	     & 08:41:59.038  & +39:20:20.88	&	1.63	   &	 2.77	  &	 19.42       &        470	   \\
	     & 08:41:59.225  & +39:20:21.16	&	0.85	   &	 1.10	  &	 25.30       &        470	   \\
	     & 08:41:56.762  & +39:20:19.79	&	0.54	   &	 0.89	  &	 20.02       &        500	   \\
	     & 08:42:02.412  & +39:20:30.94	&	0.93	   &	 1.49	  &	 20.44       &        527	   \\
	     & 08:41:57.794  & +39:20:11.63	&	1.12	   &	 1.56	  &	 23.56       &        543	   \\
*	     & 08:41:52.003  & +39:21:14.16	&	5.31	   &	 3.44	  &	 50.73       &        586	   \\
	     & 08:41:52.142  & +39:21:48.16	&	1.21	   &	 0.82	  &	 48.76       &        609	   \\
	     & 08:41:55.778  & +39:20:03.75	&	0.63	   &	 0.63	  &	 33.07       &        648	   \\
	     & 08:42:02.225  & +39:20:09.46	&	0.96	   &	 0.30	  &	105.43       &        651	   \\
	     & 08:42:02.412  & +39:20:06.74	&	0.99	   &	 0.98	  &	 32.99       &        678	   \\
*	     & 08:41:52.802  & +39:20:21.41	&	7.38	   &	 3.22	  &	 75.25       &        690	   \\
	     & 08:42:05.930  & +39:22:09.65	&	0.82	   &	 1.34	  &	 20.01       &        772	   \\
	     & 08:41:58.027  & +39:23:01.05	&	0.66	   &	 0.80	  &	 27.08       &        779	   \\
	     & 08:41:52.099  & +39:20:11.62	&	1.86	   &	 1.28	  &	 47.73       &        788	   \\
	     & 08:41:51.722  & +39:20:16.51	&	0.74	   &	 0.19	  &	126.03       &        789	   \\
	     & 08:42:07.313  & +39:21:23.42	&	0.56	   &	 0.91	  &	 20.07       &        798	   \\
	     & 08:41:49.610  & +39:21:22.58	&	0.55	   &	 0.85	  &	 21.21       &        800	   \\
	     & 08:42:05.486  & +39:22:24.88	&	0.59	   &	 0.65	  &	 29.80       &        805	   \\
	     & 08:41:51.209  & +39:20:17.33	&	0.68	   &	 0.30	  &	 75.60       &        821	   \\
	     & 08:41:51.701  & +39:20:08.90	&	0.62	   &	 0.68	  &	 30.16       &        829	   \\
	     & 08:41:59.340  & +39:23:07.04	&	0.61	   &	 0.80	  &	 25.07       &        829	   \\
	     & 08:42:07.829  & +39:21:32.66	&	0.92	   &	 1.10	  &	 27.59       &        849	   \\
	     & 08:41:51.701  & +39:20:01.01	&	0.92	   &	 1.33	  &	 22.82       &        872	   \\
	     & 08:41:50.482  & +39:20:10.25	&	0.67	   &	 0.66	  &	 33.42       &        907	   \\
	     & 08:41:51.067  & +39:20:02.09	&	0.96	   &	 0.11	  &	282.90       &        907	   \\
	     & 08:41:54.461  & +39:23:11.92	&	0.91	   &	 0.97	  &	 30.68       &        936	   \\
	     & 08:41:49.982  & +39:22:30.30	&	0.66	   &	 0.85	  &	 25.42       &        937	   \\
	     & 08:42:08.441  & +39:22:05.84	&	1.71	   &	 1.33	  &	 42.09       &        964	   \\
	     & 08:42:07.826  & +39:20:09.72	&	0.73	   &	 0.69	  &	 34.54       &       1010	   \\
	     & 08:42:10.267  & +39:21:13.89	&	0.70	   &	 0.60	  &	 38.54       &       1065	   \\
	     & 08:41:51.482  & +39:23:15.72	&	0.62	   &	 0.92	  &	 22.36       &       1093	   \\
	     & 08:42:10.759  & +39:21:24.49	&	0.66	   &	 0.25	  &	 87.20       &       1109	   \\
	     & 08:42:10.010  & +39:22:10.45	&	2.21	   &	 0.27	  &	268.26       &       1110	   \\
	     & 08:42:12.590  & +39:21:39.16	&	2.09	   &	 1.83	  &	 37.44       &       1281	   \\
	     & 08:42:13.382  & +39:20:05.34	&      11.69	   &	 4.20	  &	 91.60       &       1468	   \\
	     & 08:42:11.702  & +39:23:13.80	&	1.11	   &	 0.73	  &	 49.85       &       1481	   \\
	     & 08:42:12.840  & +39:19:12.32	&	0.82	   &	 0.62	  &	 43.35       &       1638	   \\
	     & 08:42:03.629  & +39:17:19.50	&	2.27	   &	 1.16	  &	 64.47       &       1935	   \\
	     & 08:41:52.363  & +39:17:15.40	&	1.35	   &	 1.44	  &	 30.75       &       1988	   \\
	     & 08:42:03.677  & +39:17:02.64	&	1.22	   &	 0.65	  &	 61.92       &       2064	   \\
	     & 08:42:12.223  & +39:16:41.66	&	2.61	   &	 2.83	  &	 30.30       &       2502	   \\
\hline														    
\end{tabular*}}


\noindent
{\bf Table~\tablae: Properties of the candidate LAEs around the f/g
quasar.}  {\footnotesize
Here $f_{\rm Ly\alpha}$ is the \mlya\ line
flux in $10^{-17}$ erg s$^{-1}$ cm$^{-2}$, $f_{\lambda}$ is the flux
density in the continuum in units of $10^{-19}$ erg s$^{-1}$
cm$^{-2}$ \AA$^{-1}$ estimated from the $V$-band, $W_{\rm
Ly\alpha}$ is the rest-frame Ly$\alpha$ equivalent width, and
$R_{\perp}$ the proper projected distance from the f/g quasar in
kpc. LAEs are selected using the procedure described in
\S~4.1, namely they are required to have $W_{\rm Ly\alpha}>20$\AA\,
and $f_{\rm Ly\alpha} > 5.4\times10^{-18}$ erg s$^{-1}$.  Objects
which were used to measure the overdensity profile around the f/g
quasar in Fig.~\figcluster\ are denoted by a $*$ in the column Tag,
where the name of the source is also indicated. These sources had to
satisfy a brighter flux limit, namely $f_{\rm Ly\alpha} >
4.0\times10^{-17}$\,erg s$^{-1}$ cm$^{-2}$, and lie within
$2.08^{\prime}$ of the f/g quasar.}

\clearpage

{\footnotesize
\begin{tabular*}{\textwidth}{@{\extracolsep{\fill}}lccccccc}
\hline
\hline
Name & Type & RA & DEC & z & Filter FWHM (\AA) & EW$_{\rm limit}$
(\AA) & Reference\\
\hline
MRC 2048$-$272  & HzRG & 20 51 03.5 & $-$27 03 04.1 & 2.06 & 73  & 20 & 1\\ 
MRC 1138$-$262  & HzRG & 11 40 48.2 & $-$26 29 09.5 & 2.16 & 65  & 20 & 1\\
LABd05          & LAB  & 14 34 11.0 & $+$33 17 32.6 & 2.66 & 201 & 40 & 2\\
MRC 0052$–$241  & HzRG & 00 54 29.8 & $-$23 51 31.1 & 2.86 & 66  & 20 & 1\\
MRC 0943$-$242  & HzRG & 09 45 32.7 & $-$24 28 49.7 & 2.92 & 68  & 20 & 1\\ 
LAB1            & LAB  & 22 17 26.0 & $+$00 12 37.7 & 3.10 & 80  & 20 & 3\\
MRC 0316$-$257  & HzRG & 03 18 12.0 & $-$25 35 10.8 & 3.13 & 59  & 20 & 1\\
TN J2009$-$3040 & HzRG & 20 09 48.1 & $-$30 40 07.4 & 3.16 & 59  & 20 & 1\\

\hline
\end{tabular*}

\vskip 0.01in
\noindent
{\bf Table~\tabhzrg: Properties of objects analyzed in our
protocluster-LAE correlation analysis.}  {\small The nature of the
source (HzRG or LAB), coordinates, redshift $z$, the FWHM of the NB
filter used (\AA), the limiting EW of the LAE search ($W_{\rm
limit}$), and the reference for the observations are indicated.
References used are: (1) Venemans et al. 2007 \cite{Venemans07};
(2) Prescott et al. 2008 \cite{prescott08};
(3) Nestor et al. 2011 \cite{nestor11}.}

\clearpage

\clearpage
\begin{tabular*}{\textwidth}{@{\extracolsep{\fill}}cccccccc}
\hline
\hline
Ion & $\lambda$ & $\log f$
& $v_{int}$ & $W_\lambda$ & $\log N$ & [X/H]\\
& (\AA) & & ($\kms$) & (m\AA) \\
\hline
\ion{H}{1}\\
&1215.6701 &$ -0.3805$&$[-225,1449]$&$3750.0 \pm  50.0$&$19.2 \pm 0.3$&$$\\
\ion{C}{2}\\
&1334.5323 &$ -0.8935$&$[ 355,1315]$&$1110.5 \pm  34.7$&$>14.7$&$>-1.19$\\
\ion{C}{4}\\
&1548.1949 &$ -0.7194$&$[ 355,1315]$&$< 331.5$&$<13.9$&$< 0.72$\\
&1550.7700 &$ -1.0213$&$[ 355,1315]$&$< 340.4$&$<14.2$&$< 1.03$\\
\ion{N}{2}\\
&1083.9900 &$ -0.9867$&$[ 480, 850]$&$ 216.3 \pm  31.8$&$>14.3$&$>-1.00$\\
\ion{O}{1}\\
&1302.1685 &$ -1.3110$&$[ 355, 985]$&$ 672.9 \pm  14.0$&$>15.0$&$>-0.95$\\
\ion{Al}{2}\\
&1670.7874 &$  0.2742$&$[ 355,1315]$&$ 335.0 \pm 110.0$&$>12.9$&$>-0.97$\\
\ion{Si}{2}\\
&1304.3702 &$ -1.0269$&$[ 431,1315]$&$ 259.3 \pm  17.5$&$>14.3$&$>-0.81$\\
\ion{Si}{3}\\
&1206.5000 &$  0.2201$&$[ 355,1315]$&$ 951.0 \pm  43.6$&$>13.6$&$>-0.80$\\
\ion{Si}{4}\\
&1393.7550 &$ -0.2774$&$[ 355,1315]$&$<  96.2$&$<13.0$&$<-0.27$\\
&1402.7700 &$ -0.5817$&$[ 355,1315]$&$<  93.7$&$<13.3$&$< 0.02$\\
\hline
\end{tabular*}
\newline
\newline

\noindent
{\bf Table~\tababs: Absorption line measurements.}  {\small The
columns indicate the Ion, rest wavelength $\lambda$, oscillator
strength $\log f$, velocity interval $v_{\rm int}$ used for the
calculation of the rest equivalent widths, rest-equivalent width
$W_\lambda$, ionic column density $N$, and elemental abundances
[X$\slash$H] implied by each metal absorption line measurement. The
column densities $N$ reported assume the linear curve-of-growth
approximation. The metal abundances [X$\slash$H] are computed using
ionization corrections from our favored Cloudy model presented in
Section~S10, and are reported according to the standard convention of
logarithmic relative to Solar (i.e.\ 0 = solar abundance).}

\clearpage

\bibliographystyle{Science}


\end{document}